\documentclass{aa}
\usepackage{multirow}
\usepackage{amsmath}
\usepackage{graphicx,subfig} 
\usepackage{lscape}
\usepackage{longtable}
\usepackage{txfonts}
\usepackage{color}
\usepackage{booktabs}
\usepackage{soul}
\usepackage{rotating}

\makeatletter
\renewcommand*\aa@pageof{, page \thepage{} of \pageref*{LastPage}}
\usepackage{hyperref}
\hypersetup{colorlinks=true,allcolors=[rgb]{0,0,0.8}}
\hypersetup{
  colorlinks=true,
  allcolors=[rgb]{0,0,0.8},
  pdftitle={eROSITA All-Sky Survey Cluster Catalog},
  pdfauthor={Bulbul et al.},
}

\newcommand{\rosi}{eROSITA\xspace}
\newcommand{\xmm}{\textit{XMM-Newton}\xspace}
\newcommand{\chandra}{\textit{Chandra}\xspace}
\newcommand{\rosat}{ROSAT\xspace}

\newcommand{\eromapper}{\texttt{eROMaPPer}\xspace}
\newcommand{\esass}{\texttt{eSASS}\xspace}
\newcommand{\erass}{eRASS1\xspace}
\newcommand{\extlike}{$\mathcal{L}_{\rm ext}$}
\newcommand{\detlike}{$\mathcal{L}_{\rm det}$}
\newcommand{\ext}{\texttt{EXT}}
\newcommand{\mlflux}{$\texttt {ML\_FLUX}$}
\newcommand{\mbproj}{\texttt{MBProj2D}}
\newcommand{\lx}{$L_{X}$}
\newcommand{\kt}{$kT$}
\newcommand{\lognlogs}{log~$N\,-\,$log~$S$}

\usepackage{natbib,twoopt}
\usepackage{xcolor}

\makeatletter
\newcommandtwoopt{\citeads}[3][][]{\href{http://adsabs.harvard.edu/abs/#3}%
    {\def\hyper@linkstart##1##2{}%
     \let\hyper@linkend\@empty\citealp[#1][#2]{#3}}}
  \newcommandtwoopt{\citepads}[3][][]{\href{http://adsabs.harvard.edu/abs/#3}%
    {\def\hyper@linkstart##1##2{}%
     \let\hyper@linkend\@empty\citep[#1][#2]{#3}}}
  \newcommandtwoopt{\citetads}[3][][]{\href{http://adsabs.harvard.edu/abs/#3}%
    {\def\hyper@linkstart##1##2{}%
     \let\hyper@linkend\@empty\citet[#1][#2]{#3}}}
  \newcommandtwoopt{\citeyearads}[3][][]%
    {\href{http://adsabs.harvard.edu/abs/#3}
    {\def\hyper@linkstart##1##2{}%
     \let\hyper@linkend\@empty\citeyear[#1][#2]{#3}}}
\makeatother
\usepackage{graphicx}
\usepackage{multirow}
\usepackage{siunitx}  %unit
\usepackage{txfonts}

\begin{document}
\title{The SRG/eROSITA All-Sky Survey}
\subtitle{The first catalog of galaxy clusters and groups in the Western Galactic Hemisphere}

\author{E.~Bulbul\inst{1}, 
A.~Liu\inst{1}, 
M.~Kluge\inst{1}, 
X.~Zhang\inst{1}, 
J.~S.~Sanders\inst{1}, 
Y.~E.~Bahar\inst{1},  
V.~Ghirardini\inst{1},
E.~Artis\inst{1},  
R.~Seppi\inst{1},  
C.~Garrel\inst{1},
M.~E.~Ramos-Ceja\inst{1}, 
J.~Comparat\inst{1}, 
F.~Balzer\inst{1}, 
K.~B\"ockmann\inst{2},
M.~Br\"uggen\inst{2}, 
N.~Clerc\inst{3}, 
K. Dennerl\inst{1},
K.~Dolag\inst{4}, 
M. Freyberg\inst{1},
S.~Grandis\inst{4, 5},
D.~Gruen\inst{4},
F.~Kleinebreil\inst{5, 8},
S.~Krippendorf \inst{4,6},
G.~Lamer\inst{7}, 
A.~Merloni\inst{1}, 
K.~Migkas\inst{8},
K.~Nandra\inst{1},  
F.~Pacaud\inst{8},
P.~Predehl\inst{1}, 
T.~H.~Reiprich\inst{8}, 
T.~Schrabback\inst{5, 8}, 
A.~Veronica\inst{8},
J.~Weller\inst{1,4},
S.~Zelmer\inst{1}
}
\institute{
Max Planck Institute for Extraterrestrial Physics, Giessenbachstrasse 1, 85748 Garching, Germany 
\and
Hamburg Observatory, University of Hamburg, Gojenbergsweg 112, 21029 Hamburg, Germany
\and
IRAP, Universite de Toulouse, CNRS, UPS, CNES, F-31028 Toulouse, France
\and
Universit\"ats-Sternwarte, LMU Munich, Scheinerstr. 1, 81679 M\"unchen, Germany
\and
Universit\"at Innsbruck,  Institut f\"ur Astro- und Teilchenphysik, Technikerstr. 25/8, 6020 Innsbruck, Austria
\and
Arnold Sommerfeld Center for Theoretical Physics, LMU Munich, Theresienstr. 37, 80333 M\"unchen, Germany
\and
Leibniz-Institut f\"ur Astrophysik Potsdam (AIP), An der Sternwarte 16, 14482 Potsdam, Germany
\and
Argelander-Institut f\"ur Astronomie (AIfA), Universit\"at Bonn, Auf dem H\"ugel 71, 53121 Bonn, Germany
}

\date{Jan 2024}
\titlerunning{\erass\ Catalog of Galaxy Clusters and Groups }
\authorrunning{Bulbul et al.}

\abstract{
Clusters of galaxies can be used as powerful probes to study astrophysical processes on large scales, test theories of the growth of structure, and constrain cosmological models. The driving science goal of the SRG/\rosi\ All-Sky Survey is to assemble a large sample of X-ray clusters with a well-defined selection function to determine the evolution of the mass function and, hence, the cosmological parameters. We present here a catalog of 12,247 optically confirmed galaxy groups and clusters detected in the 0.2--2.3~keV as extended X-ray sources
in a 13,116~deg$^2$ region in the western Galactic half of the sky, which \rosi \ surveyed in its first six months of operation. The clusters in the sample span the redshift range $0.003 < z < 1.32$. The majority (68\%) of these clusters, 8,361 sources, represent new discoveries without known counterparts in the literature. The mass range of the sample covers three orders of magnitude from $5\times10^{12} M_{\rm sun}$ to $2\times10^{15}M_{\rm sun}$. We construct a sample for cosmology with a higher purity level ($\sim95$\%) than the primary sample, comprising 5,259 securely detected 
and confirmed clusters in the 12,791~deg$^{2}$ common footprint of \erass\ and the DESI Legacy Survey DR10. We characterize the X-ray properties of each cluster, including their flux, luminosity and temperature, 
the total mass, gas mass, gas mass fraction, and mass proxy $Y_{X}$. These are determined within two apertures, 300~kpc, and the overdensity radius $R_{500}$, and are calculated by applying a forward modeling approach with a rigorous X-ray background treatment, $K-$factor, and the Galactic absorption corrections. Population studies utilizing \lognlogs, the number of clusters detected above a given flux limit, and the luminosity function show overall agreement with the previous X-ray surveys after accounting for the survey completeness and purity through the selection function. The first \rosi\ All-Sky Survey provides an unprecedented sample of galaxy groups and clusters selected in the X-ray band. The \erass\ cluster catalog demonstrates the excellent performance of \rosi\ for extended source detection, consistent with the pre-launch expectations for the final all-sky survey, eRASS:8.}


\maketitle

\section{Introduction}
The discovery of the accelerating expansion of the Universe represents one of the most important discoveries in modern physics \citep{Riess1998, Perlmutter1999}. Over the last two decades, tremendous observational progress has been made in measuring the density of dark energy, the name given to the component responsible for the accelerating expansion, and the other features of the new standard cosmological model. Besides early Universe probes, like the Cosmic Microwave Background (CMB) measurements, and geometrical probes like the baryon acoustic oscillations (BAO) and supernovae type Ia (SNe Ia), measures of the local amplitude of matter fluctuations, as well as their growth, provide a complementary test for the standard cosmological model. However, several `tensions,' that is, disagreements between inferred model parameters, have emerged when comparing different cosmological experiments \citep[see][for a recent review]{Huterer2018, Moresco2022}.
Clusters of galaxies, the largest collapsed objects in the Universe, offer a powerful probe for testing the theories of the growth of structure, the nature of dark energy, and gravity itself. Their spatial distribution and abundance in the sky are powerful tools to constrain the parameters describing our Universe on large scales \citep[see][the most recent review on X-ray cluster cosmology]{Clerc2023}.

Clusters of galaxies and galaxy groups are filled with hot, ionized X-ray-emitting plasma enclosed within the gravitational potential of dark matter. As a substantial reservoir of baryons in the Universe, the properties of this intracluster medium (ICM) and the physical processes that drive its evolution are of great interest. 
A plethora of multi-wavelength observations are employed to study the interaction between star formation in the central galaxy and feedback from the supermassive black hole; to determine the thermodynamical properties of the ICM \citep{Ramos-Ceja2015, Sanders2018, Ghirardini2019}; to constrain models of metal production and transport \citep{Ezer2017, Mernier2018, Liu2020}, shock and cosmic ray acceleration physics \citep{vanWeeren2019,Zhang2023}, and non-thermal physical processes \citep{Hlavacek-Larrondo2015, Liu2016, Sanders2020, RojasBolivar2021}; and to perform dark matter searches \citep{Bulbul2014, Reynolds2020}. Therefore, galaxy clusters are used broadly to study cosmological and astrophysical phenomena in the Universe and the interplay of baryons embedded in deep dark matter potential wells.

Cataloging a large sample of clusters through well-planned multi-wavelength ground- and space-based surveys provides important tools for studying gravitational theory and cosmology and exploring astrophysical phenomena. Modern-day ground-based telescopes, sensitive to the optical band, are efficient at finding red galaxy concentrations in the sky, which in turn can be used to locate and catalog clusters \citep[e.g.][]{Rykoff2016, Oguri2018, Maturi2019}. These surveys can also be employed to measure the redshifts of previously identified clusters through photometric and spectroscopic observations \citep{Kirk2015, Clerc2016, Clerc2020}. While optical surveys alone are potentially powerful in constructing complete samples of galaxy cluster catalogs, they suffer from projection effects that could lead to a high level of contamination in the samples and result in biases in cosmological experiments \citep[see][]{Costanzi2019, Grandis2021, Myles2021}.

As the majority of baryonic mass in clusters is in the ICM, cluster surveys optimized to find clusters through ICM emission with X-ray and Sunyaev Zel'dovich (SZ) Telescopes offer an alternative efficient detection method with a much better-defined selection function. SZ surveys, taking advantage of inverse Compton scattering of CMB photons, are used to construct mass-limited samples of galaxy clusters \citep{Planck2016, Bleem2020, Hilton2021}. On the other hand, surveys performed in the X-ray band are sensitive to the direct X-ray emission of the ICM in the keV band and have the potential to yield the largest samples of galaxy groups and clusters covering a wide redshift and mass range. 

The first imaging X-ray all-sky survey has been conducted with the \rosat\  \citep{Truemper1982, Voges1999} satellite. Several catalogs of clusters have been compiled from the \rosat\ All-Sky survey (RASS) data, for instance, in the Northern (NORAS) \citep{Boehringer2000_noras} and Southern (REFLEX) hemispheres \citep{Boehringer2004_reflex} with the aid of the dedicated optical follow-up programs. More recently, several surveys with smaller sky coverage have been performed, for instance, 2~deg$^{2}$ \chandra\ COSMOS \citep{Finoguenov2007}, $\sim$50~deg$^{2}$ \xmm\ XXL \citep{Pierre2016}, finding several hundred to thousand clusters of galaxies. These surveys build a bridge between the \rosat\ surveys and the new generation, wide area X-ray surveys.
  
The soft X-ray telescope on board the Spectrum-Roentgen-Gamma (SRG) mission \citep{Sunyaev2021}, the extended ROentgen Survey with an Imaging Telescope Array (\rosi), was launched on July 13, 2019, from the Baikonur Cosmodrome in Kazakhstan \citep{Predehl2021}. \rosi\ with its large collecting area in the soft X-ray band (1365~cm$^2$ at 1~keV) and its moderate angular resolution averaged over the FoV ($\sim30\arcsec$~ half-energy width at 1.49~keV), provides an unprecedented view of the X-ray sky. The primary science goal of \rosi\ is to construct the most extensive samples of clusters of galaxies with a clean and well-understood selection function. When completed and complemented with weak lensing data for the mass calibration, the cosmological parameters obtained through cluster abundances will reach a statistical power complementary to CMB probes \citep{Merloni2012}. The \rosi\ X-ray All-Sky survey is complemented by large photometric and spectroscopic optical follow-up programs with the Sloan Digital Sky Survey (SDSS-V) and 4MOST for redshift measurements to maximize the science return for \rosi\ \citep{Kollmeier2017, Finoguenov2019}.

The first cluster catalogs of \rosi\ are compiled from the \rosi\ Final Equatorial-Depth Survey (eFEDS). eFEDS \citep{Brunner2022} is a performance verification survey executed over a 140~deg$^{2}$ area with a nominal depth of around 2.2ks, similar to the final depth of the full 8-pass eROSITA all-sky survey (eRASS:8). 
The eFEDS science program has demonstrated the survey capabilities of \rosi\ and returned high-impact science that can be achieved with X-ray-selected samples once the selection effects are accounted for properly. In the eFEDS field, we detect 542 cluster candidates above the flux of $10^{-14}$~erg~s$^{-1}$~cm$^{-2}$. Four hundred seventy-seven are confirmed with photometric and spectroscopic surveys in the redshift range of 0.01 to 1.3 \citep{Liu2022, Bulbul2022, Klein2022}. The characterization of the dynamical state of the populations of clusters detected by \rosi\ demonstrates a smooth transition from the cool core to non-cool core states and from relaxed to disturbed states \citep{Ghirardini2022}, and shows no significant selection biases \citep{Seppi2023, Ramos2022}. We provide a proof-of-concept study of scaling laws between X-ray observables and cluster mass by incorporating Hyper Suprime-Cam (HSC) weak lensing data to be used in the exploitation of the \rosi\ data in cosmology \citep{Bahar2022, Chiu2022}. An alternative cluster mass estimation method through neural networks is developed and successfully applied to the data \citep{Krippendorf2023}. Taking advantage of multi-wavelength coverage with LOFAR and HSC, we study the distribution of large-scale structure \citep{Ghirardini2021, Liu2022}, X-ray luminosity and radio power correlation for constraining AGN feedback \citep{Pasini2022}, galaxy distribution at the splashback radius of eFEDS clusters \citep{Rana2023}, and difference of gas distribution in optically and X-ray-selected groups and clusters \citep{Ota2023, Popesso2023}.

After the performance verification phase was completed, \rosi\ started its All-Sky Survey program on December 12, 2019. The first All-Sky Survey was successfully executed on June 11, 2020, after 184 days. The data and publication rights of the Western Galactic half of the \rosi\ All-Sky survey (359.9442~deg~$> l >$~179.9442~deg) belong to the German \rosi\ consortium. In this work, we present the catalog of the galaxy groups and clusters of galaxies and their X-ray properties detected and confirmed in the first Western All-Sky Survey of \rosi\ (\erass\ hereafter). We base our catalog on the detection properties of the X-ray sources in the soft band (0.2--2.3~keV) provided in \citet{Merloni2024}. In the cluster catalog, we aim to maximize the discovery space of \rosi\ by applying a less strict X-ray extent likelihood selection (\extlike~$>$~3\footnote{The symbols \extlike\ and \detlike\ correspond to \texttt{EXT\_LIKE} and \texttt{DET\_LIKE\_0} in \citet{Merloni2024}.}; see Section~\ref{sec:mainsample}), as suggested in \citet{Bulbul2022}, and relying on the optical follow-up observations for cleaning of the sample as presented in \citet{Kluge2024}. 
In addition to the primary catalog, we construct a sample to be employed in the study of cosmology (cosmology sample, hereafter). The cosmology sample, assembled with a stricter X-ray selection (\extlike~$>$~6), reaches a higher purity level and relies less on optical information. In the follow-up work, we provide the selection function \citep{Clerc2024} of both samples, morphological and thermodynamical properties of the clusters and groups \citep{Sanders2024, Bahar2024}, supercluster and large-scale structure studies \citep{Liu2024, Zhang2024}, scaling relations between X-ray observables and mass (Ramos-Ceja et al. 2024 in prep., Pacaud et al. 2024 in prep.), and weak lensing mass calibration for cosmology \citep{Grandis2024, Kleinebreil2024}. Cosmological studies exploiting this sample are presented in \citet{Ghirardini2024}, \citet{Artis2024}, \citet{Seppi2024}, and Garrel et al. (2024, in prep.). The cross-calibration studies for selected clusters are presented in \citet{Migkas2024}. We will provide another catalog of clusters, and groups misclassified in the \erass\ point source sample and discuss the selection and the identification method in Balzer et al. (2024, in prep.).

This paper is organized as follows: in Section~\ref{sec:catalog}, we describe the selection, cleaning, and optical confirmation of the catalog. In Section~\ref{sec:mainsample}, we provide the primary cluster catalog and cosmology sample and their sample completeness and purity. Section~\ref{sec:crossmatch} presents the cross-matches with public surveys to identify the unique detection and discoveries in \erass. X-ray properties of the clusters and groups in the catalog are given in Section~\ref{sec:xrayprop}. Additionally, the sample properties, such as \lognlogs, the number of clusters detected above a certain flux, and luminosity function and distribution, are presented in Section~\ref{sec:sampleprop}. Quoted error bars correspond to a 1-$\sigma$ confidence level unless noted otherwise.

\section{Construction of the \erass\ galaxy groups and clusters catalog}
\label{sec:catalog}
\subsection{Source detection}
\label{sec:DetectionChain}

A detailed description of the \erass\ source catalogs is provided in \cite{Merloni2024}. Here, we summarize the main features. The \erass\ data is processed with the standard \rosi\ Science Analysis Software System  \citep[\esass,][]{Brunner2022}\footnote{Version {\tt eSASSusers\_211214\_0\_4}.}. The data (calibrated event lists, images, exposure maps, etc.) are sorted into overlapping sky tiles of size $3.6\,\times\,3.6$~deg$^2$. 
In comparison with the data processing version c001 from the \rosi\ Early Data Release\footnote{\url{https://erosita.mpe.mpg.de/edr/}}, the event calibration in the 010 processing of \erass\ has a more robust telescope module (TM) specific noise suppression of double and triple events, a better computation of the subpixel position, a corrected flagging of pixels next to bad pixels, and improved accuracy of projection \citep[for further details see][]{Merloni2024}. 
The event lists are filtered after determining good time intervals, dead times, corrupted events and frames, invalid patterns, bad pixels, and all events outside a circular detection mask of radius $0.516$~deg. Using star-tracker and gyro data, celestial coordinates are assigned to the reconstructed X-ray photons, which can then be projected into the sky to produce images and exposure maps. All valid pixel patterns are selected, namely, single, double, triple, and quadruple events. All the sky tile images and exposure maps have a pixel size of $4\arcsec$ and a size of $3240$~pixels~$\times$~3240~pixels.

The corresponding tasks of the \esass\ package apply the source detection and characterization to each of the $3.6\times3.6$~deg$^2$ sky tiles images. The source detection first consists of running a sliding-cell algorithm over the data to determine a source candidate list. A second step uses this source list to create a background map. This background is used again in the sliding-cell task to build a source list. This process is iterated two times to improve the background determination and the separation of the sources. A point spread function (PSF) fitting algorithm characterizes each source in the final list; namely, the source characterization algorithm selects source candidates according to the statistics of fitting the sources with a PSF-convolved model ($\beta$-model or $\delta$-function). The source characterization algorithm adopted a PSF-fitting radius of 15~pixels, a multiple-source searching radius of 15~pixels ($=60\arcsec$), a detection likelihood threshold of 5, an extent likelihood threshold of 3, an extended range between 2 and 15~pixels, and a maximum of four sources for simultaneous fitting. The PSF is folded with the $\beta$ model, which has a core radius ($r_{c}$) value to indicate the extent of the source. The core radius is set free to vary between $8\arcsec$ and $60\arcsec$ for extended sources. The source detection algorithm used here, \texttt{ermldet}, is tested on the \rosi\ survey simulations and compared to the core-excised wavelet detection algorithm developed by \citep{Kaefer2020}. Comparing the completeness curves, we find that \texttt{ermldet} performs as well as the wavelet detection in the redshift range of interests for cosmology studies. It performs better in higher redshifts ($z>0.8$) than the core-excised wavelet detection algorithms, where most source counts are removed when the core is excised, resulting in a non-detection.

\begin{figure}
\begin{tabular}{c}
\includegraphics[width=0.5\textwidth]{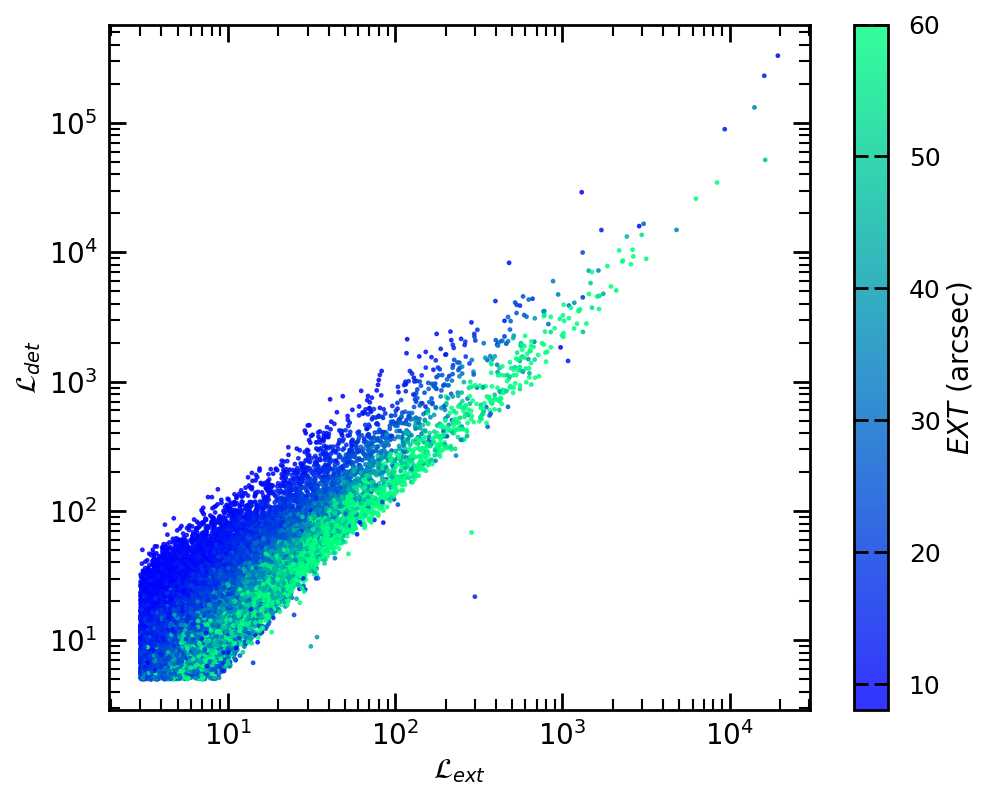} 
\end{tabular}
\caption{Extent likelihood \extlike\ and detection likelihood \detlike\ distribution of all extended sources. About 2\% of the \erass\ X-ray source catalog, totaling up to 26,682 sources, are extended with \ext~$>\,0$.\label{fig:extl_detl}}
\end{figure}

Similar to eFEDS, two \erass\ source catalogs are presented by \cite{Merloni2024}: a soft band catalog created in the 0.2--2.3~keV range and multi-band detections used to define a hard band catalog in the 2.3--5.0~keV band. Taking advantage of \rosi's superb sensitivity in the soft band, we base our cluster catalog on the single and 0.2--2.3~keV \erass\ catalog. In the main soft X-ray catalog, \cite{Merloni2024} detected 1,277,486 X-ray sources with detection likelihood (\detlike) greater than 5. Of these sources, 26,682 ($\sim$2\%) are extended sources with extent (\ext) greater than 0 and extent likelihood (\extlike) greater than 3. This selection criterion is based on the results presented in \citet{Bulbul2022} to keep high redshift clusters and compact galaxy groups with comparable angular sizes to the PSF in the cluster catalog, maximizing the source discovery potential of \rosi. The primary \erass\ galaxy cluster catalog we present in this work is selected from these 26,682 extended sources. The extent likelihood and detection likelihood distribution of all extended sources are shown in Fig.~\ref{fig:extl_detl}. The figure shows the correlation between \detlike and \extlike, as the high \extlike clusters are detected with high signal-to-noise and detection likelihood.

\begin{figure*}
\includegraphics[width=0.65\textwidth,trim={50 60 60 100}, clip]{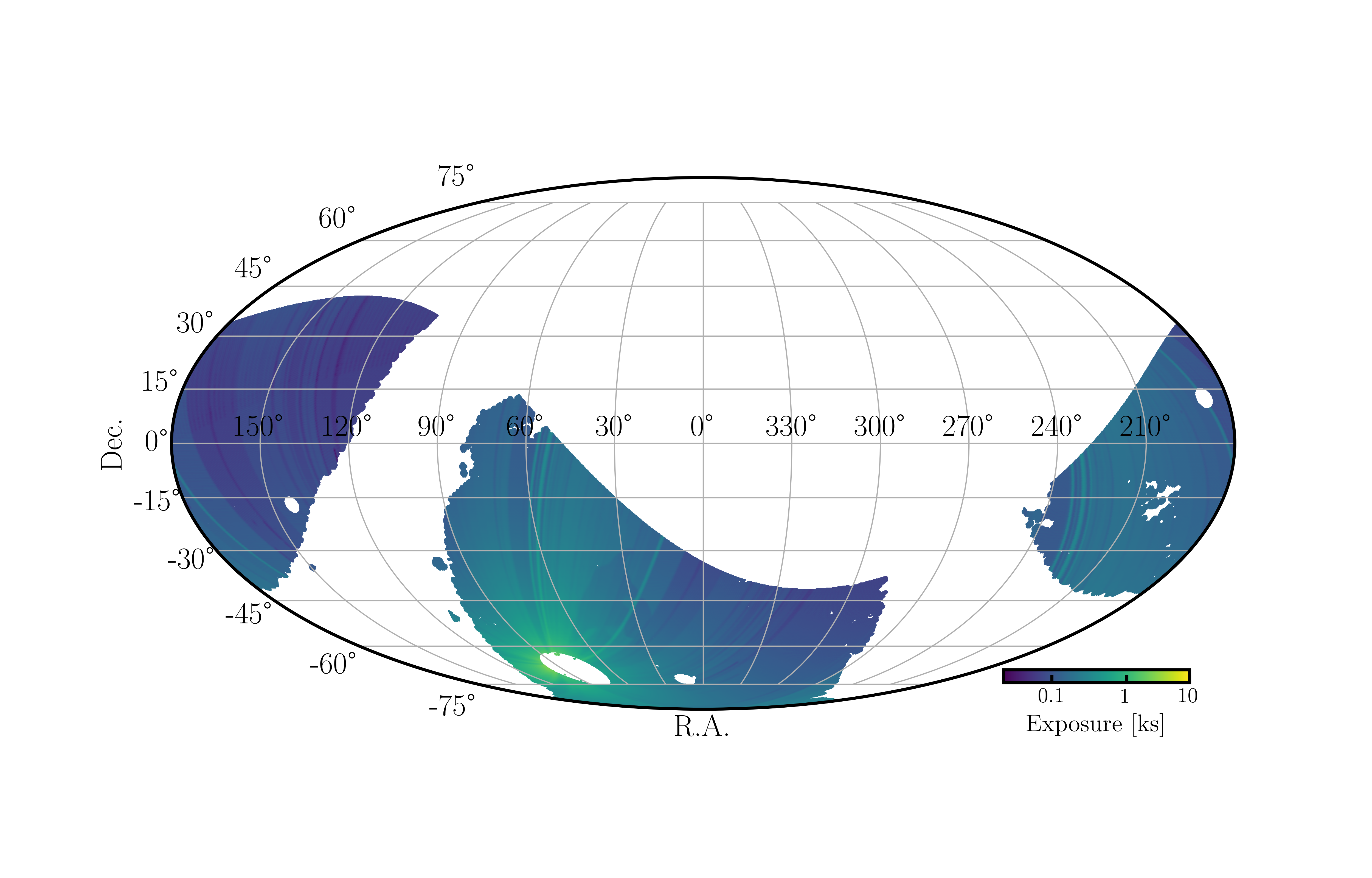}
\includegraphics[width=0.34\textwidth]{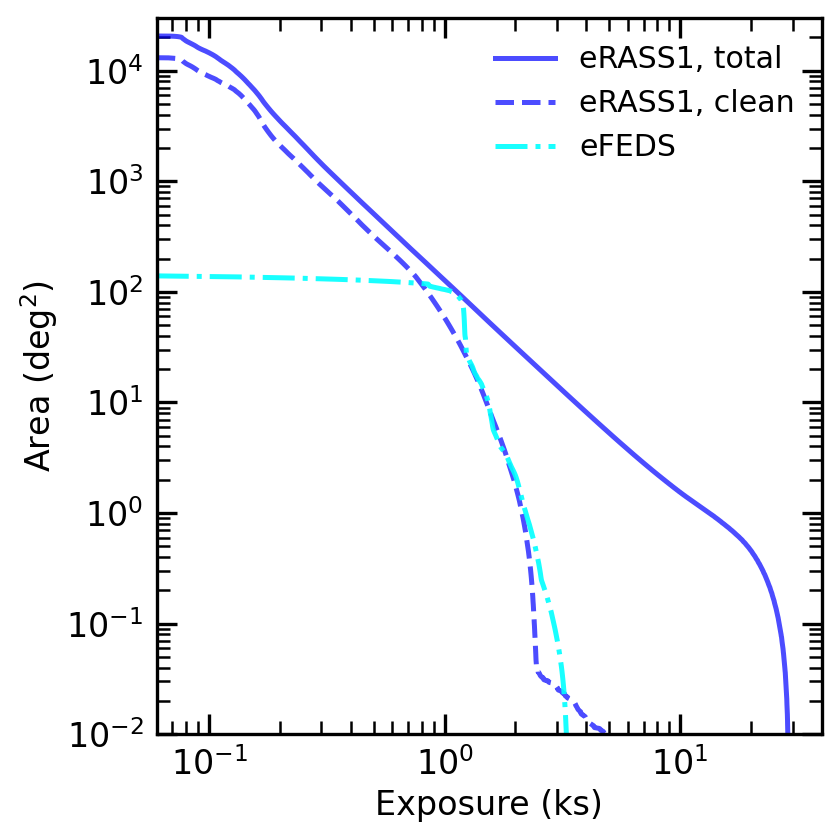} 
\caption{\label{fig:surveyarea} Survey area and exposure information of the \erass\ cluster catalog. Left panel: Vignetted exposure map in the 0.2--2.3~keV band after masking regions. The clean exposure time changes from $\sim$0.1~ks to 1.2~ks in the survey. Right panel: Cumulative \erass\ sky coverage as a function of exposure in the 0.2--2.3~keV band. Blue solid and dashed curves show the data before and after cleaning. The curve in cyan shows the depth of the eFEDS survey, as a comparison \citep{Brunner2022}.}
\end{figure*}
\subsection{Catalog cleaning}
\label{sec:cleaning}
A few sky areas should be excluded for cluster detection, mostly those contaminated by Galactic or other foreground X-ray sources. Our strategy to obtain the `clean' sky for the \erass\ galaxy cluster catalog is described as follows. First, since the optical follow-up of the cluster catalog is performed using the DESI Legacy Survey (LS) Data Release 9 (DR9) and Data Release 10 (DR10) data, we consider only the sky area within LS DR9 and DR10 footprint. Consequently, we eliminate many bright, known, extended non-cluster X-ray sources such as the Galactic disk, the Large and Small Magellanic Clouds (LMC and SMC), and most of the bright supernova remnants (SNRs) in our Galaxy. Additionally, we exclude several other geometrical regions containing high Galactic-latitude SNRs, including SN1006 \citep{Winkler2014}, Hoinga \citep{Becker2021}, G296.5+10.0 \citep{Giacani2000}, and G330.0+15.0 \citep{Leahy2020}. Some known X-ray binaries \citep{Liu2006, Liu2007} and globular clusters \citet{Harris2010} are also excluded, although they only occupy a negligible sky area ($<$1~deg$^2$). The Virgo cluster region is excluded due to its large extent in the sky, but in this case, it is added back into the cluster catalog afterward. After performing the above cleaning procedures, the effective survey area is 13,116~deg$^2$, shown on the exposure map in Fig.~\ref{fig:surveyarea}. 

The \erass\ vignetted exposure time in the soft band ranges from $<\,0.1$~ks to $\sim$~29~ks, from the ecliptic equatorial region to the ecliptic pole. The latter is located close to the LMC. However, a significant fraction of the pole area with deep exposure is excluded in this work due to the lack of LS optical coverage in this region. This effect can be seen from the exposure map and the depth curve. The depth curve of eFEDS with its uniform vignetting-corrected effective exposure of 1.2~ks is plotted in the right panel of Fig.~\ref{fig:surveyarea} for comparison.
Higher-exposure regions ($>\,$1.2~ks) in eFEDS are due to the waiting time of the spacecraft before inverting its scanning direction in the scan observing mode. Through the comparison between \erass\ and eFEDS, we note that \erass\ has a comparable survey coverage as eFEDS in the deeper exposures ($>\,$1~ks) and is more than two orders of magnitude larger than eFEDS in the shallower regime as shown in the right panel of Fig.~\ref{fig:surveyarea}.

\begin{figure*}[ht]
\begin{centering}
\includegraphics[width=0.35\textwidth]{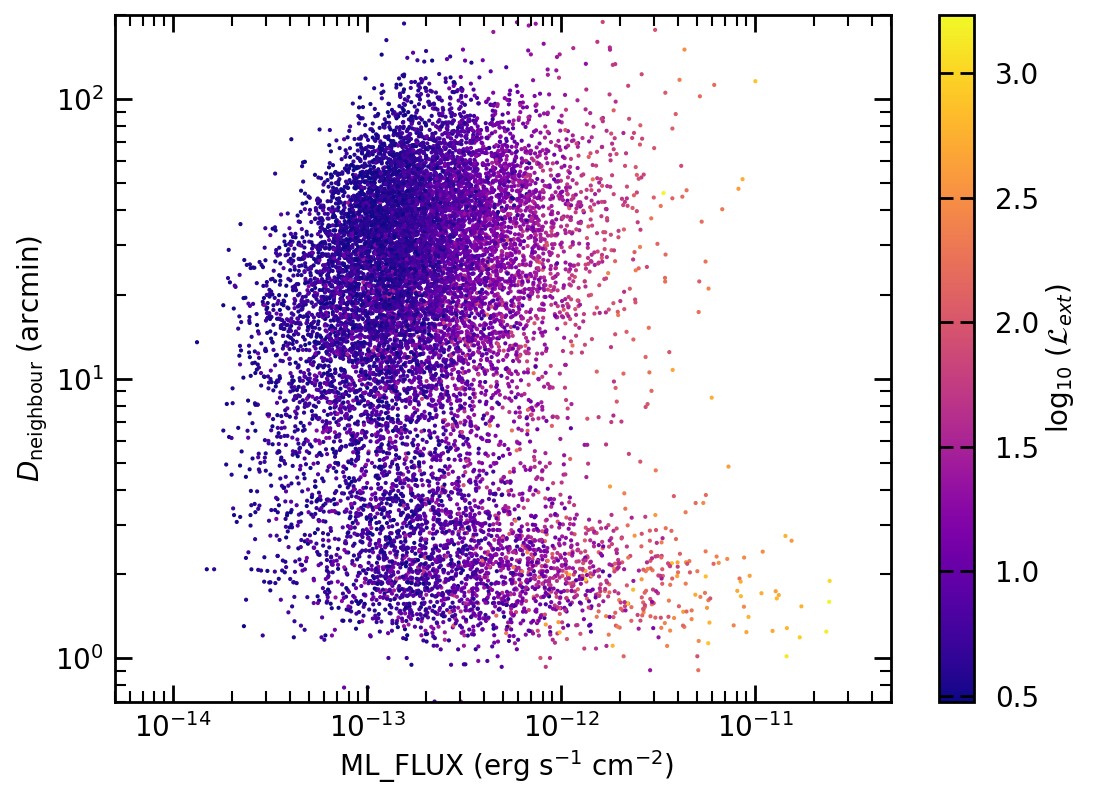} 
\includegraphics[width=0.35\textwidth]{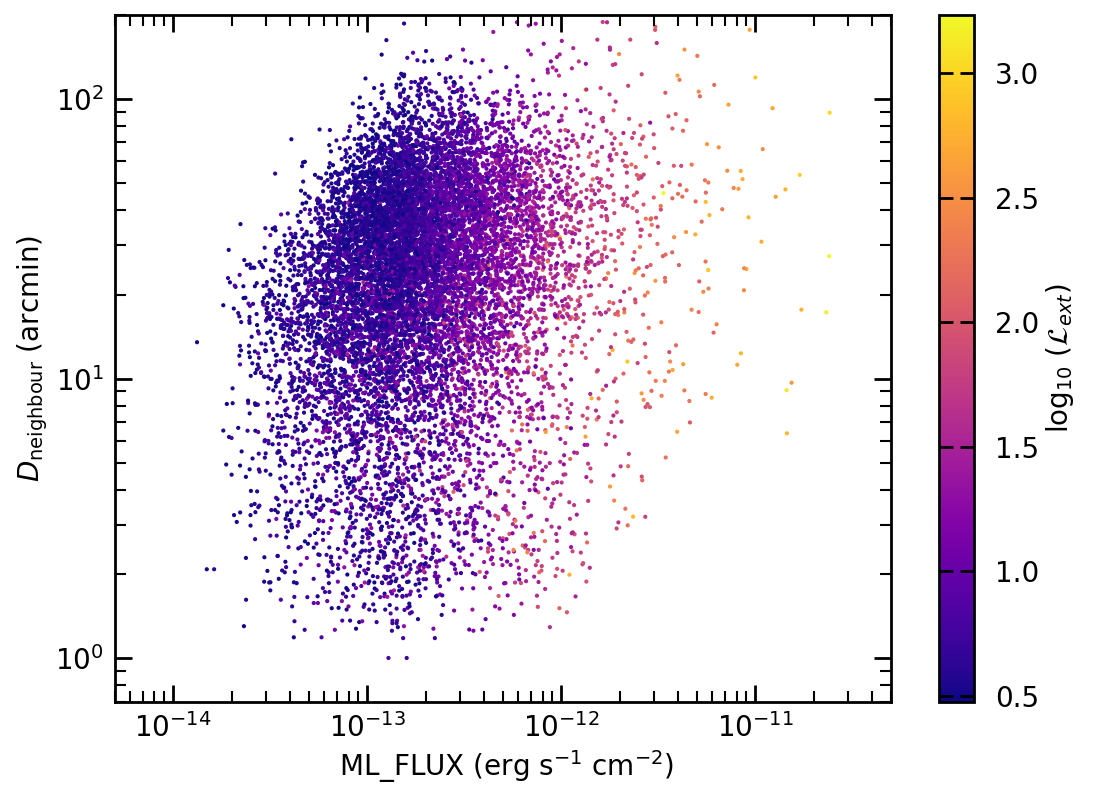}
\includegraphics[width=0.278\textwidth,trim={0 -20 0 0}, clip]{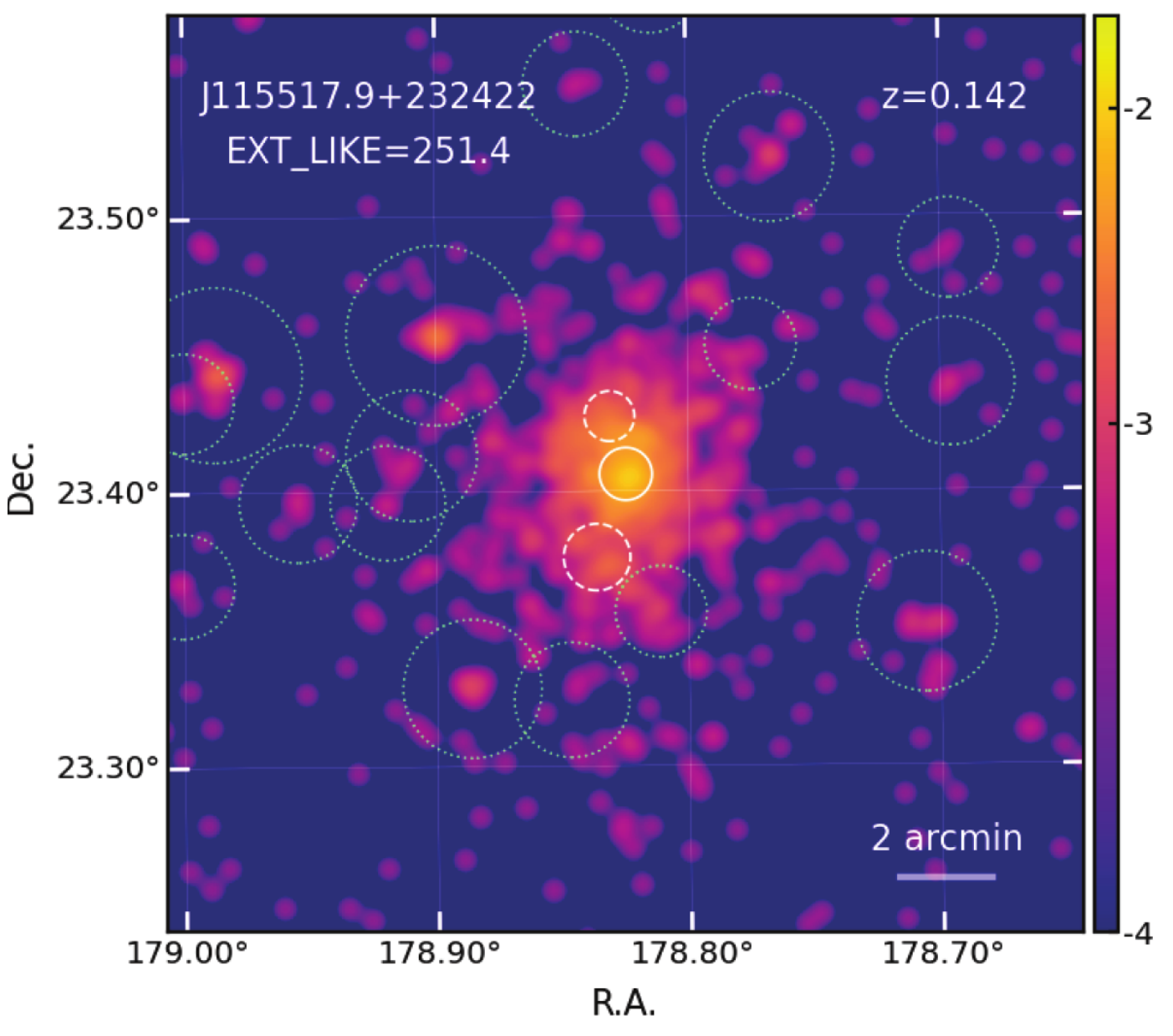} 
\caption{Distance to the closest neighbor for each source is plotted against its \mlflux, demonstrating the general properties of the split sources on the sample level before and after cleaning on the right and middle panels. The color code denotes the extent likelihood \extlike. Note the removal of the sources in the lower right corner. On the right panel, the figure shows an example of the split sources (dashed white circles) in a nearby cluster 1eRASS~J115517.9+232422 at the redshift of 0.14. After cleaning, the source center is shown in a solid white circle. The circles in green are the point sources detected by \rosi\ in the field. \label{fig:split}}
\end{centering}
\end{figure*}

The raw source catalog contains a class of spurious sources, which are aggregations close to bright extended sources, including galaxy clusters \citep[see][for further details]{Merloni2024}. The detection algorithm might split such sources into multiple contiguous sources. One clear example of this is the Virgo cluster. Due to the large extent of the cluster, the source detection algorithm splits the sources into a large number of point and extended sources. We apply the following approaches to the extended source catalog to exclude these spurious sources. First, we identify the extended sources which are located within $\sim 0.5\times R_{500}$\footnote{In this work, $R_{500}$ is defined as the overdensity radius within which the local density of the cluster is 500 times the critical density of the Universe at the cluster's redshift.} of any previously known X-ray clusters, and flag all of them as spurious sources except for the one that has the minimum distance to the known cluster. The split sources found within $R_{500}$ are marked with the flag "{\texttt SP\_GC\_CONS}" in the soft band X-ray catalog of \citet{Merloni2024}. The list of known X-ray clusters we used in this approach is compiled from the publicly available cluster catalogs, for instance, MCXC \citep{Piffaretti2011}, a compilation which includes most known ROSAT clusters, X-CLASS \citep{Koulouridis2021}, XXL \citep{Adami2018}, XCS \citep{Mehrtens2012}, and eFEDS \citep{Liu2022}. We removed $\sim1500$ sources in this step. In the second step, we search for source pairs with a distance smaller than two times the sum of the source extensions, that is, $2\times(\ext_1+ \ext_2)$. The source with lower \extlike\ in these source pairs is removed. Around $\sim2000$ sources are removed in this second step. As the third and final step, the extended source catalog is inspected visually to remove any remaining sources that are false detections due to the extended emission of a bright source. $\sim50$ more sources are further removed in this step. The effect of the above cleaning approaches can be seen in Fig.~\ref{fig:split}, where we plot the distance to the closest neighbor for each source against its \mlflux\ (the raw flux obtained from the detection chain). In the left panel of Fig.~\ref{fig:split}, the dense distribution of sources lies in the lower-right corner, with higher fluxes but small distances from their neighbors. This contradicts the fact that cluster number density decreases with increasing flux. Therefore, they are likely `split' sources needing removal. The middle panel of Fig.~\ref{fig:split} shows that our cleaning method effectively removed all such split sources after applying the above cleaning methods. A clear case of split extended sources is shown in Fig.~\ref{fig:split}, right panel. The source detection algorithm detected the cluster outskirts as independent extended X-ray sources; our split cleaning approach finds these cases (shown in dashed white circles) and removes them from the catalog and further processing. After cleaning, we keep one survivor detection for each split source in the catalog, and the centroids are set to the coordinates of the survivor detection in these cases. The same figure shows the corrected source center in a solid white circle.

In summary, 3597 sources are identified as `split' sources, with an additional 8267 sources located within the masked regions or outside the LS footprint. These sources are excluded from the catalog. Therefore, we have 14,818 extended sources in the cleaned catalog, which we attempt to identify with optical data in the next step.

\subsection{Optical identification of \erass\ clusters}
\label{sec:opticalconf}

This section briefly describes the optical identification process of \rosi\ extended sources as clusters of galaxies. Details are provided in \citep{Kluge2024}. The optical identification is performed using the \eromapper\ algorithm \citep{IderChitham2020}, a highly parallelized version of the red-sequence matched-filter Probabilistic Percolation cluster finder {\texttt redMaPPer} \citep{Rykoff2014, Rykoff2016}, adapted for the identification of X-ray-emitting clusters and groups near \rosi\ X-ray centroids. To identify overdensities of passive red galaxies around these locations, we utilize the public DESI Legacy Surveys Data Release North (LS DR9N) at ${\rm Dec} \gtrsim 32.5\deg$ and Data Release (LS~DR10) at ${\rm Dec} \lesssim 32.5\deg$ \citep{Dey2019}\footnote{\url{https://www.legacysurvey.org/}}.

We begin with the 14,818 \erass\ X-ray extended sources (\extlike~$>3$) within the LS footprint after applying the geometrical masking, as described above. 
Of these cluster candidates, we find optical counterparts for 12,554 using \eromapper. The algorithm calculates richnesses, optical centers, and other optical properties, as well as photometric cluster redshifts $z_\lambda$ for all optically identified clusters based on the colors of their member galaxies on the red sequence. For this task, we use LS~DR9N and DR10 $g,r,z$ filter bands and (in rarer cases) DR10 $g, r, i, z$ filter bands in the photometric redshift range of $z_\lambda<0.8$. In contrast, for the high redshift range beyond $z_{\lambda}>0.8$, we use the LS~DR10 $g,r, i, z, W1$ and (in rarer cases) $g, r, z, W1$, and LS DR9N $g, r, z, W1$ filter bands to measure the photometric redshifts more accurately. 

As an alternative to the photometric redshifts, we use, when possible, more precise spectroscopic redshifts $z_{\rm spec}$ and adopt literature redshifts $z_{\rm lit}$ of matched clusters from public catalogs. The best redshift $z_{\rm best}$ is then chosen using a prioritization scheme. The highest priority is assigned to $z_{\rm spec}$ because it has the lowest uncertainty. The spectroscopic cluster redshifts are calculated using a compilation of public spectroscopic galaxy redshifts and a dedicated spectroscopic follow-up program with the Hobby Eberly Telescope (Balzer et al., in prep.). Photometric redshifts have the second-highest priority, while literature redshifts are only assigned to clusters without optical identification by \eromapper\ 
or with photometric redshifts outside of the red-sequence calibration range. We note a few exceptions to this prioritization scheme, described in \cite{Kluge2024}.
The red sequence in our follow-up algorithm \eromapper\ is not calibrated below redshifts $z<0.05$. The extended galaxies, members of nearby groups, and clusters are excluded in the LS photometry and, hence, are missing in our galaxy catalogs, which we utilize to calibrate \eromapper. In this redshift range, the clusters and groups are inspected visually, and their photometric redshifts are replaced with the redshifts found in the literature if they are more accurate \citep[see][for further details]{Kluge2024}. 

\begin{figure*}
\begin{center}
\includegraphics[width=0.32\textwidth,trim={0 40 60 60}, clip]{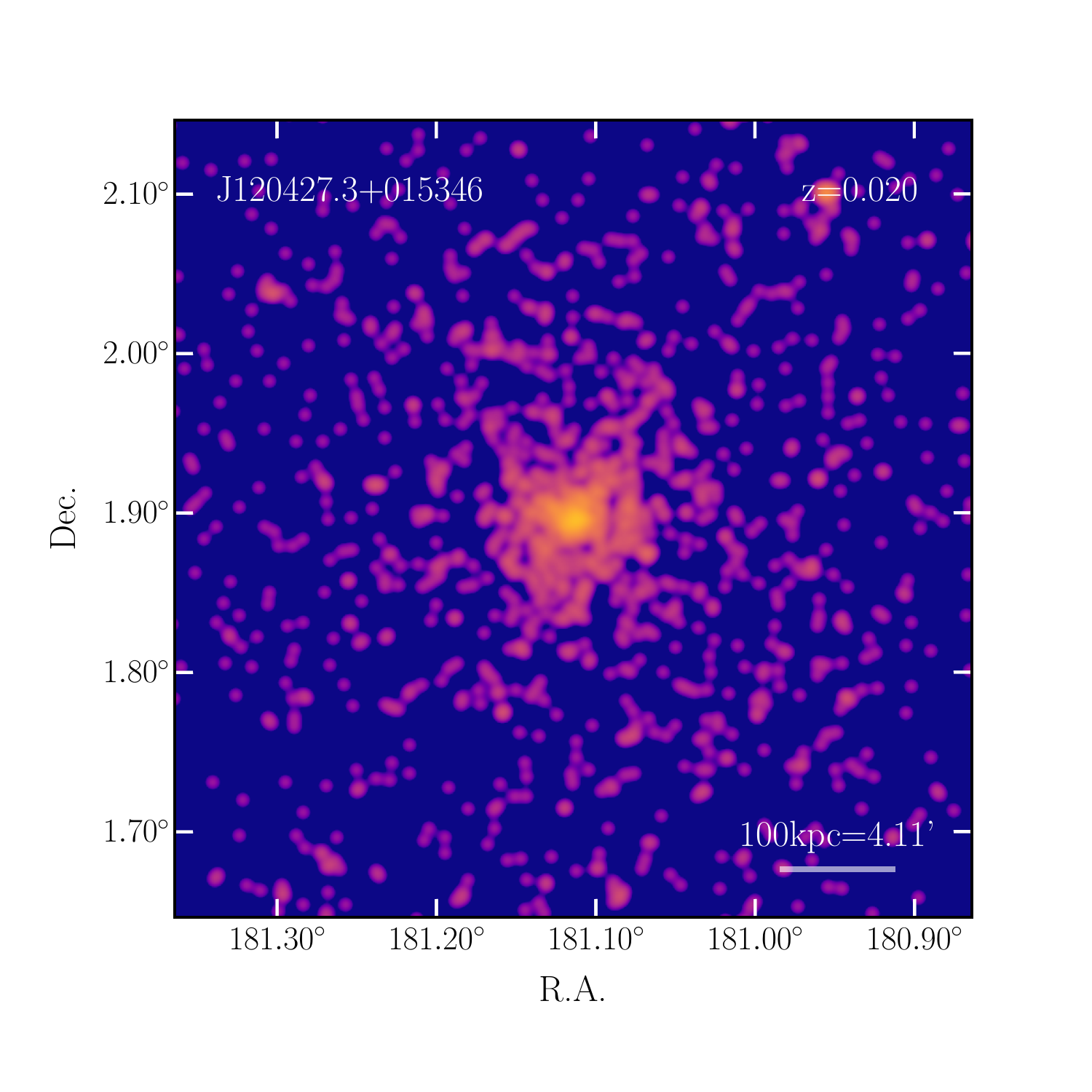} 
\includegraphics[width=0.32\textwidth,trim={0 40 60 60}, clip]{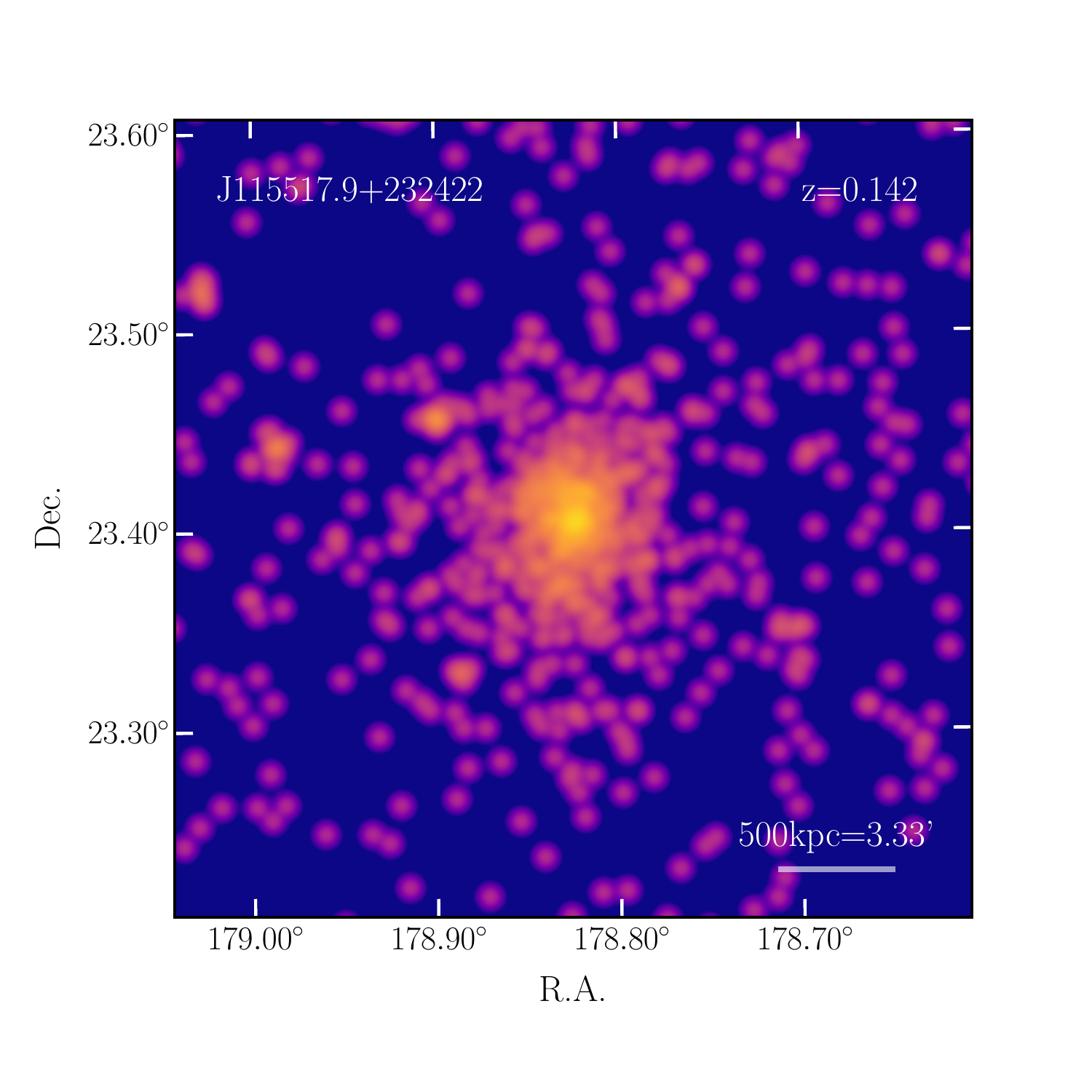} 
\includegraphics[width=0.32\textwidth,trim={0 40 60 60}, clip]{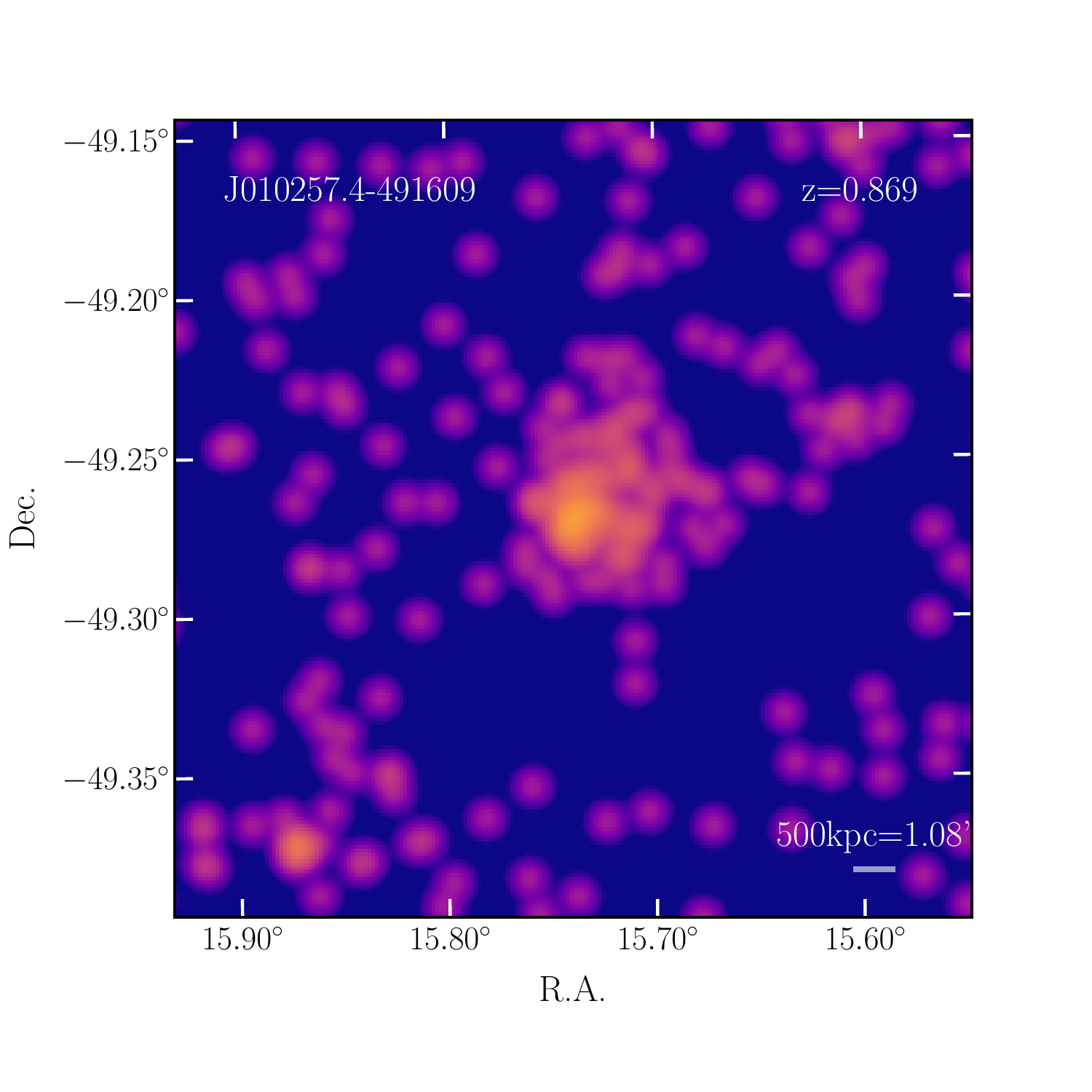} 
\includegraphics[width=0.32\textwidth,trim={0 40 60 60}, clip]{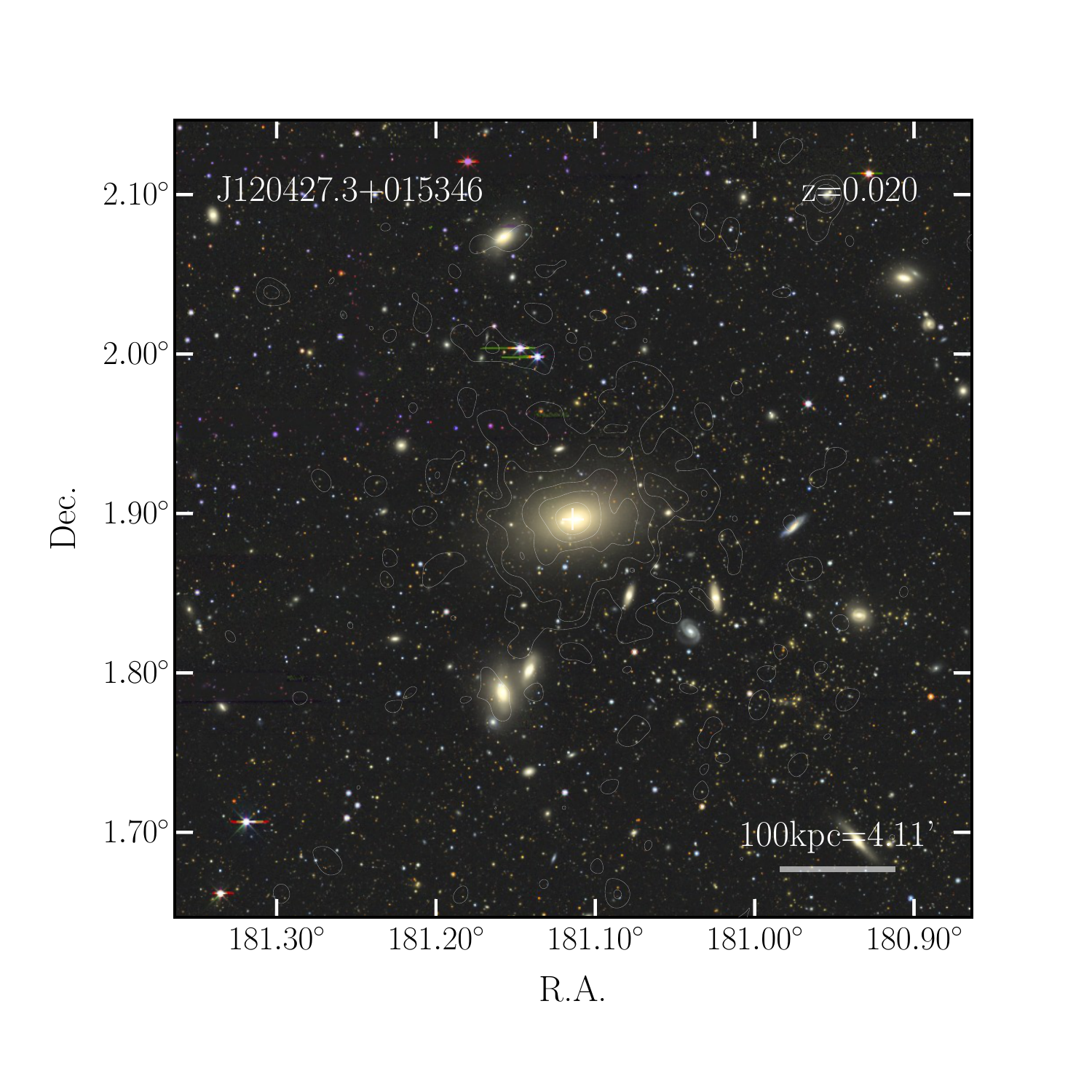} 
\includegraphics[width=0.32\textwidth,trim={0 40 60 60}, clip]{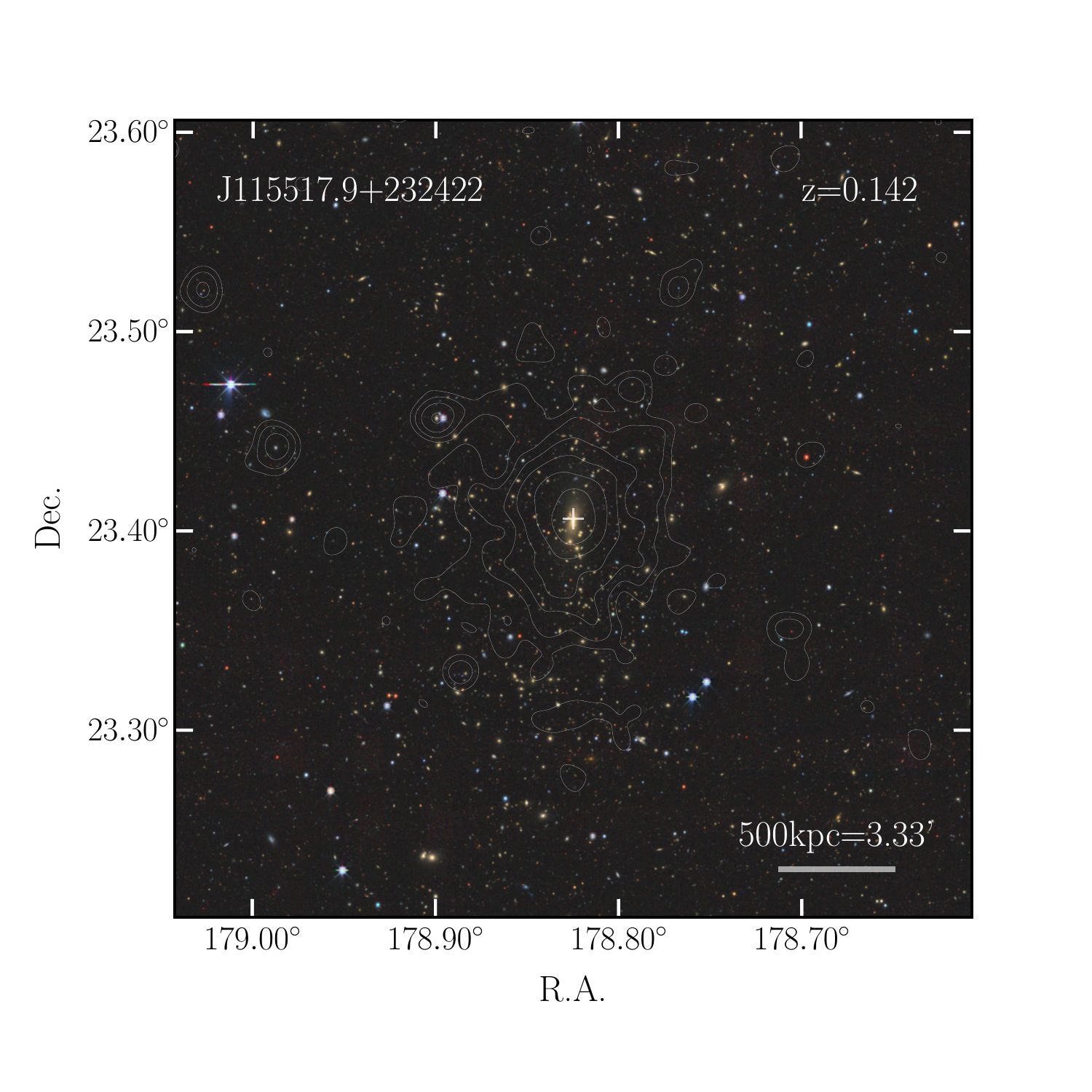} 
\includegraphics[width=0.32\textwidth,trim={0 40 60 60}, clip]{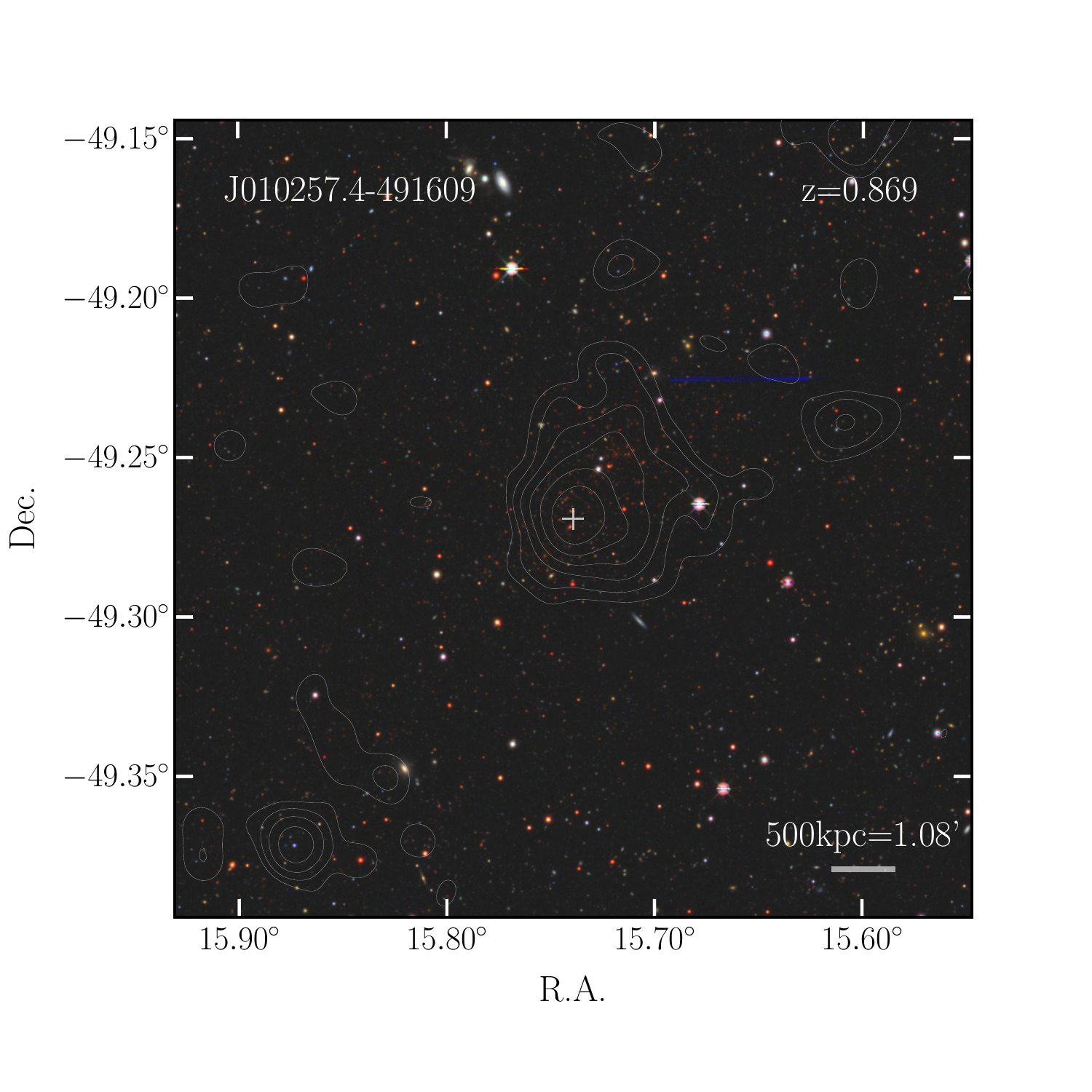} 
\end{center}
\caption{Examples of \erass clusters. From left to right: 1eRASS~J120427.3+015346 (MKW4) at $z=0.020$, \erass~J115517.9+232422 at $z=0.142$, and \erass~J010257.4-491609 ({\sl El~Gordo}) at $z=0.869$. The upper panels show \erass X-ray image in the 0.2--2.3~keV band, smoothed with a Gaussian of $\sigma=12\arcsec$. LS~DR10 {\sl grz} images are shown in the lower panels, overlaid with X-ray contours. \label{fig:images}}
\end{figure*}

In total, we optically identify 12,554 clusters and groups, of which 11,888 are in LS~DR10 (98.25\%) and 212 in the LS DR9N (1.75\%) area. An additional 151 sources are matched with known cluster catalogs from the literature, and we include their redshifts in the catalog. With these additional clusters, we identify 12,705 clusters and groups. Fig.~\ref{fig:images} shows some example X-ray and LS~DR10 images of clusters found using \eromapper\ at low, intermediate, and high redshifts. We note that the redshifts provided in this catalog are heliocentric.

\subsection{Contamination estimation via a mixture model}
\label{sec:cleaning}

Our realistic \erass\ simulations \citep{Seppi2022} show that identifying contaminants purely based on X-ray selection is challenging. We developed a mixture model that considers both X-ray and optical properties to estimate the overall contamination fraction in the \erass\ cluster catalog. Bright AGN and spurious sources due to background fluctuations are expected to constitute most of the contamination in our X-ray extended source catalogs \citep{Seppi2022}. Additionally, on the optical side, line-of-sight projections and misidentification of stars as galaxies can lead to spurious overdensities of apparently red extended objects near the \erass\ extended sources. The mixture model identifies contaminants on both X-ray and optical sides by comparing the distribution of redshift--richness--X-ray count rate properties of random sources (RS) and AGN to the distribution of the 12,705 clusters.

Random line-of-sight projections and the positions of spurious X-ray sources are expected to appear relatively homogeneously in the sky. The contribution of these contaminants is therefore determined by running \eromapper\ on one million random points in the LS footprint. We require these points to be at least five optical cluster radii $D>5 R_\lambda$ away from the extended \erass\ sources. Of the remaining $\sim$710,000 random points,
we identify an overdensity of red-sequence galaxies in $\sim$250,000 cases and measure their photometric redshifts and richness. The other major contaminants, AGN mischaracterized as extended sources, are expected to reside in slightly more overdense regions due to their clustering \citep{Comparat2019}. We estimate their contribution by running \eromapper\ on the X-ray centroids of $\sim850,000$ of all
\erass\ point source detections within the LS footprint. Of these sources, 
we find an identifiable red sequence with a measured photometric redshift and richness in $\sim$490,000 cases. With these distributions, we compute a contamination estimator $P_{\rm cont}$ as the ratio between the kernel density estimates of the probability $P$ of random sources (RS), AGN, over cluster candidates ($C$). This procedure is repeated for all filter band combinations, and the contamination estimator is calculated for all of them independently before merging as a catalog. For the 291 clusters, for which the best redshift is adopted from the literature, we manually set their $P_{\rm cont}=0$.

The mixture model can then be expressed in the following form:

\begin{align}
P(\hat{C}_R, \hat{z}, \hat{\lambda}, \hat{\mathcal{H}}_i) &= (1-f_{\rm RS}-f_{\rm AGN}) \cdot P(\hat{C}_R, \hat{z}, \hat{\lambda}, \hat{\mathcal{H}}_i | {\rm C}) + \nonumber \\ 
& \quad + f_{\rm RS} \cdot P(\hat{C}_R, \hat{z}, \hat{\lambda}, \hat{\mathcal{H}}_i | {\rm RS}) +  \nonumber\\
& \quad + f_{\rm AGN} \cdot P(\hat{C}_R, \hat{z}, \hat{\lambda}, \hat{\mathcal{H}}_i | \rm AGN)
\label{eqn:mixture_model1}
\end{align}

\noindent where $f_{\rm RS}$ and $f_{\rm AGN}$ are the global fractions of contamination by random sources and AGN, respectively, and $\hat{C}_R$, $\hat{z}$, $\hat{\lambda}$, $\hat{\mathcal{H}}_i$ are our set of observables, count rate, redshift, richness, and sky position.

This model expresses that the probability density function (PDF) for the measured observables of our sources is the sum of the PDF of the three different components: clusters, AGN, and random sources. The constant in front of each term represents the fraction of that kind of source; here, we note that $f_{\rm C} = (1-f_{\rm RS}-f_{\rm AGN})$ is the cluster fraction. The PDFs are weighted so that when integrated over the entire parameter space, the total PDF is normalized to 1.

\begin{figure*}
\begin{centering}
\begin{tabular}{c}
\includegraphics[width=0.98\textwidth,trim={0.1cm 1.5cm 0.1cm 1.5cm}, clip]{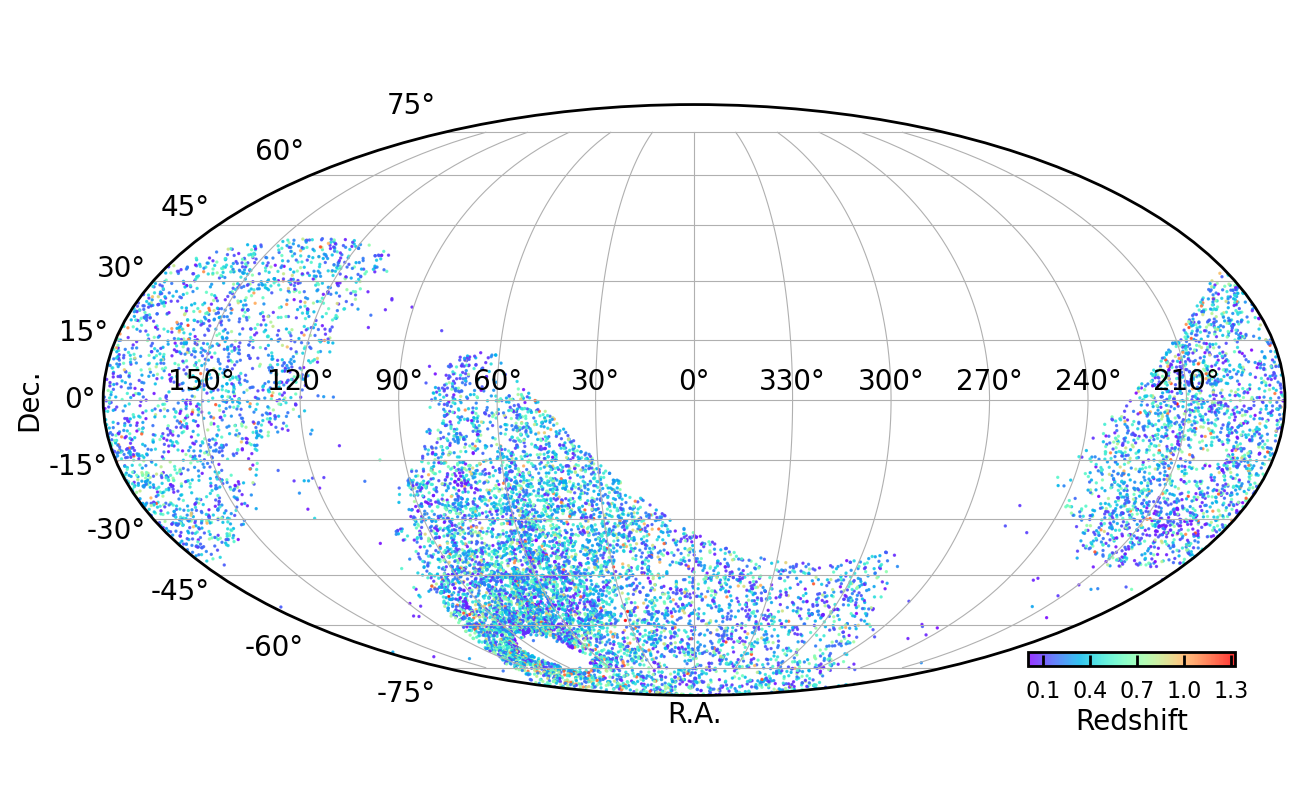} 
\end{tabular}
\end{centering}
\vspace{-0mm}
\caption{The projected locations of the 2,247 clusters and groups in the primary catalog in the \rosi\ and LS DR9N and DR10 13,116~deg$^2$ common footprints are shown. The detections outside the footprint with $|b|<15$~deg are the redshifts of the clusters added from the literature. The redshift confirmed by the follow-up algorithm \eromapper\ is color-coded \citep{Kluge2024}.\label{fig:projimage}}
\end{figure*}

We can simplify the previous equation by dividing both sides by $P(\hat{C}_R, \hat{z}, \hat{\mathcal{H}}_i)$, the entire sample probability density, thus isolating optical richness, which is the quantity that truly allows the mixture model to disentangle the three populations in our catalog. The Equation~\eqref{eqn:mixture_model1} becomes:

\begin{align}
P(\hat{\lambda} | \hat{C}_R, \hat{z}, \hat{\mathcal{H}}_i) &= 
\bigg( 1 - f_{\rm RS} \cdot \frac{P(\hat{C}_R, \hat{z}, \hat{\mathcal{H}}_i | {\rm RS})}{P(\hat{C}_R, \hat{z}, \hat{\mathcal{H}}_i)} - \nonumber \\
& \quad - f_{\rm AGN} \cdot \frac{P(\hat{C}_R, \hat{z}, \hat{\mathcal{H}}_i | {\rm AGN})}{P(\hat{C}_R, \hat{z}, \hat{\mathcal{H}}_i)} \bigg) \cdot P(\hat{\lambda} | \hat{C}_R, \hat{z}, {\rm C}) + \nonumber \\
& \quad + f_{\rm RS} \cdot P(\hat{\lambda}, \hat{z} | {\rm RS}) \cdot \frac{P(\hat{C}_R, \hat{\mathcal{H}}_i | {\rm RS})}{P(\hat{C}_R, \hat{z}, \hat{\mathcal{H}}_i)} + \nonumber \\
& \quad + f_{\rm AGN} \cdot P(\hat{\lambda}, \hat{z} | {\rm AGN}) \cdot \frac{P(\hat{C}_R, \hat{\mathcal{H}}_i | {\rm AGN})}{P(\hat{C}_R, \hat{z}, \hat{\mathcal{H}}_i)},
\label{eq:mixture_model2}
\end{align}

\noindent where $P(\hat{C}_R, \hat{z}, \hat{\mathcal{H}}_i | {\rm RS})$ and $P(\hat{C}_R, \hat{z}, \hat{\mathcal{H}}_i | {\rm AGN})$ are estimated using a kernel density estimator in our realistic \erass\ simulations \citep{Seppi2022}, where we can identify precisely the different contaminants, while $P(\hat{\lambda}, \hat{z} | {\rm AGN})$ and $P(\hat{\lambda}, \hat{z} | {\rm RS})$ are obtained as specified above using \eromapper. Finally, we define $P_{\rm cont}$ as the ratio of the sum of the last two terms in Equation~\eqref{eq:mixture_model2} to the total probability. The total contamination fraction ($f_{\rm tot}=f_{\rm RS}+f_{\rm AGN}$) is used to calculate the purity of the sample (1-$P_{\rm cont}$), which is given in Table~\ref{table:catcontam}. 

Conceptually, this approach might look similar to the optical contamination estimators, which apply a simple redshift-dependent richness cut relying solely on optical data \citep[e.g.][]{Klein2019}. The main advantage of our method is that it assigns a probability of being a contaminant that is informed by optically determined richness and redshift, as well as by the X-ray properties, namely, count-rate for each X-ray detection, including the candidate clusters and contaminants.
Using X-ray information, which is not affected by the projection issues found with the optical data, helps keep clusters with low X-ray flux and low richness at higher redshifts in the sample. In traditional richness cleaning methods, these sources would be marked as contaminants and removed from subsequent catalogs. Our approach instead allows the user to decide on the preferred $P_{\rm cont}$ cut. In the primary \erass\ cluster catalog, we provide the probability of being a contaminant for each source based on our mixture model. The $P_{\rm cont}$ cuts can be used to further clean the catalog based on the science application. In Table~\ref{table:catcontam}, we provide several selection criteria and the overall estimated purity in the samples after the cleaning is applied.

Cleaning through $P_{\rm cont}$ effectively suppresses the AGN contamination in the sample. Bright AGN, with high-count rates and low-richness values, are outliers in the richness versus count-rate parameter space and, therefore, have high $P_{\rm cont}$ values. The mixture model can identify the contaminants, for instance, the random association of background galaxies incorrectly associated with an X-ray source, as the majority of these cases will be outliers in the count-rate and richness scaling laws. These cases can be identified and removed efficiently by setting a $P_{\rm cont}$ cut. However, there are some instances where the mixture model cannot accurately estimate a detection's $P_{\rm cont}$ value. In contamination fraction calculations, the count-rate measurements of random locations in the sky are chosen to be away from \erass\ detected clusters. However, these locations may have extended sources just below our flux limit. Therefore, in these random locations, the count-rate measurements in the catalog, produced by the source detection algorithm, would be higher than the post-processed count rate measured by \mbproj, our forward modeling tool for measuring cluster properties (see Sec.\ref{sec:xrayanalysis}) after more careful treatment of the background. The mixture model's kernel-density computed in our realistic \erass\ simulations \citep{Seppi2022} and the probability density function (PDF) of contaminants cannot be estimated accurately when the discrepancy between the \mbproj\ count rate and the count rate in the catalog is significant. In these cases, the PDF values calculated from that combination of parameters approach zero. In the primary catalog, we find a small fraction of such cases (458 objects; 3.6\%) by visually inspecting the sources where the discrepancy occurs. We remove these cases in the primary catalog by hand; see Section~\ref{sec:legacyample}. Because of these caveats, we recommend users clean the cluster catalog with strict X-ray selection, for instance, \extlike$>6$ or \extlike$>10$, when employing the cluster catalog for population studies and cosmology.

\section{Samples of clusters of galaxies and galaxy groups}
\label{sec:mainsample}
\subsection{The primary galaxy clusters and groups catalog}
\label{sec:legacyample}

Our previous study based on the eFEDS survey shows that compact galaxy groups or high-redshift clusters with angular sizes comparable to the PSF of \rosi\ may be classified as point rather than extended sources. The X-ray selection applied through \extlike\ plays a vital role in keeping or excluding these sources in the cluster sample \citep[see][for details]{Bulbul2022}. Applying a high \extlike\ cut in the X-ray selection process efficiently cleans the sample by removing the majority of contaminants, for instance, AGN, at the expense of removing many clusters \citep{Seppi2022}. 

For example, in the eFEDS sample of 542 cluster candidates, an \extlike\ threshold of 6 was adopted. Increasing this from 6 to 12 would remove $\sim$60 contaminants and $\sim$150 real clusters. Unless purity is a significant concern for a given scientific application, a practical solution to reach a higher purity and completeness level in the sample without losing many real clusters is to use the mixture model method described in Section~\ref{sec:cleaning}. 
Here, we adopt a relatively low X-ray selection threshold for the primary cluster sample on \extlike\ (namely, \extlike\ $>3$).
This achieves higher completeness compared to an \extlike$>6$ sample like eFEDS, increasing the discovery space of \rosi,
while relying on the optical identification process confirmation to improve the reliability of the sample further (see Section~\ref{sec:cleaning}). 
With this \extlike\ threshold, we find 12,247 clusters of galaxies or galaxy groups after removing the 458 contaminants described above, mainly in the low-count regime and due to background fluctuations, and where the mixture model failed. The total effective survey area covered by this, which we designate as the primary \erass\ cluster sample, is 13,116~deg$^2$ in the combined LS DR9N and DR10 footprint. The distribution of these clusters in the \rosi\ sky is shown in Fig.~\ref{fig:projimage}. The number of clusters detected in the first All-Sky Survey is consistent with the pre-launch predictions \citep{Merloni2012, Pillepich2012}, demonstrating the superb performance of \rosi. 

\begin{figure}
\begin{tabular}{c}
\includegraphics[width=0.49\textwidth, clip]{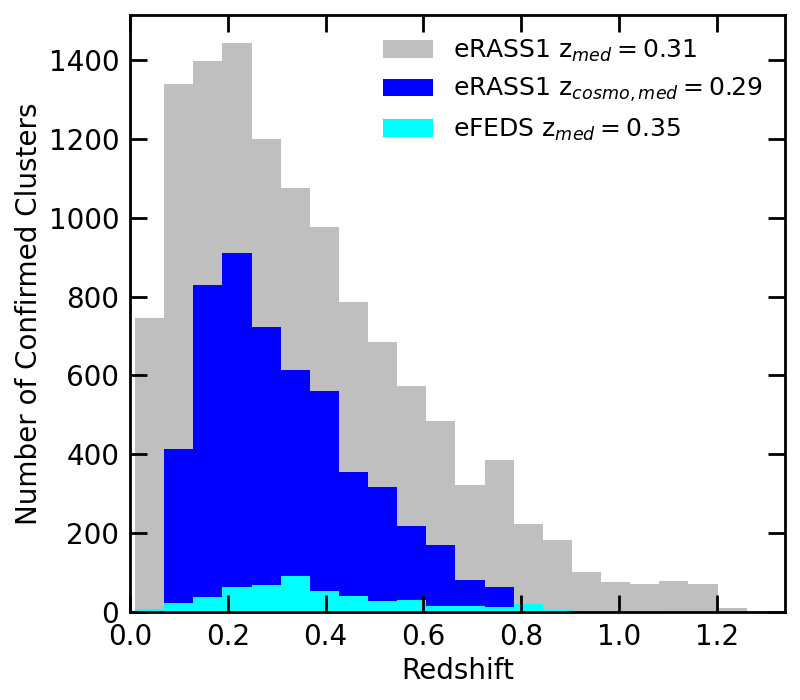} 
\end{tabular}
\caption{Redshift distribution of the 12,247 confirmed \erass\ clusters and groups. Shown in gray is the \erass\ cluster sample with \extlike$>3$, compared to those of the cosmology sample in blue, and the redshift distribution of the 477 clusters confirmed in the eFEDS field in cyan \citep{Liu2022}. The median redshift of the \erass\ cluster catalogs is slightly lower than that of the eFEDS clusters ($z_{\rm med}=0.35$).\label{fig:hist_z}}
\end{figure}

Each cluster in the sample is identified by \eromapper or a literature redshift assigned to it and has an associated redshift measurement.
We present the redshift distribution of the primary cluster catalog and the cosmology sample in Fig.~\ref{fig:hist_z}. The redshift range, populated in the $z_{\rm best}$ column in the catalog, spans from 0.003 to 1.32 with a median sample value of 0.31, slightly lower than the median redshift of 0.35 of the 477 clusters confirmed in the 140~deg$^{2}$ eFEDS field \citep{Liu2022}. 
The reason for the lower median redshift of the two samples is twofold: eFEDS, although comprised of a small area, has higher exposure and hence greater sensitivity to high-redshift clusters. Secondly, the greater depth of the follow-up HSC survey, compared to LS and WISE surveys, has a significant impact on the identification of the higher redshift clusters. 

In the catalog, the majority of redshifts (8790, 69\%) are determined through photometric measurements. A large fraction of the sample (29\%, 3510 clusters) of the sources have spectroscopic redshifts, and 247 of the redshifts are from the literature (291 clusters with literature reshifts before visual cleaning) \citep[see][]{Kluge2024}. 
The catalog includes 1451 nearby galaxy clusters and groups at $z<0.1$, including well-known examples such as the Virgo, Fornax, and Centaurus clusters. Most of the sample, 10,074  (83\%) galaxy clusters, and groups lie in the redshift range of $0.1<z<0.8$. The sample comprises 722 clusters at high redshift $z>0.8$. The highest redshift cluster, 1eRASS~J020547.4-582902, is located at redshift 1.32 with \extlike$\,=\,4.1$, showing \rosi's capability to detect high redshift clusters. This cluster is cross-matched to a detection of a cluster in the Atacama Cosmology Telescope (ACT)~DR5 catalog at the same redshift. Of the 12,247 clusters and groups 
in the primary sample, 8,361, corresponding to 68\% of the sample, are detected for the first time and are new discoveries. The remaining 3,886 clusters can be found in other cluster catalogs in the literature. The details of cross-matching with published cluster catalogs in the literature are described in Section~\ref{sec:crossmatch}. 

\begin{figure}
\includegraphics[width=0.5\textwidth,trim={0 0 0 0}, clip]{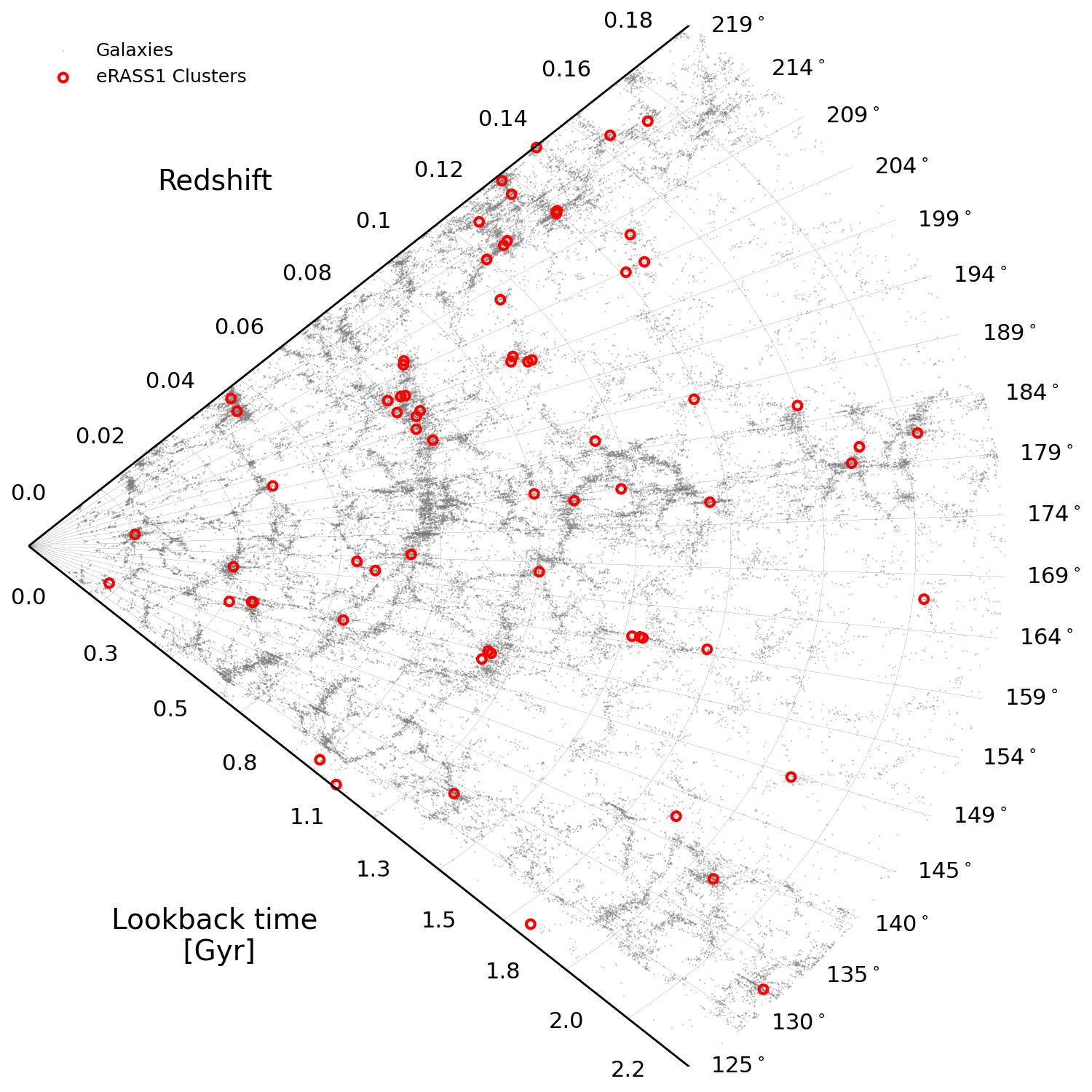} 
\caption{Distribution of nearby \erass\ galaxy clusters along the equator (red circles) illustrating the large scale structures of the Universe. Positions are projected along declination ($|Dec.|<2$~deg) within redshift 0.2. Galaxies from the GAMA and SDSS survey are represented with grey dots \citep{Driver2022, Almeida2023}.
}
\label{fig:lss}
\end{figure}

Located at the intersection of the cosmic filaments, the detection of a large number of massive galaxy clusters and groups with \rosi\ can be used to extract information about the clustering pattern and to map the large-scale structure of the Universe. Historically, the local universe has been studied in detail by the galaxy surveys in the optical band, showing a number of large-scale patterns \citep{Geller1989, Colless2001}. The distribution of 97,952 galaxies in the GAMA and SDSS spectroscopic surveys in gray circles in Fig.~\ref{fig:lss} provide maps of the large-scale structures in the Universe \citep{Driver2022, Almeida2023}—the filaments stretching several tens of Mpc can be seen in the figure. The galaxy groups and clusters located at the intersection of the cosmic web detected by \rosi\ in the first All-Sky survey are shown in red circles. The apparent correlation between the location of the connecting knots of galaxy surveys and \rosi\ detections demonstrates the potential of \rosi\ in mapping the large-scale structure \citep[see also][]{Ghirardini2022, Liu2022}. This correlation will become more apparent as the survey gets deeper and \rosi\ detects a more significant number of low-mass group-size haloes as part of the large-scale structure. An accompanying paper by \citet{Liu2024} provides a catalog of superclusters and their member profiles.

\subsection{The cosmology sample}
\label{sec:cosmosample}

The selection methodology for our primary cluster sample emphasizes completeness over purity. Different samples can nonetheless be constructed with higher levels of purity by applying various cuts, for instance, in \extlike\ or $P_{\rm cont}$ (see Table~\ref{table:catcontam}).
Highly pure samples of clusters are more suitable for accurate determination of the cosmological parameters and testing of cosmological models. Here we describe the construction of the \erass\ cosmology sample, which is used in the work of \citep{Ghirardini2024} and the follow-up weak lensing mass calibration papers \citep{Grandis2024, Kleinebreil2024}.

We first select a sample with a stricter \extlike\ threshold (\extlike$>6$), resulting in 11,141 extended sources in the clean \erass cluster catalog. This selection efficiently removes a large fraction of contaminants that are not true extended X-ray sources, such as AGN and background fluctuations.  To ensure the uniformity of the optical identification process and to track any biases in the redshift measurements, we use only the LS~DR10 region and exclude the LS~DR9N area. Specifically, we limit our area to uniformly processed LS~DR10-South region with $DEC\lesssim32.5\degr$.
This limits the survey area to 12,791~deg$^{2}$. Within this LS~DR10-South footprint, at \extlike~$>6$, there are a total of 8,129 extended X-ray sources. Of these, \eromapper\ is able to identify 7,077 candidate clusters in the LS~DR10-South footprint by combining g, r, and z filter bands to ensure homogeneity of the richnesses, redshifts, and contamination estimation. Of the 7,077 candidate clusters in the LS~DR10-South footprint, we can identify the optical counterparts of 6,562 securely detected clusters.

To select only clusters with photometric redshift lower than the limiting redshift at that sky position, we apply the flag {\texttt{IN$\_$ZVLIM==True}} \citep[see][for further details]{Kluge2024}. 
For the excluded clusters due to the limiting redshift cut, the detection of faint galaxies just above the limiting luminosity $L>0.2$~$L_{*}$ of the corresponding Legacy surveys
becomes uncertain, and the measured richness artificially increases because of Eddington bias \citep{Kluge2024}. As weak lensing mass bias calculations rely on the richness values \citep[see][for further details]{Grandis2024}, we limit our sample to a regime where the member selection is most reliable. Furthermore, the photometric redshift range is kept limited to $0.1\,<\,z\,<\,0.8$, where the photometric redshifts are the most reliable \citep[see][for details]{Kluge2024}. The final cosmology sample comprises 5259 galaxy clusters in the 12,791~deg$^{2}$ LS~DR10-south area. The redshift distribution of the sample is displayed in Fig.~\ref{fig:hist_z} with a sample median of 0.29, slightly lower than the primary cluster catalog due to the elimination of high redshift sources. 

\begin{table}
\caption{The number of confirmed clusters with a set of selection on \extlike\ and mixture model property $P_{\rm cont}$. The resulting purity levels of the samples are given after the cuts are applied.}
\label{table:catcontam}
\begin{center}
\begin{tabular}[width=0.5\textwidth]{lcccc}
\hline\hline 
X-ray and Optical   & No of Clusters  &  Purity & $z_{med}$  \\
 Selection Criteria&               &            \% & \\ \hline

\multicolumn{4}{l}{The Primary Galaxy Clusters and Groups Catalog} \\
\hline
\extlike$>3$, no $P_{\rm cont}$ cut & 12247 & 86  & 0.31 \\
\extlike$>3$, $P_{\rm cont} < 0.80$ & 11522 & 91  & 0.30 \\
\extlike$>3$, $P_{\rm cont} < 0.50$ & 10865 & 95  & 0.30 \\
\extlike$>6$, no $P_{\rm cont}$ cut & 6752 & 94  & 0.27 \\
\extlike$>6$, $P_{\rm cont} < 0.80$ & 6604 & 95  & 0.26 \\
\extlike$>6$, $P_{\rm cont} < 0.50$ & 6474 & 97  & 0.26 \\
\extlike$>10$, no $P_{\rm cont}$ cut & 4171 & 97  & 0.24 \\
\extlike$>15$, no $P_{\rm cont}$ cut & 2646 & $>$99   & 0.22 \\
\extlike$>20$, no $P_{\rm cont}$ cut & 1896 & $>$99   & 0.21 \\
\extlike$>25$, no $P_{\rm cont}$ cut& 1449 & $>$99  & 0.20 \\
\hline
\multicolumn{4}{l}{The Cosmology Sample}\\

\hline 
\extlike$>6$, no $P_{\rm cont}$ cut & 5259 & 95 & 0.29 \\
\extlike$>6$, $P_{\rm cont} < 0.80$ & 5170 & 96  & 0.29 \\
\extlike$>6$, $P_{\rm cont} < 0.50$ & 5087 & 97  & 0.29 \\
\hline\hline
\end{tabular}
\end{center}
\end{table}
\subsection{Purity and completeness}

To compute the completeness of the sample, we use the digital twin of \erass\ the details of which are described in \citet{Seppi2022}. We briefly summarize the main characteristics here. In the digital twin, a dark matter halo light cone extending to redshift six is generated from the UNIT1i simulation \citep{Chuang2019}. The X-ray emission is predicted using accurate models of AGN and cluster emissivity from \citet{Comparat2019, Comparat2020}. The X-ray background is generated by re-sampling the actual \erass\ data. X-ray events are generated with the \texttt{SIXTE} simulator \citep{Dauser2019} with the ancillary response file (ARF), the redistribution matrix file (RMF), and the attitude file of the spacecraft in the \rosi\ Data Release 1 calibration database consistently. We match the resulting source catalog to the input catalog by tracing the origin of each photon. The sources are grouped into several classes. A source with a primary counterpart of a simulated point source (AGN or a star) or a simulated extended source (cluster) is included in the catalog. Additionally, sources that are secondary counterparts of simulated point sources or extended sources are also added to the catalog. We also consider false detections due to random background fluctuations, where the source catalog entry is not associated with any simulated source.

\begin{figure}
\begin{center}
\includegraphics[width=0.49\textwidth]{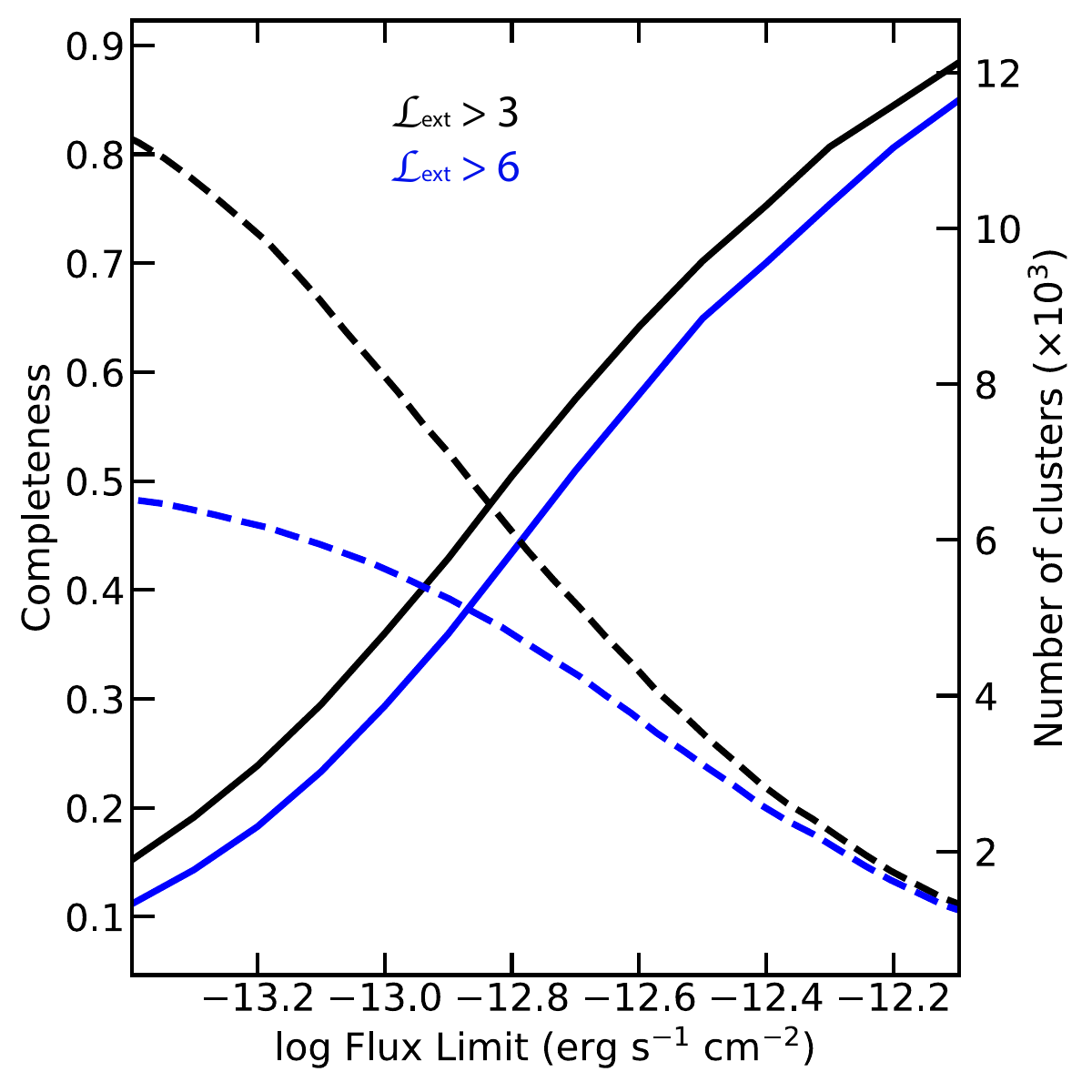}
\caption{Completeness as a function of intrinsic unabsorbed flux limit for two \extlike\ thresholds: the cluster catalog \extlike~$>3$ (in black) and the cosmology sample \extlike~$>6$ (in blue). The dashed curves show the corresponding number of clusters for both samples. The trade-off between completeness and the number of clusters in the primary cluster sample is clear in the figure.}
\label{fig:comp}
\end{center}
\end{figure}

Fig.~\ref{fig:comp} shows the completeness and number of clusters as functions of flux limit for two \extlike\ thresholds: \extlike~$>$3 (black) and \extlike~$>$6 (blue). X-ray-selected samples can reach high completeness levels when higher flux cuts are applied. For instance, above the flux limit of $8\times10^{-13}$~ergs~s$^{-1}$~cm$^{-2}$, the primary cluster catalog has a completeness level of $\sim$90\%; 
however, the number of clusters ($\sim1300$) above that flux is relatively low. Choosing a lower flux limit of $4\times10^{-14}$~ergs~s$^{-1}$~cm$^{-2}$ would increase the sample size to $>10,000$ while the completeness declines to a level of 13.3\% in the primary sample (\extlike>3 selection). The cosmology sample with (\extlike>6), above the flux limit of $10^{-13}$~ergs~s$^{-1}$~cm$^{-2}$, has a completeness of $\sim$30\%. 
In general, the trade-off between completeness, purity, and the number of clusters should be decided based on scientific application.  

We compute the sample's purity by integrating the clusters' $P_{\rm cont}$ values (see Table~\ref{table:catcontam}). Based on our estimates with the mixture model, the primary cluster catalog with 12,247 sources in the 13,116~deg$^2$ survey area has a purity level of 86\%. Applying either higher X-ray and optical selections through $P_{\rm cont}$ or both would further reduce the contamination of this sample. For instance, selecting clusters with a lower probability of being a contaminant, namely, $P_{\rm cont}<0.5$, removes most of the contamination in the sample, leaving 10,865 clusters with a sample purity level of 95\% (see Table~\ref{table:catcontam}). Applying higher thresholds in the X-ray selection, for instance, \extlike~$>6$, produces samples with 94\% purity in the cost of removing 5,495 extended sources. Increasing the \extlike\ threshold to 10 removes the majority of the contamination and produces highly pure samples at $>98$~\% purity level with a sample size of 4171. Above the X-ray \extlike~$>10$ threshold, X-ray selection is so efficient that additional optical selection is unnecessary.  

The cosmology sample optimized for high purity with a different set of X-ray (\extlike~$>6$) and optical selections has 5259 clusters candidates in the LS~DR10 survey area of 12,791~deg$^{2}$ and has a purity level of 95\%. Applying $P_{\rm cont}<0.5$ cut increases the purity level to 97\%, leaving 5087 securely confirmed clusters in the sample. X-ray selection is a more reliable way of cleaning samples than optical selection as projection effects impact X-rays less; we recommend users apply stricter X-ray selections, for instance,\extlike~$>\,10$ or 15, when necessary to generate purer cluster samples.

\begin{table*}
\caption{Public Cluster Catalogs Cross-matched with \erass}
\label{table:match}
\begin{center}
\begin{tabular}[width=0.5\textwidth]{llllllll}
\hline\hline 
External Catalog &  Common &  Total No of  & No of Clusters: & Median & Median & Reference\\
                 & Footprint & Clusters   & In the Footprint/ &   Off-set& Redshift & \\               
                 & (deg$^2$) &    & Cross-matched & (arcsec) & &\\
\hline 
MCXC & 13116 & 1743 & 681 / 588 & 31.6 & 0.11 & \citet{Piffaretti2011} \\
MARD-Y3 & 2733 & 2959 & 2331 / 904 & 37.4 & 0.23  & \citet{Klein2019} \\
CODEX & 3508 & 10382 & 3106 / 574 & 50.2 & 0.22 & \citet{Finoguenov2020} \\
RXGCC & 8784 & 944 & 395 / 256 & 55.3 & 0.08 & \citet{Xu2022} \\
XCS DR1& N/A\tablefootmark{b}  & 503 & 184 / 36 & 19.4 & 0.23  & \citet{Mehrtens2012} \\
XXL 365 & 23 & 302 &137 / 11 & 14.1 & 0.32 & \citet{Adami2018} \\
XCLASS &  N/A\tablefootmark{b}  & 1559 &800 / 304 & 16.9 & 0.14  & \citet{Koulouridis2021} \\
eFEDS & 140 &542 &531 / 63 & 17.1 & 0.29 & \citet{Liu2022} \\
PSZ2 & 10281 & 1653 & 633 / 439 & 62.5 & 0.21 & \citet{Planck2016} \\
SPT-2500~deg$^2$ & 2288 & 677 & 649 / 352 & 21.3 & 0.44 & \citet{Bocquet2019} \\
SPT-ECS & 1798 & 470 & 381 / 241 & 23.9 & 0.41 & \citet{Bleem2020} \\
SPT-pol 100 deg$^2$ & 78 & 89 & 89 / 20 & 16.3 & 0.38 & \citet{Huang2020} \\
ACT-DR5 & 6877 & 4195 & 2565 / 1176 & 24.1 & 0.41  & \citet{Hilton2021} \\
Abell & 8784\tablefootmark{a} & 1059 & 466 /152 & 52.5 & 0.06 & \citet{Abell1989} \\
DES-Y1 & 3644 &  6729 & 6075 / 781 & 17.7 & 0.38 &\citet{McClintock2019} \\
NEURALENS &  9833 & 1312 & 697 / 81 & 14.8 & 0.40& \citet{Huang2021} \\
Wen \& Han High-z & 11369 & 1959 &508 / 11 & 14.8 & 0.75 & \citet{Wen2018} \\
MaDCoWS & 7436 &  2839 & 1309 / 27 & 27.3 & 0.95   & \citet{Gonzalez2019} \\
GOGREEN-GCLASS & N/A\tablefootmark{b} & 26 & 11 / 3 & 14.2 & 1.13 & \citet{Balogh2021} \\

\hline\hline
\end{tabular}
\tablefoot{
\tablefoottext{a}{The Galactic latitude cut of \citet{Abell1989} catalog is not strict, this value is calculated using $|b|>20$~deg as an approximation.}
\tablefoottext{b}{Targeted observations.}
}
\end{center}
\end{table*}
\section{Comparisons with overlapping cluster surveys}
\label{sec:crossmatch}

\begin{figure}
\includegraphics[width=0.5\textwidth,trim={0 0 0 0}, clip]{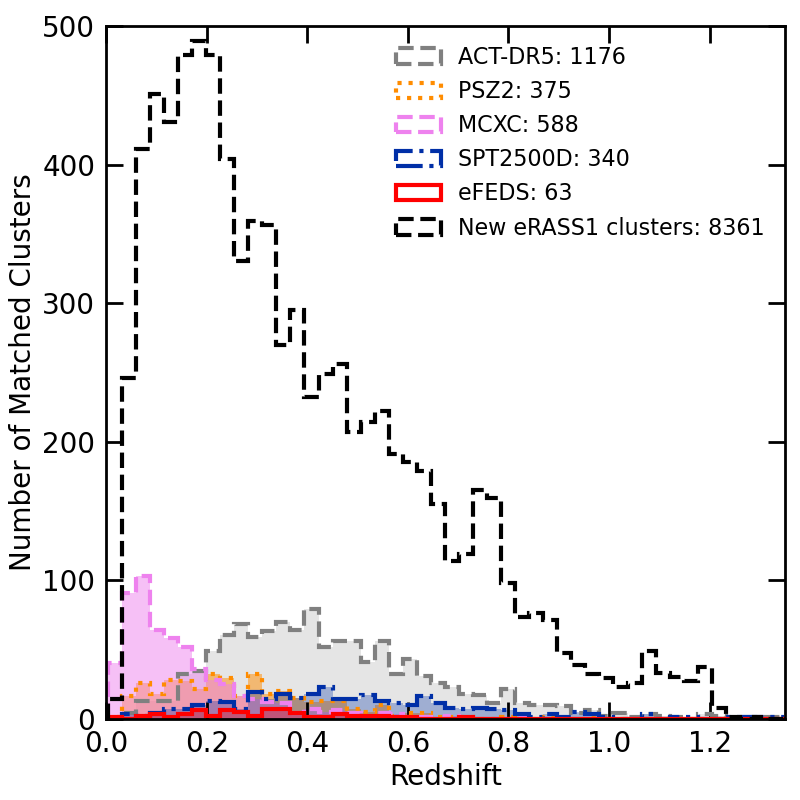} 
\caption{Redshift histogram of the matched \erass\ clusters in the \rosat, {\sl Planck} All-Sky Surveys, ACT, and SPT-SZ surveys and the newly discovered \erass\ clusters. The redshifts indicated here are adopted from the literature catalogs that are matched. For the \erass\ catalog, $z_{best}$ is used in the histogram. The largest overlap of clusters (1176) is between RASS and ACT~DR5 surveys due to the large common area between the two surveys with \rosi.}
\label{fig:hist_matched}
\end{figure}

In this section, we provide the counterparts of eRASS1 clusters found in several previously published cluster catalogs compiled from X-ray, SZ, and optical surveys. The matching radius we use for this work is set to $2\arcmin$; this radius is ideal for identifying close associations for \rosi. Using larger matching distances causes the clusters to be associated with the surrounding large-scale structure, for instance, infalling haloes, for most surveys, except for the {\it Planck} survey as demonstrated in \citet{Bulbul2022}. The identifiers of the matched clusters are given in the catalog under the column ``Match Name''. In total, we find 3,886 clusters have counterparts in various cluster catalogs. 68\% of our sample, a total of 8,361 clusters, are not matched with any other "optically confirmed"  source in the published catalogs. Therefore, they are newly discovered clusters. If we only consider clusters in X-ray and SZ catalogs, namely, ICM-based catalogs, then 9795 clusters corresponding to 80\% of the \erass\ cluster sample are detected and identified for the first time. We calculated the purity level of the new detections in the primary sample to be 81\%, whereas the newly discovered clusters in the cosmology sample have a purity level of 92\%.
We provide the details of the matched catalogs and the overlapping area with \rosi\ Western Galactic Half in the following subsections. A summary of the results and the distribution of redshifts of the cross-matched samples are given in Table~\ref{table:match} and Fig.~\ref{fig:hist_matched}. For comparisons, we over-plot the redshift histogram of the newly discovered \erass\ clusters in the figure. For \erass\ clusters, we use the $z_{best}$ in the histogram.

\subsection{Surveys in the X-ray band}

The first true imaging all-sky survey in the soft X-ray band was performed by \rosat\ in the period of 1990--1991 \citep[RASS,][]{Voges1999}. The RASS was shallower and had a worse angular resolution, with the half-power radius being $84\arcsec$ \citep{Boese2000}, compared to the \rosi\ All-Sky Survey. Nonetheless, many of our clusters were already discovered with ROSAT. Several sub-catalogs have emerged over the past two decades. The \rosat\-ESO Flux Limited X-ray Galaxy Cluster Survey \citep[REFLEX,][]{Boehringer2004_reflex} was compiled in a region with a declination of $\leq + 2.5$~deg and excluding the Milky Way via a Galactic latitude cut of \mbox{$\b\ \le -20$}~deg in the southern sky, consequently covering a survey area of 13,924~deg$^{2}$ down to a flux limit of $3\times10^{-12}$~erg~s$^{-1}$~cm$^{-2}$. A total of 441 clusters of galaxies are confirmed in the REFLEX catalog. The Northern \rosat\ All-Sky galaxy cluster survey \citep[NORAS,][]{Boehringer2000_noras} was based on RASS data excluding the same region around the Galactic plane but covering the northern sky. In this catalog, 437 clusters were confirmed based on a combination of count rate ($>$0.06 cts~s$^{-1}$ in the 0.1--2.4~keV band) and a source extent likelihood. Another cluster catalog that was based on RASS is the \rosat\ Brightest Cluster Sample (BCS) \citep{Ebeling1998}, consisting of 
201 X-ray-brightest clusters of galaxies in the northern hemisphere ($\delta\ge 0$~deg), at high Galactic latitudes of $|b|\ge 20$~deg. Cluster surveys with more limited area but greater depth have also been performed with \rosat. For instance, \citet{Cruddace2002} performed a cluster survey in a region of 56~deg$^{2}$ around the South Galactic Pole, down to a flux limit of $1.5\times10^{-12}$~erg~s$^{-1}$~cm$^{-2}$. The 400 Square Degree \rosat\ PSPC Galaxy Cluster Survey \citep[][400SD]{Burenin2007} covered an area of 397~deg$^{2}$ with high Galactic latitude $|b|> 25$~deg with a flux limit of $1.4\times10^{-13}$~erg~s$^{-1}$~cm$^{-2}$. All these \rosat\ catalogs were compiled by \citet{Piffaretti2011} into a meta catalog of 1743 confirmed clusters, with which we cross-match the coordinates in our catalog ($RA$, $DEC$) with $2\arcmin$ matching radius. We apply the same procedure to match the subsequent catalogs.

The MCXC survey has the largest area in common with \erass\ due to its full sky coverage; the cross-match results in 588 clusters in a sample of 681 clusters in the common footprints of both surveys. The cross-matched sample has a sample median redshift of 0.11. The median centroid offset is $31\farcs6$, consistent with \rosi\ and \rosat's PSF size.

The CODEX \citep[COnstrain Dark Energy with X-ray clusters][]{Finoguenov2020} sample is also based on the \rosat\ All-sky Survey, identifying 10,382 X-ray sources with red sequence counterparts in the $10,382$~deg$^{2}$ SDSS area reaching down to an X-ray flux limit of $10^{-13}$~ergs~s$^{-1}$~cm$^{2}$. The majority of the CODEX survey area is in the northern sky with a limited common footprint with eROSITA in $3508~deg^{2}$. Of the 3106 objects in the CODEX catalog, we detect and confirm only 574 of these sources, with a median redshift and offset of $0.22$ and $50\farcs2$. The large offset between the \rosat\ based MCXC and \erass\ sample, still within the PSF size of \rosat, could be due to the different detection algorithms, namely, wavelet filtering, employed to detect clusters. The MARD-Y3 catalog, on the other hand, is constructed from the DES-Y3 gold catalog using the priors from the \rosat\ All-Sky survey catalog \citep{Klein2019} $-$ the cross-match results in 904 associations with a median redshift of 0.23. The low matching rate may indicate a large contamination level in the CODEX and MARD-Y3 catalogs based on shallower ROSAT data. It is worth noting that CODEX and similar catalogs, such as MARD-Y \citep{Klein2019} and RASS-MCMF \citep{Klein2023}, are constructed based on the X-ray catalog with no extent selection and no ICM selection; therefore, they are subject to higher levels of contamination and significant optical selection effects. The significance of this catalog is that we provide an initial selection of the extent likelihood in X-rays that significantly reduced the contamination in the sample.

RXGCC is one of the most recent catalogs based on the \rosat\ data, performed with a wavelet-based detection algorithm and strict extension likelihood ($>25$) and extent ($>0\farcm67$) selection, includes clusters down to flux of $2.49\times10^{-12}$~erg~s$^{-1}$~cm$^{-2}$ \citep{Xu2022}. This catalog is optimized to find very extended groups missed in the previous \rosat\ catalogs. We find 256 of their most nearby groups and clusters ($z_{\rm med}=0.08$) in our sample, with a large centroid offset of $55\farcs2$, consistent with the other \rosat\ samples. The larger offsets are likely due to the considerable of these clusters and groups.

Deep X-ray surveys focused on scanning smaller areas have been performed with \xmm, resulting in cluster catalogs reaching lower flux limits than \rosat. Most relevantly, in the 50~deg$^{2}$ \xmm\ XXL survey with roughly 10~ks exposure time, a total of 365 clusters are detected down to flux limits of 10$^{-15}$~erg~s$^{-1}$~cm$^{-2}$ \citep{Pierre2016, Adami2018}. Of these 365 clusters, 11 are detected in \erass\ observations with a small central offset of $14\farcs1$. The median redshift of the cross-matched sample is consistent with the median redshift of the XXL sample. Other relevant \xmm\ surveys include the \xmm\ Cluster Survey \citep[XCS,][]{Mehrtens2012} and \xmm\ Cluster Archive Super Survey \citep[X-CLASS,][]{Koulouridis2021}, consisting of 503 and 1559 optically confirmed galaxy clusters detected using \xmm\ archival data. Of the 1559 clusters in the X-CLASS sample, 304 have \erass\ counterparts, with a median redshift of 0.14. XCS has 36 close associations in \erass. The median centroid offset, $\sim17\arcsec$, is well within the \rosi\ survey averaged PSF and consistent with other \xmm\ surveys. In general, the clusters observed with \xmm\ have small offsets with respect to their \erass\ counterparts, which is not surprising given the $6\arcsec$ FWHM PSF of \xmm\ and $\sim5\arcsec$ localization accuracy of \rosi \citep{Brunner2022}\footnote{\url{https://www.cosmos.esa.int/web/xmm-newton/technical-details-epic}}. Although the number of \rosi\ counterparts of the clusters detected in \xmm\ surveys seem to be small in number due to the shallower survey depth and slightly worse angular resolution of \rosi, the size of the cross-matched sample will increase as \rosi\ survey gets deeper. The main advantage of \rosi\ based cluster surveys is the large and contiguous area coverage compared to the \xmm\ data. 

The first survey of \rosi\ is the eFEDS, which covered a 140~deg$^2$ region in the Equatorial strip, with approximately $10\times$ deeper exposure than \erass. In the eFEDS field, we confirm 477 of the 542 cluster candidates with \extlike$>6$. We find 63 common detections with a median redshift of 0.29. The details of the detection probabilities of these commonly detected clusters are provided in \citet{Clerc2024}. The median offset is $17\farcs1$, most likely caused by high Poisson shot noise due to the low exposure time of \erass. We make in-depth comparisons of X-ray properties of eFEDS clusters with \erass\ in Section~\ref{sec:chandracomp}.

\subsection{Sunyaev Zel'dovich effect surveys}

Another efficient way of detecting clusters through their ICM emission is by searching for the SZ effect signatures in the cosmic microwave background. The SZ effect signal is independent of redshift, making SZ surveys sensitive to clusters at higher redshifts. A few independent ground- and space-based SZ cluster catalogs have become available in the last decade. However, due to their wide beam sizes or limited area coverage, the catalogs of extended sources remain limited to a few hundred to a few thousand clusters. Specifically, we compare our catalog with those from the {\it Planck} satellite, ACT, and the South Pole Telescope (SPT).

{\it Planck}, launched in 2009, assembled its final All-Sky cluster catalog of 1,653 sources in 2016. The \rosi\ primary cluster catalog presented here and {\it Planck} have 10,281~deg$^{2}$ common footprint area with 439 counterparts in our catalog with a median sample redshift of 0.21. {\it Planck} detected clusters show the largest offset with the \rosi\ counterparts compared to other surveys cross-matched with our primary cluster catalog; this is most likely due to the large beam size \citep[$9\farcm65$ at 100~GHz][]{Planck2014}.

The ACT DR5 catalog consists of 4,500 clusters of galaxies collected from the runs of the ACTPol survey between 2008 and 2018 in the 13,211~deg${2}$. The sample is mass-limited with a high completeness level of 90\% and a signal-to-noise ratio (S/N) threshold of $>4\sigma$. However, the false detection rate can be as high as 30\% in this sample \citep[see Fig.~6 in][]{Hilton2021}. When the (S/N)~$>5\sigma$ selection is applied, the contamination fraction in their catalog drop to a few percent. Regardless, we provide the \erass\ counterparts of all detections reported in their catalog for completeness. ACT has a smaller beam size than {\it Planck}; therefore, it is more sensitive to detecting clusters at higher redshifts. This is reflected in the cross-matched sample's median redshift ($z_{\rm med}\,=\,0.41$). ACT has the largest number of associations with \erass: 1,176 common clusters in the 6,877~deg$^2$ common footprint. Of these common clusters, 81 are at high redshifts ($z>$~0.80). Three of the highest redshift clusters in the \erass\ catalog (1eRASS~J020547.4-582902, J064017.1-511255, J015608.9-554159) are matched to ACT counterparts of ACT~CL~J0205.7-5829, ACT-CL~J0640.2-5113, ACT-CL~J0156.1-5542 with spectroscopic redshifts of 1.32, 1.32, and 1.28, respectively. We find an agreement for these clusters in our spectroscopic and photometric measurements. One of our highest redshift clusters, 1eRASS~J044237.2-590529, has a counterpart in the ACT catalog (ACT-CL~J0442.6-5905) with a relatively small angular separation of $11\farcs7$. However, the photometric redshift reported in \citet{Hilton2021}, $z_{\lambda}\,=\,1.48\,\pm\, 0.035$, is in disagreement with our "bias-corrected photometric redshift measurement of $z_{\lambda}\,=\,1.17^{+0.02}_{-0.03}$ based on LS DR 10 $g,\, r,\, z,\, W1$ data.

The SPT, a 10-meter diameter telescope located at the South Pole Station, has performed deeper surveys in a more confined area than ACT in the western Galactic half of the sky. The first SPT survey, SPT-SZ hereafter, covering a region of 2500~deg$^{2}$, was completed in 2011. The sample based on SPT-SZ data comprises 516 confirmed clusters with a detection significance of 4.5$\sigma$ \citep{Bleem2015, Bocquet2019}. Another survey performed with the upgraded SPT-pol receiver covers a 100~deg$^{2}$ area. The survey catalog contains 89 candidates detected with a signal-to-noise ratio $>4.6$, 81 confirmed via optical and infrared follow-up observations \citep{Huang2020}. Later, the SPT-pol receiver was used to scan a larger area in 2013, 2014, and 2015, covering a 2700~deg$^2$ region. In the SPT-pol Extended Cluster Survey (SPT-ECS), a total of 470 optically confirmed clusters are reported down to a signal-to-noise $>4$ \citep{Bleem2020}. We find 353, 241, and 21 counterparts in the \erass\ catalog with mean redshifts between 0.38 and 0.44 and centroid offsets between $16\arcsec$ and $21\arcsec$, comparable to ACT centroid shifts.

\begin{table*}
\caption{Description of the Columns in the Primary Cluster Catalog FITS File}
\label{table:catcolumns}
\begin{center}
\begin{tabular}{lll}
\hline\hline 
Column &   Units & Description  \\
\hline
\multicolumn{3}{l}{The Primary Galaxy Clusters and Groups Catalog} \\
\hline
DETUID &  & Unique X-ray Detection No. This column should be used for matching sources with  \\
& & the main \erass\ X-ray catalog by \citet{Merloni2024}  \\
Name &  & Cluster Name in the IAU format \\
$RA$ & degrees (J2000) & R.A.  of the X-ray detection \\
$DEC$ & degrees (J2000)  & Dec  of the X-ray detection \\
$RA\_XFIT$ & degrees (J2000) & R.A. of the cluster after the image is fit with MBProj2D \\
$DEC\_XFIT$ & degrees (J2000) & Dec in degrees (J2000) of cluster after the image is fit with MBProj2D \\
\extlike & & Extent likelihood of the X-ray detection by ermldet\\
\detlike & & Detection likelihood of the X-ray detection by ermldet\\
Exp. && Average exposure in seconds \\
z$_{\rm best}$ & & Best available cluster redshift. \\
& & Spectroscopic, or photometric, or literature values \citep[see][for details]{Kluge2024} \\
P$_{\rm cont}$ & & Probability of the detection being contamination assigned by the mixture model\\
CR~[$300~{\rm kpc}, R_{500}$] & Counts s$^{-1}$  & Count Rate in 0.2--2.3~keV within 300~kpc and $R_{500}$ \\
CTS~[$300~{\rm kpc}, R_{500}$] & Counts & Total X-ray Counts in 0.2--2.3~keV within 300~{\rm kpc} and $R_{500}$ \\
$F_{X}$~[$300~{\rm kpc}, R_{500}$] & ergs s$^{-1}$ cm$^{-2}$ & X-ray Flux in 0.2--2.3~keV within 300~kpc and $R_{500}$ \\
\lx~[$300~{\rm kpc}, R_{500}$] &  ergs s$^{-1}$ & Integrated X-ray Luminosity  in 0.2--2.3~keV within 300~kpc and $R_{500}$ \\
CR$_{0520}$~[$300~{\rm kpc}, R_{500}$] & Counts s$^{-1}$  & Count Rate in 0.5--2.0~keV within 300~kpc and $R_{500}$ \\
CTS$_{0520}$~[$300~{\rm kpc}, R_{500}$] & Counts & Total X-ray Counts in 0.5--2.0~keV within 300~kpc and $R_{500}$ \\
$F_{0520}$~[$300~{\rm kpc}, R_{500}$] & ergs s$^{-1}$ cm$^{-2}$ & X-ray Flux in 0.5--2.0~keV within 300~kpc and $R_{500}$ \\
$L_{0520}$~[$300~{\rm kpc}, R_{500}$] &  ergs s$^{-1}$ & Integrated X-ray Luminosity in 0.5--2.0~keV within 300~kpc and $R_{500}$ \\
$L_{bol}$ [$R_{500}$]&  ergs s$^{-1}$ & Bolometric Luminosity within $R_{500}$ \\
\kt\ & keV  & ICM temperature \\
M$_{{\rm gas}, R_{500}}$ & M$_{\rm sun}$ & Calibrated cluster gas mass within R$_{500}$ \\
$Y_{X, R_{500}}$ & M$_{\rm sun}\times$ keV & The mass proxy in $R_{500}$  \\
M$_{500}$ & M$_{\rm sun}$ & Cluster total mass \\
M$_{500}$\_PDF\_array & M$_{\rm sun}$ & PDF of cluster total mass \\
$f_{{\rm gas},500}$ &  & Cluster gas mass fraction \\
$R_{500}$ & kpc & Overdensity radius  \\
Match Name & & Name of the cluster if matched to a known confirmed source within $2\arcmin$ radius\\
\hline
\multicolumn{3}{l}{The Cosmology Sample} \\
\hline
z$_{\lambda}$ & & Photometric redshift of the cluster based on the LS~DR10 data\\
CR  & Counts s$^{-1}$  & Count Rate measured at $z_\Lambda$ \\
$L_{X}$ &  ergs s$^{-1}$ & Integrated X-ray Luminosity within $R_{500}$ at $z_\Lambda$ \\
\kt\ & keV  & ICM temperature at $z_\Lambda$ \\
M$_{{\rm gas}}$ & M$_{\rm sun}$ & Cluster gas mass within $R_{500}$ at $z_\Lambda$ \\
$Y_{X}$ & M$_{\rm sun}\times$ keV & The mass proxy in $R_{500}$ at $z_\Lambda$ \\
\hline\hline
\end{tabular}
\end{center}
\end{table*}
\subsection{Optical surveys}

Surveys executed in the optical band results in large catalogs of galaxy clusters. We compare our X-ray cluster sample with the following catalogs of optical galaxy concentrations: Abell, GOGREEN, DES, Wen \& Han, and MaDCoWS. The Abell catalog contained 1,059 nearby rich clusters with redshift measurements \citep{Abell1989, Andernach1991}; 466 are in our \erass\ footprint. We find only 152 clusters in our catalog within $2\arcmin$ of these Abell clusters. The low matching fraction is partly due to the misclassification and projection effects in the Abell catalog \citep{vanHaarlem1997}. One of the largest cluster catalogs compiled through the {\texttt RedMaPPer} algorithm is from the DES-Y1 data \citep{McClintock2019}. We find 781 close associations in our catalog with DES-Y1 clusters with a median sample redshift of 0.38 and a median separation of $17\farcs6$. One likely reason for the low matching ratios between \erass\ and optical catalogs, such as the Abell catalog, is the offset between the centering of the optically selected clusters. The offset between the X-ray centroid and optical centers is often larger than our matching radius of $2\arcmin$. 1eRASS~J103943.5-084115 at a redshift of 0.08 is a good example for this case. The optical centroid of A1069 is $6\arcmin$ away from the \rosi\ cluster center, although the match is clear due to the agreement in redshifts between two catalogs. Additionally, the limited red-sequence calibration ($z>0.05$) and flagged LS~DR10 photometry for very extended galaxies makes the identification of low-redshift \erass\ clusters ($z<0.05$) difficult, leading to incomplete \erass\ catalog in this redshift regime. The identification process and limitations are explained in detail in \citet{Kluge2024}.

The \rosi\ cluster catalog has a significant number of detections at high redshifts $z>0.8$. At these redshifts, optical identification through red sequence cluster finding tools in $g, r, i, z, W1$ bands becomes challenging due to the limited depth of the LS.
Dedicated infrared cluster surveys are helpful in the identification of high-redshift candidate clusters. \citet{Wen2018} compiled a catalog of 1,959 high redshift clusters in the redshift range of $0.7 < z <1.0$ based on the SDSS and Wide-field Infrared Survey Explorer (WISE). The common area of this survey with \erass\ is small, but we detect 10 of their high redshift clusters. GOGREEN (Gemini Observations of Galaxies in Rich Early ENvironments) and GCLASS (Gemini CLuster Astrophysics Spectroscopic Survey) surveys selected a sample of the {\it Spitzer} Adaptation of the Red Cluster Sequence (SpARCS) Survey \citep{Balogh2021, Wilson2009}, and SPT clusters in the redshift range of $0.8<z<1.5$, and have spectroscopically confirmed 26 clusters in this redshift range. The \erass\ sample has two close associations with this sample \citep{Bleem2020}. Another high redshift cluster survey is based on the data from the WISE mission, the Massive and Distant Clusters of the WISE Survey (MaDCoWS) \citep{Gonzalez2019}. Its main goal is to find galaxy clusters at $0.7<z<1.5$. We find 12 clusters in common associations in \erass. In the near future, the combination of {\it Euclid} and \rosi\ surveys will help confirm the majority of the \rosi\ high redshift candidates with the inclusion of $Y, J,$ and $H$ bands in the \eromapper\ pipeline and open this attractive redshift regime for the structure formation studies.

\begin{figure*}
\begin{center}
\includegraphics[width=0.49\textwidth ]{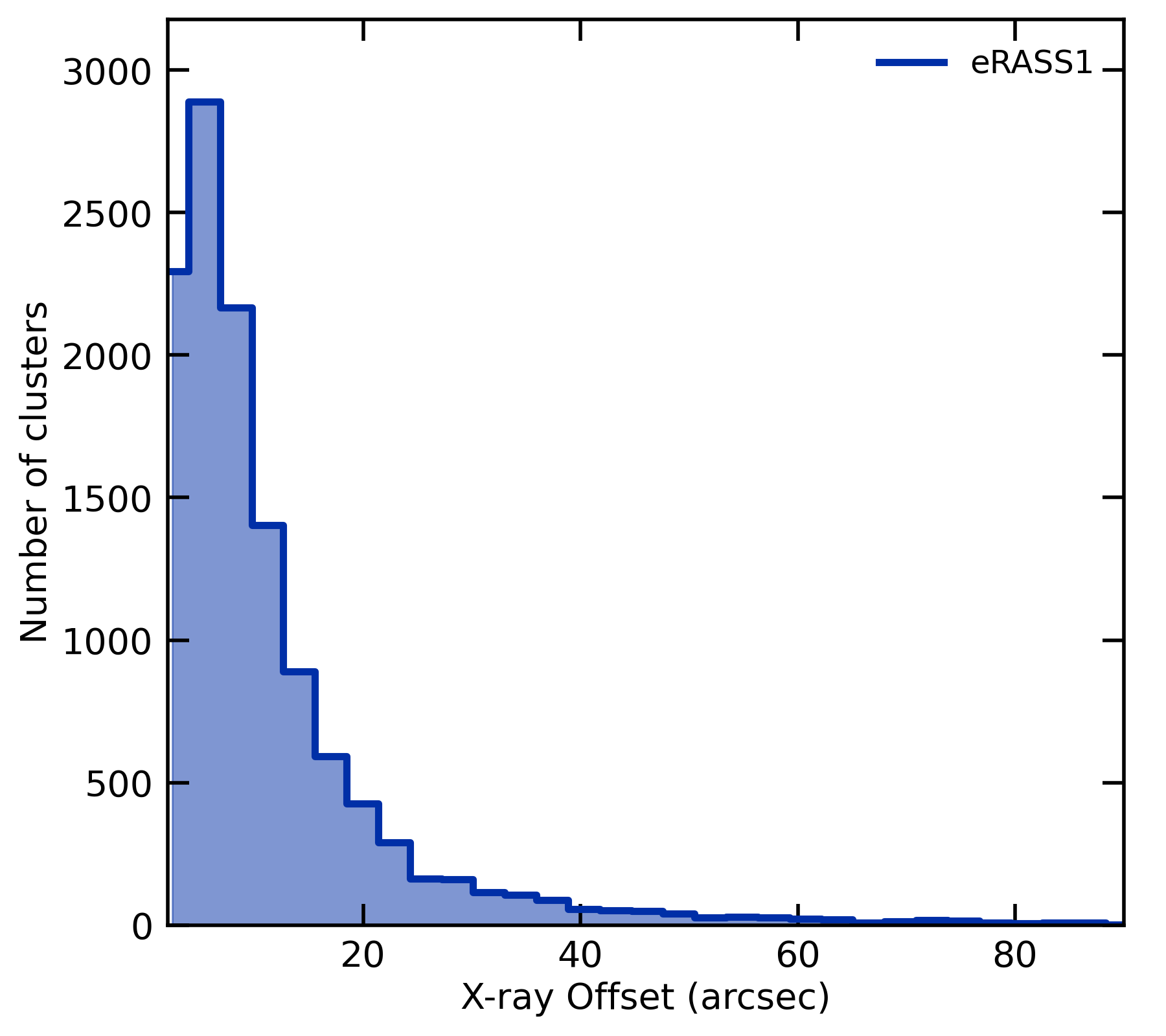}
\includegraphics[width=0.44\textwidth]{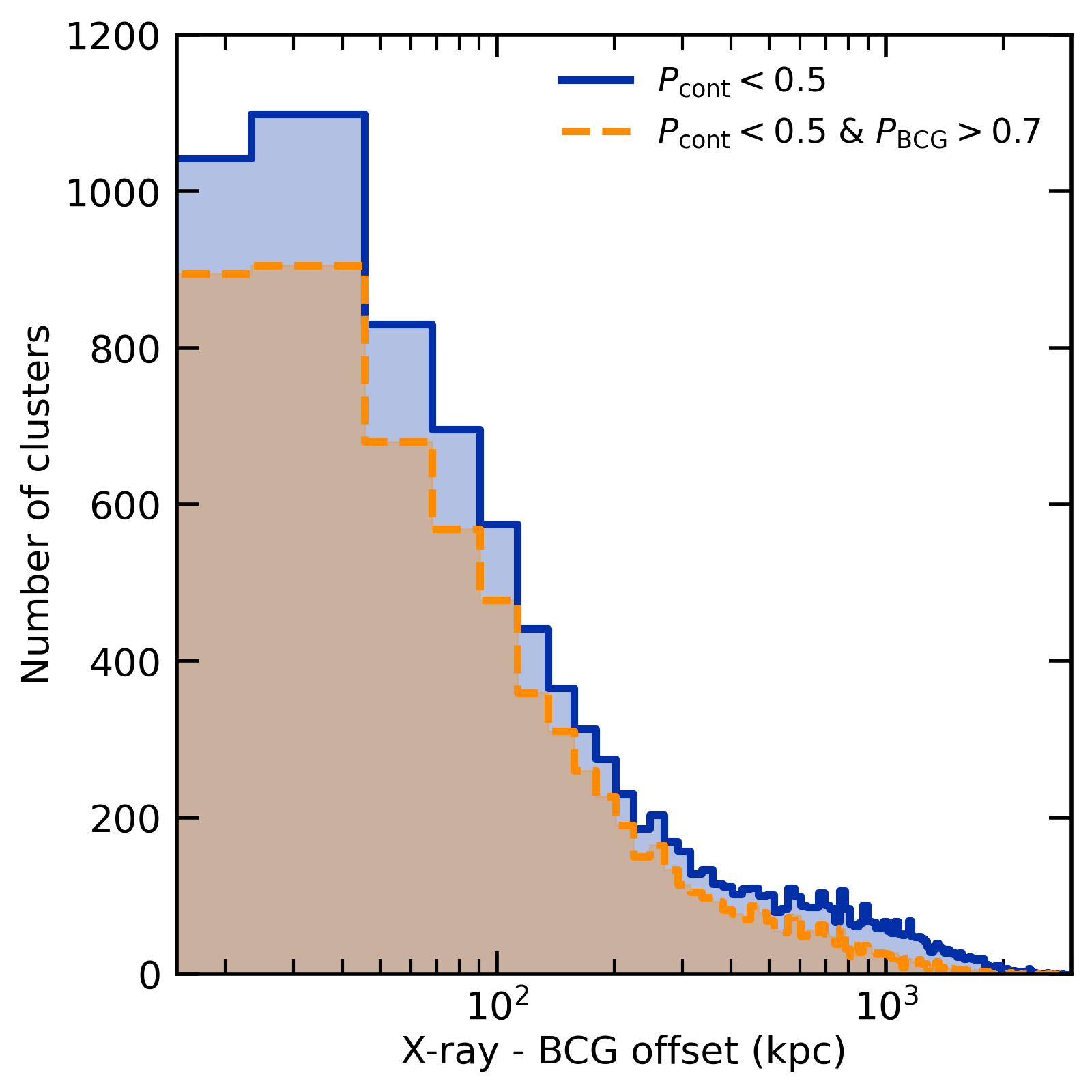}
\caption{Offset between the coordinates in the \esass detection chain ({\textit RA, DEC}) and the coordinates after post-processing of the data by \mbproj\ ($RA\_XFIT, DEC\_XFIT$), shown on the left panel, have a small median value of 7.7$^{\prime\prime}$. The tail of the distribution is due to the split extended sources in the {\it eSASS} catalog, cleaned carefully in this work. The offset between the coordinates indicated by the \mbproj\ and the BCG coordinates reported in \citet{Kluge2024} are presented on the right panel. The median offset between the X-ray center and BCG is 179~kpc for the full sample and 127~kpc after applying the selection $P_{\rm BCG}>0.7$ to the sample, selecting secure BCGs. Large offset values are an indicator of the merger status of clusters. }
\label{fig:cordoffset}
\end{center}
\end{figure*}
\section{X-ray properties of the \erass\ galaxy clusters and groups}
\label{sec:xrayprop}

One of the main goals of this paper is to provide X-ray properties of the clusters and groups detected in \erass. The initial source properties provided in \citet{Merloni2024} help define the sample, but quantities such as count rates and fluxes supplied by the source detection algorithm are only very rough estimates for clusters. Here, we improve them using more precise measurements, taking into account the corrections for Galactic absorption, $K$-factor, and a more accurate ICM and background modeling. 

\subsection{Post-Processing of the \rosi\ data}
\label{sec:xrayanalysis}

Further X-ray analysis of the \erass\ clusters is performed using the MultiBand Projector in 2D (\mbproj) tool \citep{Sanders2018}. \mbproj\ (Appendix \ref{sec:mbproj2d}) is a code that forward-models background-included X-ray images of galaxy clusters to fit cluster and background emission simultaneously. During this process, a Markov Chain Monte Carlo (MCMC) analysis is used to generate profiles of cluster physical quantities. Besides using a forward modeling approach, the main advantage of \mbproj\ is that the code allows simultaneous and self-consistent modeling of the surface brightness and temperature information and easier visual inspection for the goodness of the fit and background modeling. 

Given a sufficient number of energy bands, \mbproj\ can provide equivalent results from spatially-resolved spectral fitting, such as radial profiles of density, temperature, and metallicity. In recent work, \mbproj\ is applied to the deepest \rosi\ data currently available (eRASS:5) of the galaxy cluster SMACS~J0723.3$-$7327 \citep{Liu2023}, showing good performance. The X-ray properties determined using \mbproj\ are provided in both catalogs, primary cluster catalog and cosmology sample; see Table~\ref{table:catcolumns}.

Our \mbproj\ analysis of the \erass\ clusters is summarized as follows. We use images in multiple energy bands for the analysis to help constrain the ICM temperature. Specifically, for each cluster, we create images and exposure maps in the following energy bands (in units of keV): [0.3--0.6], [0.6--1.0], [1.0--1.6], [1.6--2.2], [2.2--3.5], [3.5--5.0], [5.0--7.0], using the {\tt evtool} and {\tt expmap} commands in \esass. The image size is $8~R_{500} \times 8~R_{500}$, which is large enough to include a local background region beyond the cluster. The initial $R_{500}$ adopted to determine the image size is a rough estimate using the $L-M$ scaling relation for eFEDS clusters \citep{Chiu2022}, where the luminosity is estimated measured from the \rosi\ X-ray data, and mass is measured from the weak lensing signal. A superior estimate of $R_{500}$ is provided later from the cosmology analysis (see Sec.\ref{sec:Xrayprop}), but we note that the fitting of the cluster properties is not very sensitive to the precise size of the image. Faint point sources with \texttt{ML\_RATE}~$<0.4$ in the 0.2$-$2.3~keV band within the image are masked, while brighter ones are fitted to ensure that the emission due to the outer wings of the PSF is properly modeled. This follows the approach adopted by \citet{Ghirardini2022}. PSF and ARF variations across different bands are adequately considered by creating separate response files for each band. 
In the fitting process, we allow the central coordinates of the source
to vary to enable more accurate measurements of the X-ray peak of the surface brightness of each cluster. We present the offset between clusters' initial coordinates provided by the source detection chain ($RA, DEC$) and the refined coordinates obtained from the \mbproj\ processing, namely, ($RA\_XFIT, DEC\_XFIT$) in the left panel of Fig.~\ref{fig:cordoffset}. The median offset of $7\farcs7$ is consistent with the $2\sigma$ range of the localization uncertainties of \rosi\ \citep{Brunner2022}. There is a sharp drop in the distribution of offset values larger than 10\arcsec. However, we observe a relatively large offset in a small fraction of the sample. This large offset is often due to the cleaning of `split' sources in the \erass\ catalog (see Section~\ref{sec:catalog} for the details of the cleaning procedure) and the improvement in the background subtraction and modeling of the ICM emission. In the post-processing with \mbproj, a more accurate background estimate is performed using a local region extracted around the cluster, with a more rigorous point source excision process. In contrast to \mbproj, the background map created by the source detection process may still include emission from bright point sources due to the wings of PSF or ICM emission in cluster outskirts, as the extended sources were excised within their extent ({\it EXT}) values in the creation of the background map. Additionally, we use a more flexible ICM electron density model with more free parameters than the $\beta$ profile with a fixed slope assumed during the source detection process. Therefore, the X-ray centers determined in our post-processing analysis are more accurate and are used in our cosmology and scaling relation analyses \citep{Ghirardini2024, Grandis2024}.

\begin{figure*}
\begin{center}
\includegraphics[width=0.97\textwidth]{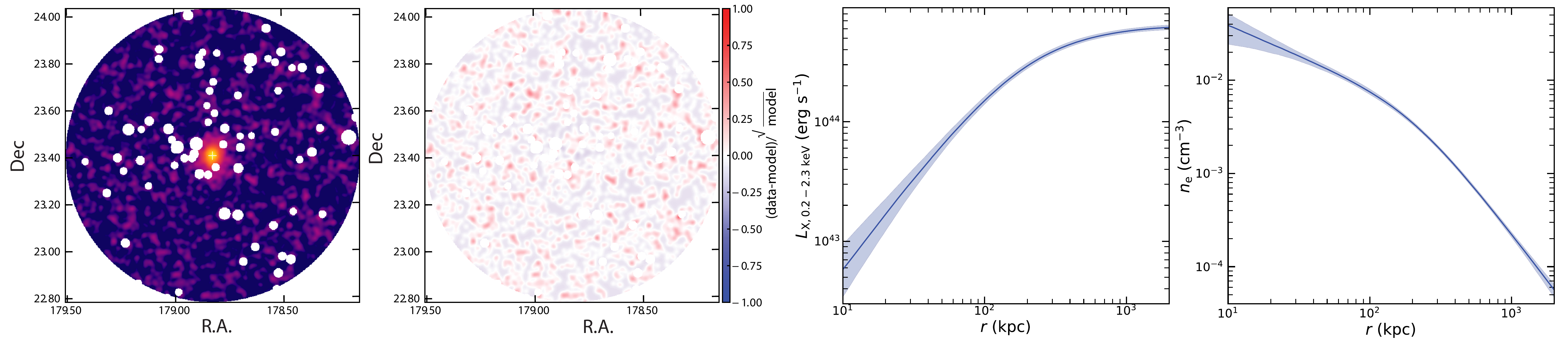}
\caption{Example of \mbproj\ analysis and results. The left panels show the 0.2--2.3~keV image of \erass~J115517.9+232422 and the residual image after \mbproj\ fitting.  Point and extended sources are masked from the image. The noise level in the residual image shows that the source emission is successfully modeled from the fitting process. The right panels are the luminosity profiles and electron density profiles.}
\label{fig:example_mbproj2d}
\end{center}
\end{figure*}

We use the electron density profile from \citet{Vikhlinin2006a}, but without the second $\beta$ component, which is not needed in our fits as most of the X-ray emission comes from the core region of the cluster:

\begin{equation}
n_{\mathrm p}n_{\mathrm e} = n_0^2 \cdot \frac{(r/r_c)^{-\alpha}}{(1+r^2/r_c^2)^{3\beta-\alpha/2}}\frac{1}{(1+r^{\gamma}/{r_s}^{\gamma})^{\epsilon/\gamma}}.
\label{eqn:density_model}
\end{equation}

\noindent The electron and proton densities are represented by $n_{\rm e}$ and $n_{\rm p}$, and we assume $n_{\rm e}=1.17n_{\rm p}$ \citep{Bulbul2010}. $\gamma$ is fixed at 3. $n_0$, $r_c$, $\alpha$, $\beta$, $r_s$ and $\epsilon$ are free parameters.

For each cluster, the MCMC chain produces posterior distributions of free parameters, such as the X-ray center, ICM temperature (\kt), and electron density model ($n_{\rm e}(r)$). X-ray properties such as flux and luminosity as functions of radius and energy bands are then derived from the chains. Most of the \erass\ clusters we analyzed in this work have fewer than a few hundred counts. Therefore, we do not attempt to constrain the 1D temperature variation but assume a constant temperature throughout the cluster. Metal abundance is fixed at 0.3$A_{\rm sun}$ due to the limited number of photons. The adopted solar abundances are from \citet{Asplund2009}. We use the HI4PI survey self-consistently to compute the Galactic neutral hydrogen column density throughout the analysis \citep{HI4PI2016}. We note that most of the confirmed clusters in our sample have relatively low $n_{\rm H}$ because they are located at high Galactic latitudes. Therefore, the effects of equivalent total hydrogen column density absorption in our results are negligible. Allowing $n_{\rm H}$ free to vary or adopting the total hydrogen density $n_{\rm H,tot}$ \citep{Willingale2013} does not change our measurements significantly. Due to the shallow nature of the survey, we are limited by the low number of counts in our analysis. Therefore, we use C-stat as the estimator for the goodness-of-the-fit to avoid any biases on the best-fit parameters \citep{Kaastra2017}. The physical properties of the clusters are computed at $z_{\rm best}$.

The offset between the fitted X-ray center and the brightest clusters galaxy (BCG) can be used to trace the dynamical state of a cluster \citep[e.g.,][]{Hudson2010, Rossetti2017, Seppi2023, Ota2023}. Shown in the right panel of Fig.~\ref{fig:cordoffset} is the projected separation of the X-ray ($RA\_XFIT, DEC\_XFIT$) and the BCG coordinates given in \citep[$RA\_BCG, DEC\_BCG$;][]{Kluge2024}. We convert the angular separation into physical scales (kpc) following the procedure described in \citet{Seppi2023}, using the $z_{\rm best}$ column. The \erass\ cluster sample with \extlike~$>3$ shows a median offset of 178~kpc. Selecting a more secure sample $P_{\rm cont}<0.5$ (see Table \ref{table:catcontam}) yields a very similar sample median offset of 179~kpc. Most of the \erass\ clusters have an offset smaller than $\sim200$~kpc, while a significant tail extending to several hundred kpcs and a slight excess around 1000~kpc can be observed in the distribution. We further cleaned the distribution by selecting galaxies based on their probability of being the BCG of the cluster ($P_{\rm BCG}$) \citep[see][for the probability $P_{\rm BCG}$ measurements]{Kluge2024}. The elimination of the galaxies that are less likely to be the BCGs, with a selection of the probability $P_{\rm BCG}<0.7$, removes many of the sources with large offsets (see Fig.~\ref{fig:cordoffset}) and brings the sample median to 127~kpc (22\arcsec). The clusters with large offsets are likely more disturbed, as pointed out in \citet{Seppi2023}. Detailed analysis of the morphology and dynamical status of the \erass\ clusters will be performed in another work (Sanders et al., 2024).

\begin{figure*}
\begin{center}
\includegraphics[width=0.49\textwidth ]{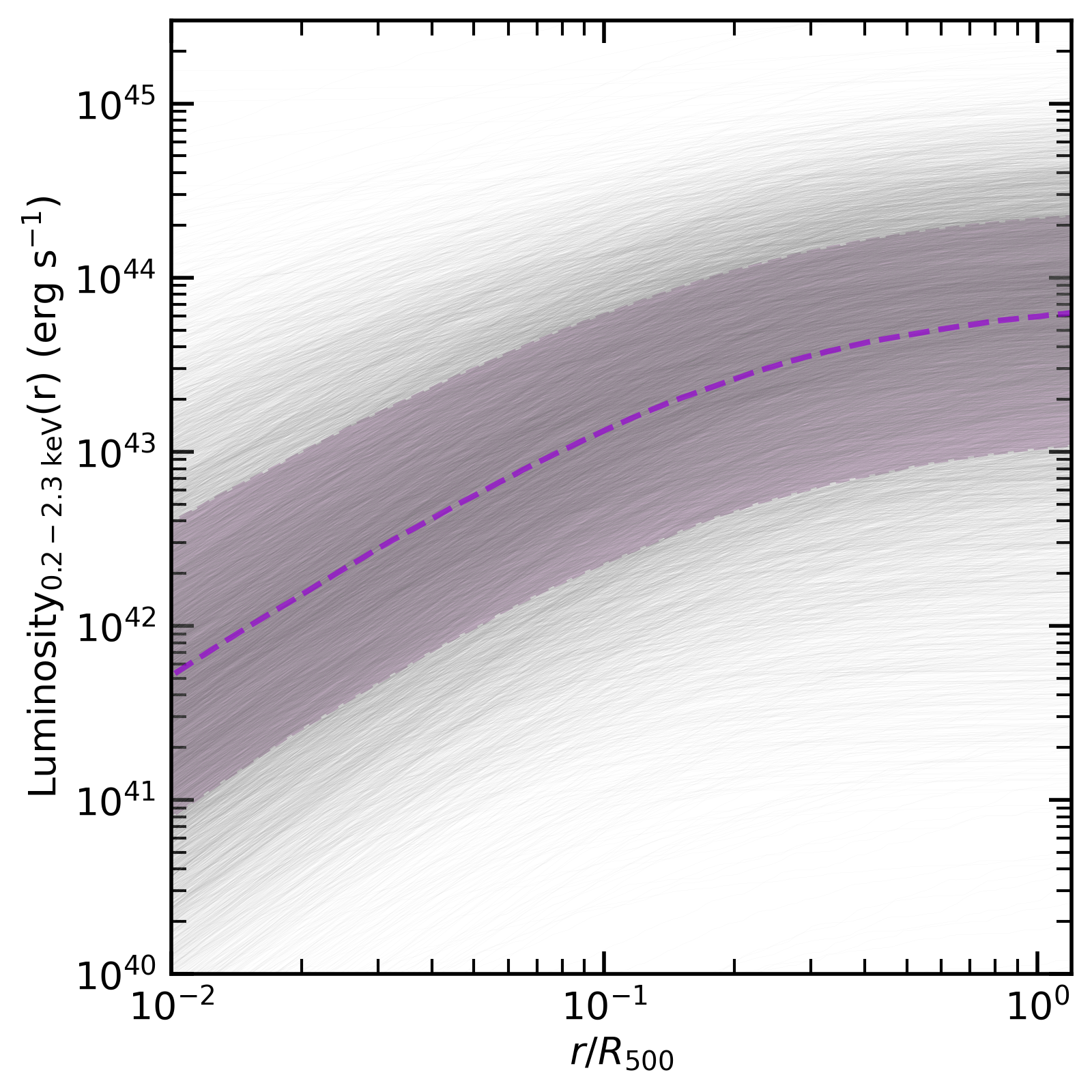}
\includegraphics[width=0.49\textwidth]{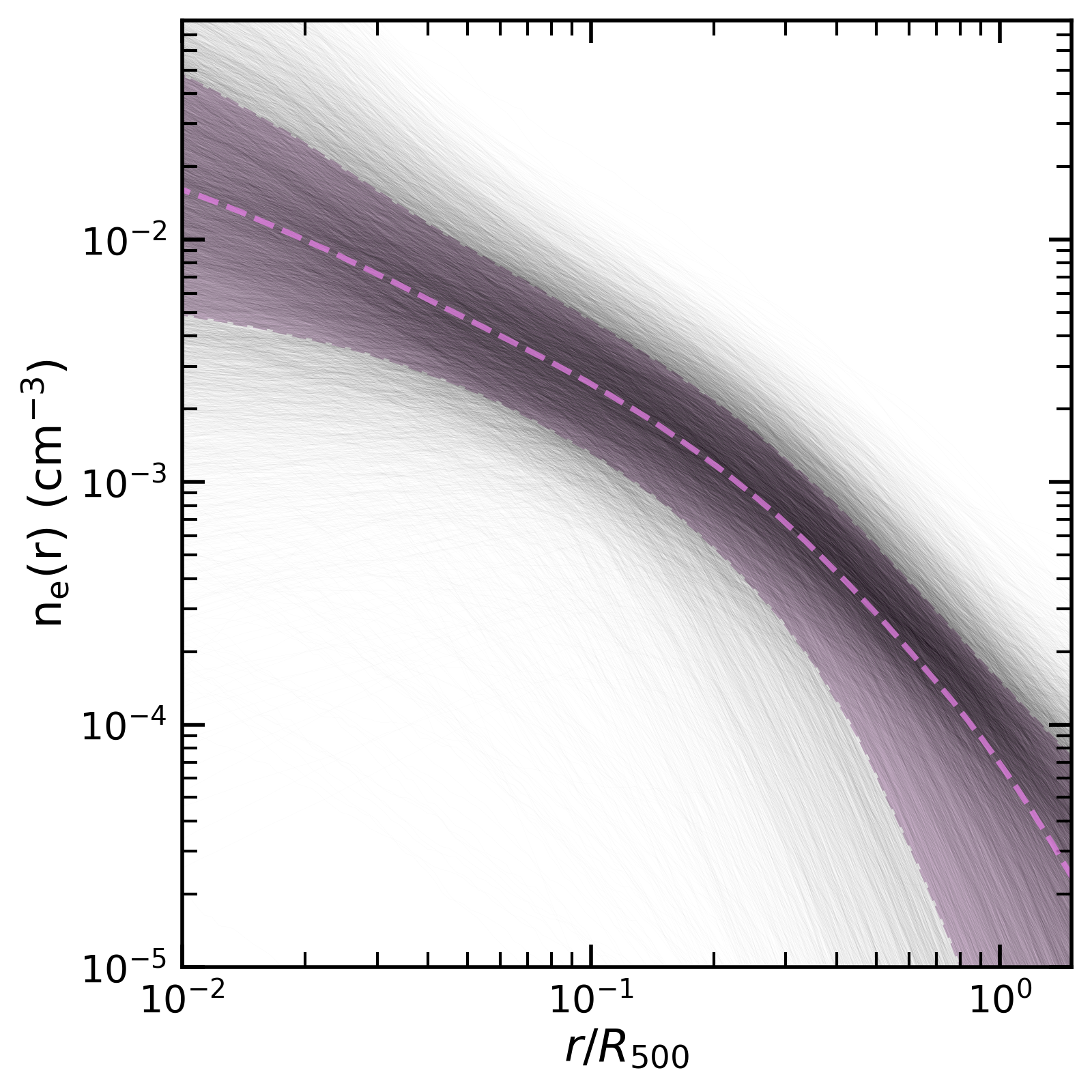}
\caption{Luminosity (left panel) and electron number density (right panel) profiles of the \erass\ clusters shown in gray curves. The dashed curve shows the median profile with the 84th and 16th percentiles plotted in shaded regions.}
\label{fig:neprof}
\end{center}
\end{figure*}

In Fig.~\ref{fig:example_mbproj2d}, we show examples of X-ray products for an example cluster (1\erass~J115517.9+232422). On the leftmost panel, the soft band image of the cluster is shown after the point sources in the field are excluded. The second panel to the left shows the residual noise image after the cluster emission is modeled. These images show that our analysis method can adequately model the emission from the contaminants and cluster itself. Visual inspections suggest that no excess emission is found in the residual plots of the \erass\ clusters. The two rightmost panels show the luminosity and electron number density profiles of the same cluster with the 68\% confidence intervals displayed in shaded regions. We note that the shallow nature of the survey does not allow us to make number density measurements at large radii, which is clear from the size of the uncertainties. The figure shows the extrapolations of the best-fit models in these regions. 

We also show the electron density profiles and luminosity profiles for all the clusters in the primary \extlike~$>\,3$ sample, after rescaling with $R_{500}$, which are plotted in Fig.~\ref{fig:neprof}. The median of the profiles is plotted in violet, with the 16 and 84 percentiles shown in shaded regions. The figure on the left panel shows that the sample consists of both cool-core clusters with high central density and non-cool-core clusters with flatter central profiles. The evolution and universal behavior of profiles will be discussed in the follow-up papers by Bahar et al. (2024, in prep.) and Moysan et al. (2024, in prep.)

\subsection{X-ray properties of \rosi\ clusters and groups}
\label{sec:Xrayprop}

In the primary catalog, we provide the X-ray properties, including the total number of counts, count-rate, flux ($F_{X}$), luminosity (\lx), temperature (\kt), $Y_{X}$ integrated over two radii (300~kpc and $R_{500}$) in two energy bands (0.2--2.3~keV and 0.5--2.0~keV) obtained from the runs with \mbproj. After obtaining the mass estimates, $R_{500}$ measurements are calculated by $R_{500}=\left( \frac{3}{4\pi} \frac{M_{500}}{500\rho_{c}}\right)^{1/3}$. The method used to calculate the total mass (M$_{500}$) is described in the following section. The critical density, $\rho_c$, is calculated at the cluster's redshift using the best-fit cosmological parameters from the \rosi\ cluster count measurements \citep[see][]{Ghirardini2024}.

In cosmological analyses utilizing cluster counts, it is essential to establish a mass calibration with scaling relations between the selection observables and cluster mass. The X-ray observable count rate offers a convenient mass proxy to describe a selection, which also has a well-established scaling with the halo mass \citep[see][]{Ghirardini2024}. However, the observable must reflect each cluster's intrinsic count rate; therefore, a similar X-ray reprocessing procedure must be applied to the cosmology sample. To estimate the intrinsic count rate of each source, the X-ray background must be treated carefully, and the Galactic column density ($N_{\rm H}$) and $K$-factor corrections must be accounted for. The only distinguishing component of this analysis is that the count rate measurements must be done at the photometric redshifts consistent with the cosmology sample, as justified in Section~\ref{sec:cosmosample}. Therefore, the only observables computed for this sample are the count rates within $R_{500}$. From hereon, we will present the properties of the primary \extlike$>3$ cluster sample.

\begin{figure}
\begin{centering}
\includegraphics[width=0.49\textwidth]{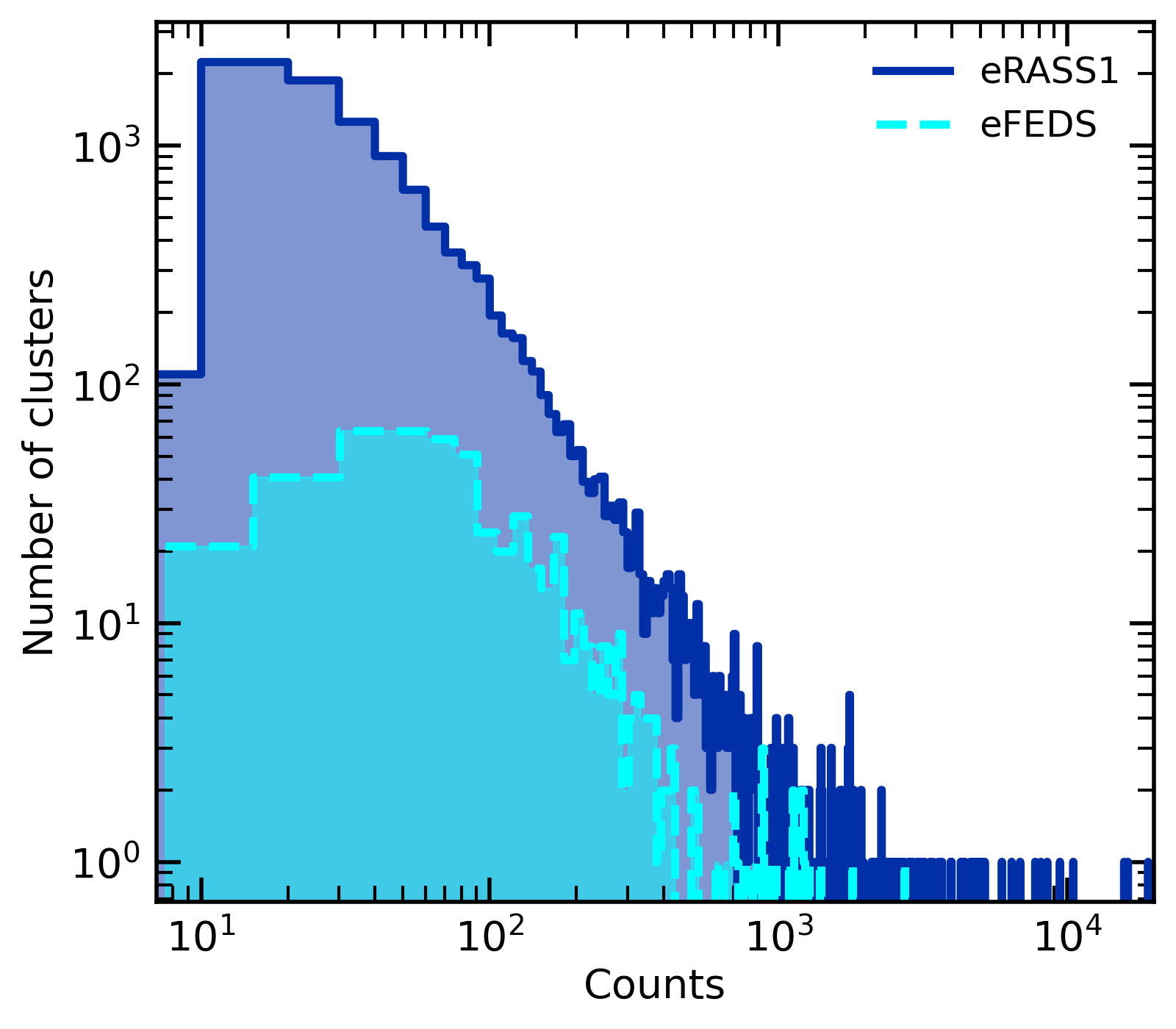} 
\end{centering}
% \vspace{5mm}
\caption{Counts distribution of the confirmed \erass\ and eFEDS clusters within $R_{500}$ in the reference band 0.2--2.3~keV \citep{Liu2022}. The majority, 10,178, of the \erass\ sample (83\%), have less than 100 counts within R$_{500}$. A small fraction, 157 clusters and groups (1\%), has over 1000 counts.  \label{fig:counts}}
\end{figure}

The shallow nature of the survey limits the number of X-ray counts collected within $R_{500}$ of each cluster. The count distribution in the 0.2--2.3~keV band of \erass\ clusters is displayed in Fig.~\ref{fig:counts}. eFEDS counts in the same band are also shown in the figure as a comparison. In the \erass\ sample, only a tiny fraction (148 clusters; 1.2\% of the sample) of clusters or groups have X-ray counts $>1000$, and they are primarily nearby clusters and groups. About 1892 clusters (15.5\% of the sample) lie in the 100-1000 count range. The majority of the sample, 10,207 clusters (83.3\%), have less than 100 counts.

\begin{figure*}
\begin{center}
\includegraphics[width=0.49\textwidth] {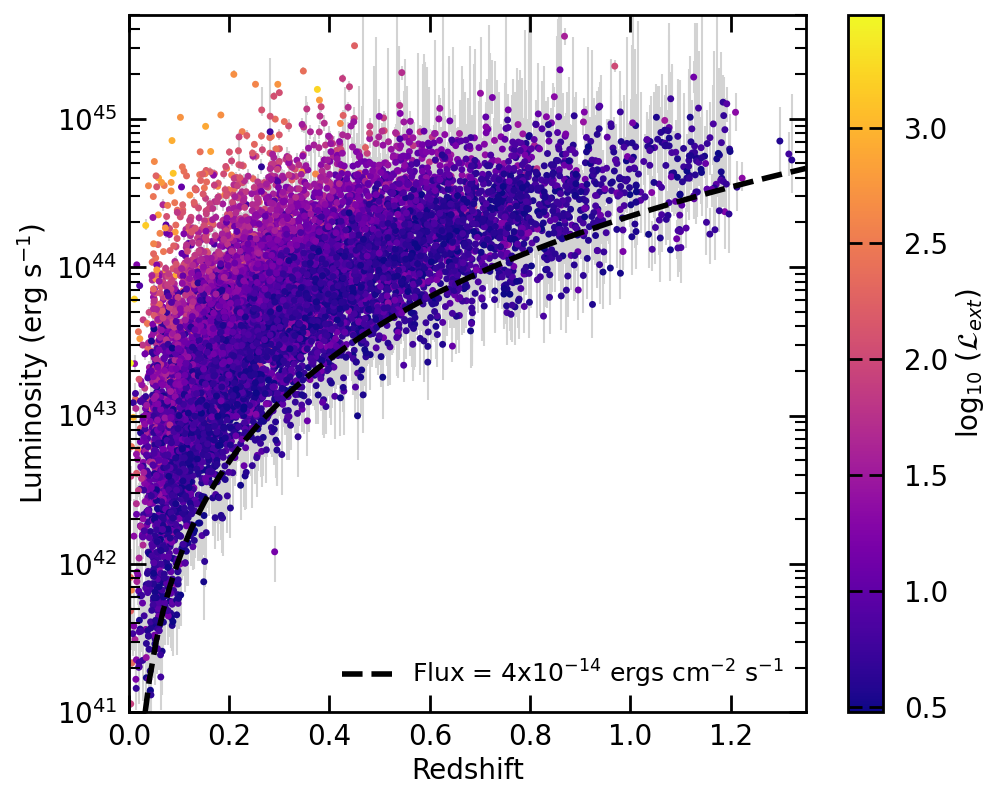} 
\includegraphics[width=0.49\textwidth]{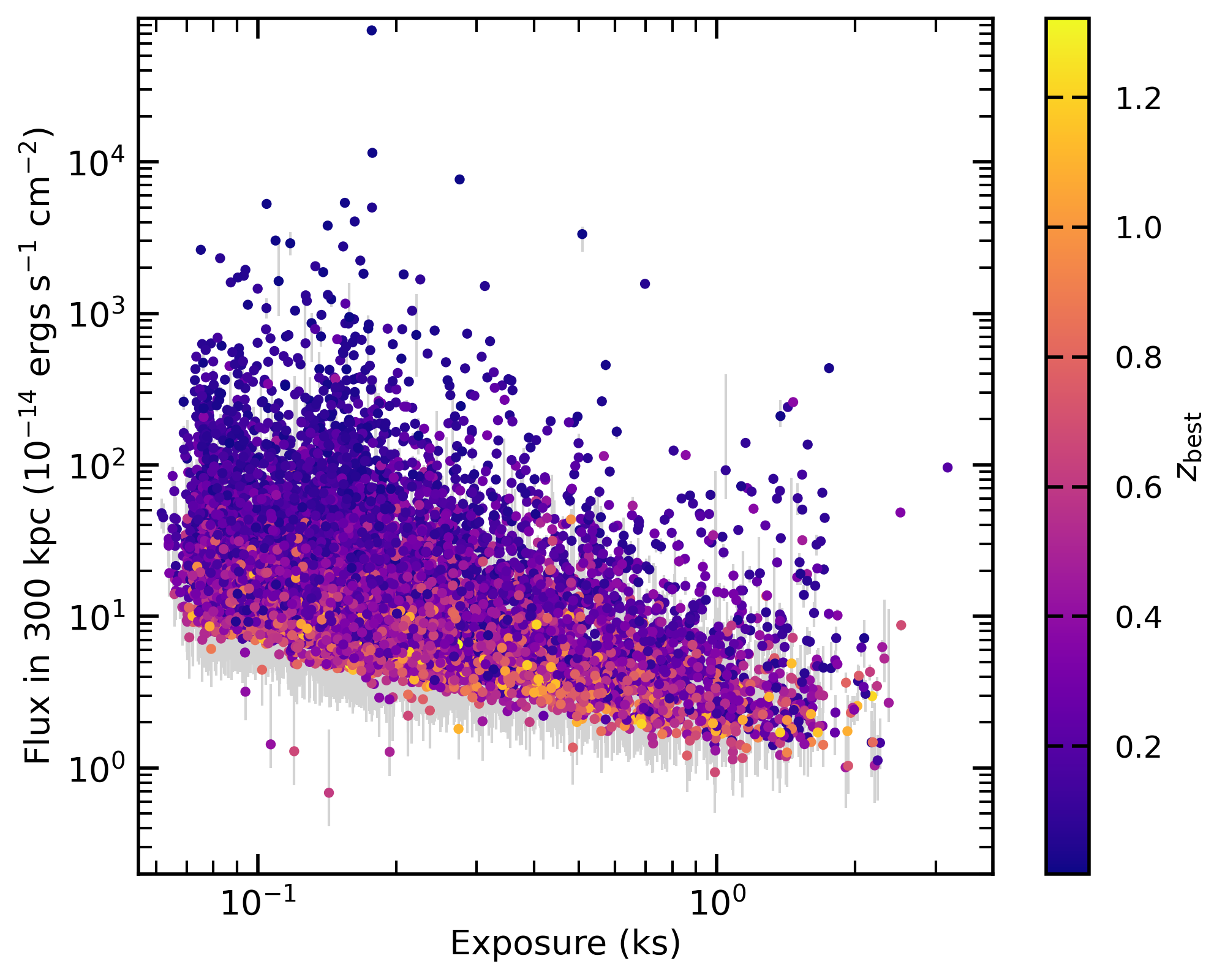}
\end{center}
\caption{Soft band luminosity distribution of the \erass\ clusters with redshift shown on the left panel. We only show the luminosities with significant detection at a $>2\sigma$ confidence for demonstration purposes. The flux value of $4\times10^{-14}$~ergs~s$^{-1}$~cm$^{-2}$ is overplotted as the dashed line. We do not use a flux limit to select clusters; the flux limit is plotted here as a reference. On the right panel, flux within 300~kpc is displayed as a function of exposure time. We only present the significant flux detections with detection confidence $>2\,\sigma$; the upper limits are excluded from the figure. The color bar indicates $z_{\rm best}$. The figure demonstrates that the detection of sources depends on the survey's depth. } 
\label{fig:L_legacy}
\end{figure*}

The commonly inferred physical property from X-ray imaging observations of galaxy clusters is the band averaged luminosity within a characteristic overdensity radius $R_{500}$. Ultimately, each cluster in the sample with a reliable redshift measurement has such a luminosity measurement. However, we only plot in Fig.~\ref{fig:L_legacy} to the left the luminosity measurements with significant measurement, namely, at a $>2\sigma$ confidence level, and exclude upper limits for display purposes. As a reference, we also show as a dashed line the estimated flux in the 0.2--2.3~keV band of $4\times10^{-14}$~erg~s$^{-1}$~cm$^{-2}$. Unlike previous \rosat\ cluster catalogs, we do not make flux cuts on the cluster sample since the \erass\ survey depth is not uniform across the sky, and the \rosi\ cluster survey is not a flux-limited survey. This effect is apparent in Fig.~\ref{fig:L_legacy} (right panel); the flux of the detected and identified clusters is a function of the depth of the survey. The dashed line nonetheless provides a rough estimate of the flux limit of the survey. 

The sample spans more than four orders of magnitude in luminosity, ranging from $1.1\times10^{41}$~ergs~s$^{-1}$ to $3.6\times10^{45}$~ergs~s$^{-1}$. The X-ray brightest cluster in the \rosi\ sky is 1eRASS~J010257.4-491609, a.k.a El Gordo, at a redshift of 0.86, followed by 1eRASS~J134730.8-114510 at a redshift of 0.45. About 84\% of the sample (10,410 clusters) has a luminosity measurement at the $>2\sigma$ confidence level or more. The most luminous 1\% of the sample, comprising 104 clusters, has luminosities larger than $8\times10^{44}$~erg~s$^{-1}$, while 3138 (25.6\%) of the sample lies in the faint end with luminosities \lx~$<1\times10^{43}$~erg~s$^{-1}$. The majority of the clusters, 73.5\% of the sample, lie between those two values. Most of the high-$z$ clusters have high luminosities above $10^{44}$~erg~s$^{-1}$ and low \extlike\ values $<10$, which is expected since we adopt a relatively low cut on \extlike, to keep the high-$z$ clusters in the \erass\ cluster sample, and rely on optical cleaning to remove AGN and other contaminants. Only bright and massive clusters are detected at high redshifts due to the `Malmquist bias,' inherent in all flux-limited surveys \citep{Malmquist1922}. The decrease in the number of bright clusters detected at low redshifts reflects the shrinking survey volume in the nearby Universe.

\begin{figure}
\begin{tabular}{c}
\includegraphics[width=0.49\textwidth]{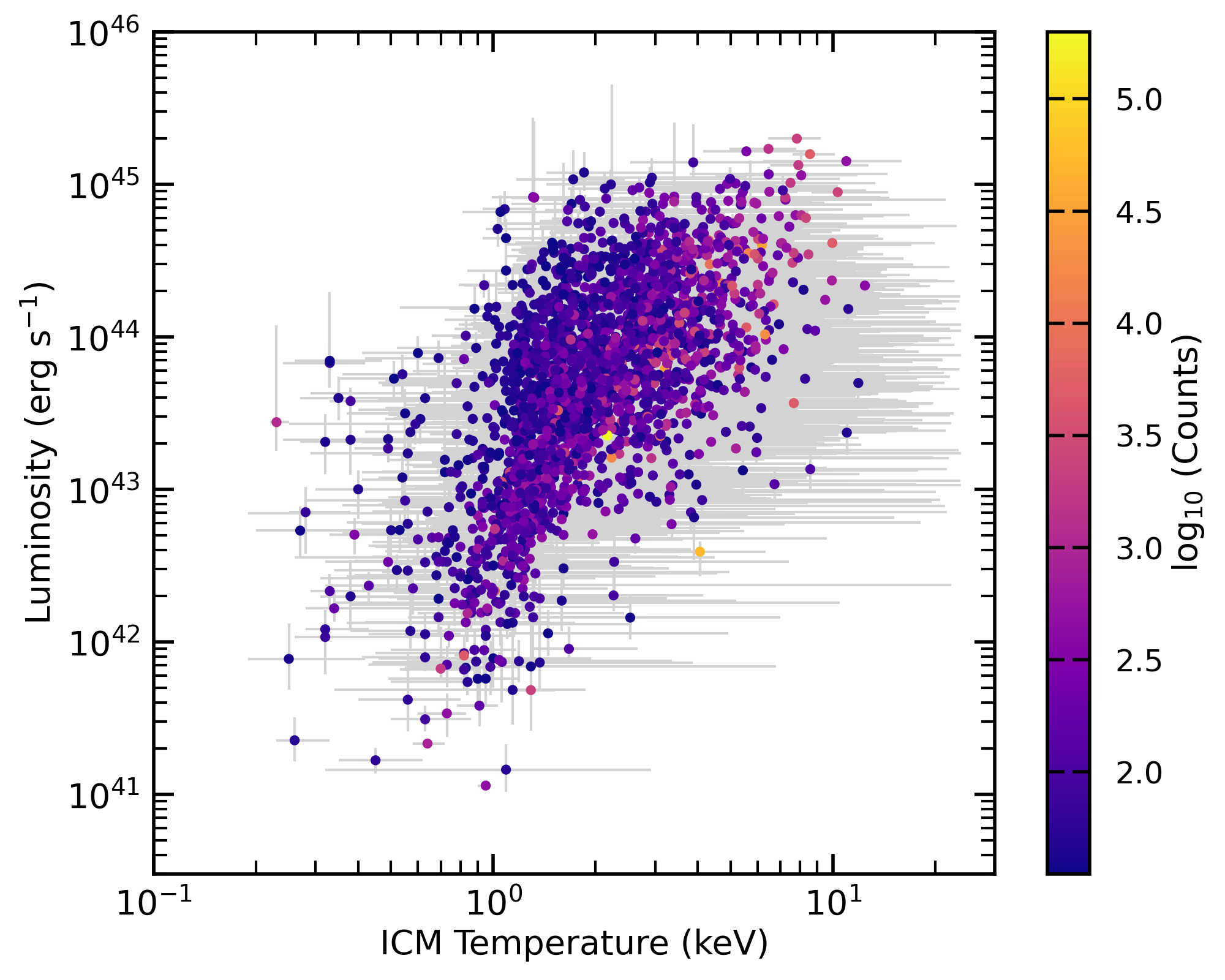} 
\end{tabular}
\caption{Soft X-ray luminosity in the 0.2-2.3~keV band and temperature relation for a subsample of clusters. 
The color bar indicates each measurement's counts within $R_{500}$. We only show a subselection of 2,443 temperature measurements with $>35$ counts in the sample. } \label{fig:lt}
\end{figure}

Naturally, due to the shallow nature of the survey, the small number of counts per cluster is insufficient for reliable direct temperature measurements. To quantitatively evaluate the reliability of temperature constraints for the sample, we examine the MCMC chain for each object. We compute a histogram of \kt\ points using 100 bins equally distributed in a logarithmic space from the minimum to the maximum point values in the chain. The computed histogram is an approximation of the marginal posterior distribution of temperature. We first calculate the 98.7\% (2.5$\sigma$) highest density interval (HDI) of the distributions. If the upper (or lower) boundary of the HDI reaches the parameter limit, we consider the \kt\ as unconstrained and only report the 68.2\% lower (or upper) HDI boundary as the lower (or upper) temperature limit. In summary, we have 142 and 8,599 clusters with only upper and lower temperature limits, respectively. We can constrain temperature for the remaining 3,506 clusters, where we take the highest peak of the distribution as the best-fit \kt, and the two 68.2\% HDI boundaries as the lower and upper limits. 
The ICM temperature measurements are provided in column \kt\ in the catalog. In Fig.~\ref{fig:lt}, we plot the temperature as a function of luminosity for 2,443 \erass\ clusters with temperature measurement and $>$35 counts within $R_{500}$. A total of 193 clusters have temperature measurements with a relative uncertainty of $<20$\%. We note that the temperature measurements of clusters with lower counts in $R_{500}$ may be unreliable. We find that in the temperature range considered in this work, we need at least 100~counts to measure \kt\ with 20\% precision, whereas 20\% precision on \kt\ can be achieved with 45~counts within $R_{500}$. The relatively large uncertainties on the temperature measurements at $>5$~keV are due to the decline of \rosi's sensitivity in the hard X-ray band \citep[see also][]{Bahar2022}. The temperatures obtained through \rosi\ spectroscopy will be presented in future work, where the trends with the number of counts and comparison with \mbproj\ results will also be investigated (Liu et al. 2024). The high observed scatter in the low-\kt\ regime at $z_{\rm best}<0.05$ could be due to the heliocentric redshifts used in measurements. The heliocentric-to-CMB corrections at very low redshift should be accounted for when the low-$z$ sample is used for astrophysical studies. The scaling relations between the X-ray observables, for instance, \lx\ and \kt, and their evolution with redshift can be employed to understand the role of gravitational physics in the formation process of galaxy clusters \citep{Pratt2019, Bulbul2019, Lovisari2021, Pratt2022}. The \erass\ sample will constrain the scaling laws between X-ray observables and halo mass both for galaxy groups (M$_{500}<10^{14}\, M_{\rm sun}$) and massive clusters (M$_{500}>10^{14} M_{\rm sun}$) self consistently by taking into account the selection effects for the first time with such a large sample  (Ramos-Ceja et al. 2024, Pacaud et al. 2024). 

\subsection{Gas mass and total mass measurements}

This section describes our method for measuring the integrated properties, including M$_{\rm gas}$ and M$_{\rm 500}$. To compute the gas mass of a cluster from the electron density profile, we use the enclosed ICM mass within a given aperture obtained by integrating the best-fit density model,

\begin{equation}
M_{\mathrm{ICM}} = 4\pi \mu_\mathrm{e} m_\mathrm{p}\int_{0}^{R} n_\mathrm{e}(r)\ r^{2}\ {\rm d}r,
\end{equation}

\noindent where we assume a metallicity of 0.3$A_{\rm sun}$ of the ICM and a mean molecular weight $\mu_{\rm e}$ of 1.18. The mass proxy $Y_{X}$ is defined as the product of the gas mass and the ICM temperature, $Y_{X}=M_{\rm gas}\times kT$.

\begin{figure}
\begin{center}
\includegraphics[width=0.49\textwidth]{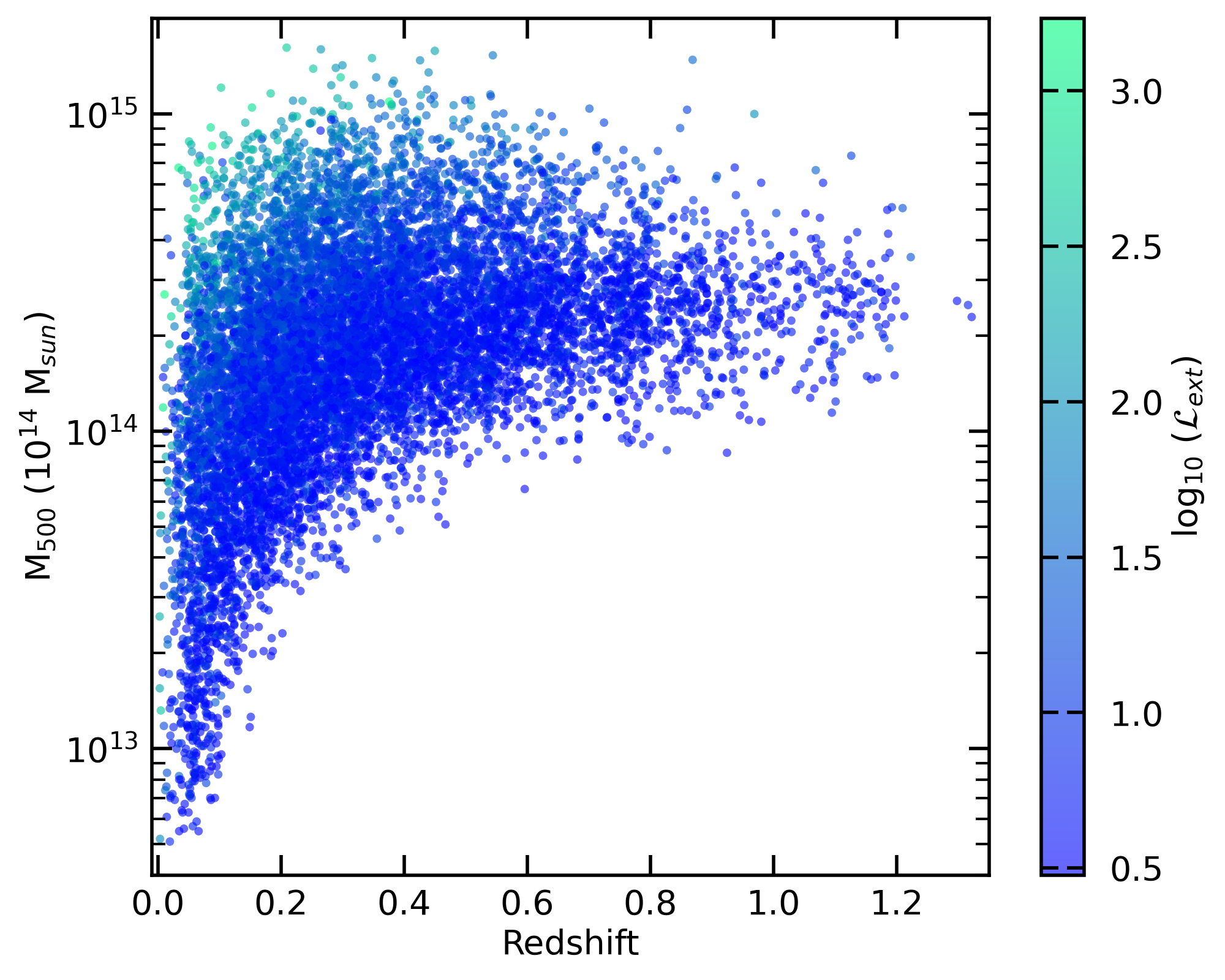} 
\caption{Right panel displays the mass-redshift relation of the \erass\ cluster sample. M$_{500}$ is estimated using the scaling relations between count-rate and shear profiles consistent with \rosi\ cosmology \citep{Ghirardini2024}.}
\label{fig:mass_z}
\end{center}
\end{figure}

Shown in Fig.~\ref{fig:mass_z} is the mass distribution of the primary \erass\ cluster sample, where the mass M$_{500}$ is estimated by our cosmology pipeline \citep{Ghirardini2024} after calibrating the count-rate scaling relation with the measured weak lensing shear data. These measurements are described in detail in \citet{Ghirardini2024}; here, we summarize the process in the cosmology pipeline. In short, we build the probability density function (PDF) of mass as a function of our observed count rate and redshift;

\begin{align}
    P(M | \hat{CR}, \hat{z}_{\rm best}, I) &= \frac{P(M, \hat{CR}, \hat{z}_{\rm best}, I)}{P(\hat{CR}, \hat{z}_{\rm best}, I)} \propto P(M, \hat{CR}, \hat{z}_{\rm best}, I) = \nonumber \\
= \iint & P(I | CR, z, \mathcal{H}_i) \cdot P(\hat{CR} | CR) \cdot P(\hat{z}_{\rm best} | z) \cdot \nonumber \\
& \quad \cdot P(CR | M, z) \cdot P(M, z)\ \ dCR\ \ dz,
\end{align}

\noindent where $P(I | CR, z, \mathcal{H}_i)$ is the selection function calculated at the sky position at each sky position $H_i$, the uncertainty on count rate and redshift are propagated through $P(\hat{CR} | CR)$ and $P(\hat{z}_{\rm best} | z)$. The employed scaling relation between count-rate and mass and the mass function is modeled with the $P(CR | M, z)$ and $P(M, z)$ terms in the equation. We note that the term $P(\hat{CR}, \hat{z}_{\rm best}, I)$ is not computed since it is just a mass-independent normalization factor. We compute the median and uncertainty on this PDF for 1000 MCMC steps of the cosmology run in the $\Lambda$CDM modeling, selecting these points randomly among the final 1000 steps with a thinning factor of 10, namely, among the best 12,000 ($\textrm{N}_{\rm steps} \times \textrm{N}_{\rm walkers}$). The total mass (M$_{\rm tot}$) and the gas mass fraction ($f_{\rm gas}= M_{\rm gas}/M_{\rm tot}$) of each cluster within $R_{500}$ is calculated using the \rosi\ cosmology given in \citet{Ghirardini2024} assuming the best-fit scaling relations between count-rate and weak lensing shear measurements.

We test the aforementioned method, inferring mass within an overdensity radius, $R_{500}$, from the scaling laws between count rate and mass on the mock observations generated for the cosmology pipeline to verify that the output masses are accurate and bias-free. We find the sum of all the PDFs follows the distribution of the true masses at any redshift bin on an ensemble level. The PDFs of the mass distribution of individual clusters are publicly available in the catalog. However, the median or mean mass values of the low-mass clusters (M$_{500}<\sim10^{14}$~M$_{sun}$) are biased high, and the median mass of high-mass clusters (M$_{500}> \sim 7\times 10^{14}$~M$_{sun}$) are biased low. This statistical issue occurs when a mass function and selection are applied to the sample, driving the masses from the count-rate versus mass scaling relations. To correct this bias, we force the mass distribution for each cluster to follow the predicted mass distribution given the cluster-specific count rate and redshift measured. This ensures that recovered mass distribution follows the expected from our modeling. It is important to note that in the cosmology pipeline, cluster masses are marginalized over a probability density function on a sample level; therefore, this observed bias does not affect the cosmology results reported in \citet{Ghirardini2024}. We only report the mass estimates where our selection function is reliable.

The masses of the clusters in the primary sample lie in the range of $5\times10^{12}$~M$_{\rm sun}$ to $10\times10^{15}$~M$_{\rm sun}$. The median mass of the primary sample within $R_{500}$ is $1.8\times10^{14}$~M$_{\rm sun}$. The 867 high mass clusters with M$_{500}>5\times10^{14}$ constitute about 7.1\% of the sample. The majority of the sample (71.6\%) lies in the intermediate mass regime between $1\times10^{14}$~M$_{\rm sun}<$~M$_{500}<$ $5\times10^{14}$~M$_{\rm sun}$. The most massive cluster in our sample is 1eRASS~150407.6-024816 (ACT-CL~J1504.1-0248, PSZ2~G355.07+46.20) with an estimated mass of $1.62^{+ 0.08}_{-0.12}\times10^{15}$~M$_{\rm sun}$ an extent likelihood of 479.6 at the redshift of 0.21. This cluster is followed by two clusters with similar mass measurements 1eRASS~J102350.1-271526 (PSZ2~G266.83+25.08, SPT ECS~J1023-2715) and 1eRASS~J134730.8-114510 RXCJ~J1347.5-1144, PSZ2~G324.04+48.79) with masses $1.59^{+0.18}_{-0.17}\times10^{15}$~M$_{\rm sun}$ at a redshift 0.2 and $1.58^{+0.10}_{-0.12}\times10^{15}$~M$_{\rm sun}$ at a redshift 0.45, respectively.

Owing to the superb sensitivity of \rosi\ in the soft X-ray band, we find a large number of galaxy groups in our catalog. Although there is no clear dividing line between clusters and groups in the literature, if we classify the latter as haloes with masses M$_{500}<10^{14}~M_{\rm sun}$, we find that about 21.3\% of the sample (2610 confirmed sources) lies in the galaxy group regime. The lowest mass group in the sample has a total mass of $5 \times 10^{12}\ M_{\rm sun}$; the groups detected in \erass\ represent the largest and cleanest sample to date and will extend our understanding of ICM thermal dynamical properties and AGN feedback to the low-mass regime \citep{Bahar2024}.

\subsection{Comparisons of X-ray properties with eFEDS and Chandra} 
\label{sec:chandracomp}

\begin{figure}
\includegraphics[width=0.49\textwidth]{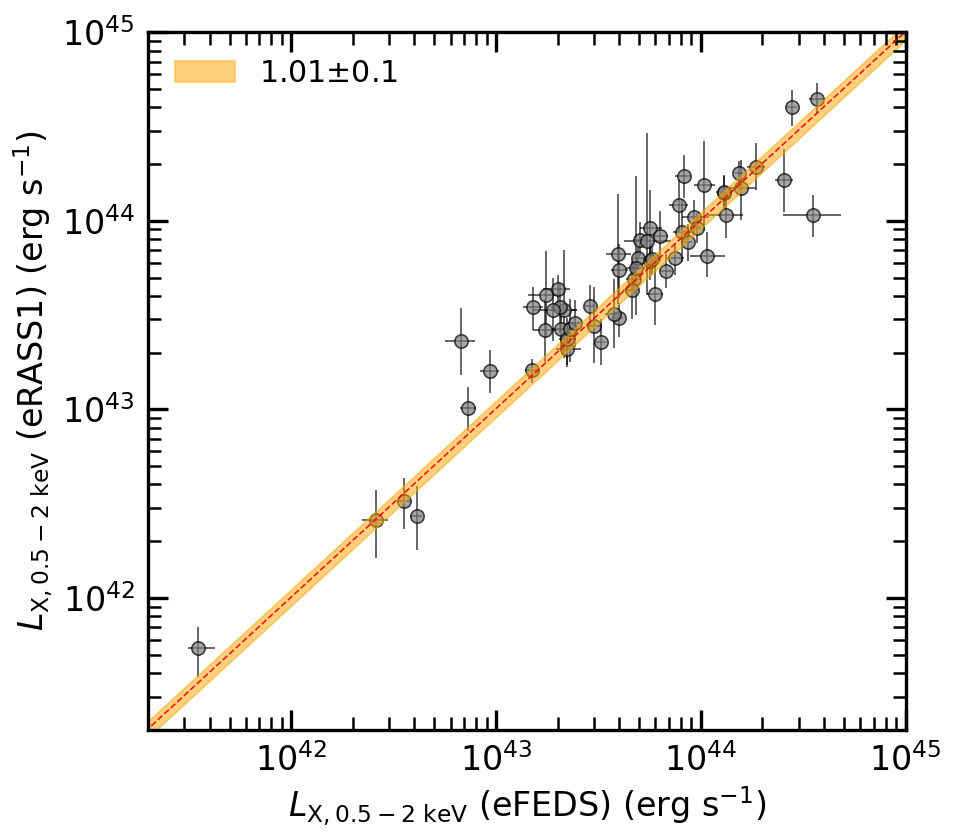} 
\caption{Comparison of the luminosities of the 62 matched clusters detected in \erass\ and eFEDS. The luminosities are measured within the same radius of 300~kpc and the same energy band of 0.5--2.0~keV. The best-fit line with slope $1.01\pm0.1$ is overplotted in the orange-shaded region. The overall agreement between eFEDS and eRASS1 results is seen regardless of the depth difference between the two surveys.\label{fig:efeds}}
\end{figure}

The eFEDS observations, approximately ten times deeper than \erass, allow us to test and compare our \erass\ results with our previous published work. Since eFEDS, we have updated the X-ray data processing software \citep[see][for details]{Merloni2024}. Additionally, our strategy for cluster analysis has changed with the addition of \mbproj\ to the analysis pipeline. 63 of 477 eFEDS clusters have counterparts in \erass\ (see Section~\ref{sec:crossmatch}). Among them, 62 have consistent redshifts within $\Delta z<0.05$, allowing for a direct comparison of the measured X-ray observables. One cluster has an inconsistent redshift as another cluster is detected in projection, leading to different redshift measurements between two different identification methods \citep[see][for further details]{Kluge2024}. We compare in Fig.~\ref{fig:efeds} the luminosities of these 62 clusters as reported in \citet{Liu2022} and measured in this work, within the same radius of 300~kpc and using the same energy band 0.5--2.0~keV. 

Although the methods used in the two works differ from each other, and the data have very different depths, the best-fit of the relation with a slope of $1.01\pm0.1$ indicates that there is no significant offset between eFEDS and \erass\ in Fig.~\ref{fig:efeds}, despite the \erass\ measurements having relatively large error bars due to the shallower data. A few clusters in the low $L_{X}$ regime have higher luminosities in the \erass\ catalog than those reported in the eFEDS catalog. In these cases, likely, the source counts are mildly boosted by a nearby AGN in the shallower \erass\ data \citep[see][for more information]{Clerc2024}. These AGN are removed from the analysis when measuring the eFEDS luminosities. In \erass, due to its shallower depth, they are not detected, their emission contributing to overall cluster \erass\ luminosity or the Eddington bias.

Deep \chandra\ archival observations of the previously known \erass\ clusters, in principle, provide essential tests for the accuracy of our flux determination method and calibration differences \citep{Liu2023}. To understand the (in)consistency in the calibrations between different instruments, comparing the luminosities obtained by \rosi\ with those produced using a different X-ray telescope is also helpful. A sound sample consists of the SPT-selected clusters observed by \chandra, which are massive, luminous, and relatively distant. We took the sample of 83 clusters analyzed by \citet{Sanders2018} and fit both the \chandra\ and \rosi\ data using \mbproj, using the same spatial region and point source mask, density parametrization, redshift, cosmology, and Galactic absorption.

In detail, we took the input \chandra\ images and background in ten bands, input masks, cluster centers, and point source positions from \citet{Sanders2018}, binning the images with 2\arcsec~pixels. Due to \rosi's PSF, we enlarged the point sources excluded in the masks to a minimum of $0\farcm35$ radius and excluded the data beyond a radius of 5~Mpc from the cluster center. These data were fitted with \mbproj, assuming the density model in Eqn.~\ref{eqn:density_model}, with an isothermal temperature with a flat prior in log space and a metallicity of 0.3A$_{\rm sun}$. As background models, we took the blank sky fields observation images, smoothed by a Gaussian with $\sigma=10$~pixels, and allowed their overall normalizations to vary freely in the analysis.

For \rosi, we construct images and exposure maps between 0.2 and 2.3~keV using data from the deeper eRASS:4 survey to obtain better statistics, process with the 020 version of the \esass\ pipeline \citep{Merloni2024}, 
using Telescope Modules (TMs) 1, 2, 3, 4, and 6. The data are filtered using the \esass\ tool {\tt evtool} \footnote{\url{https://erosita.mpe.mpg.de/edr/DataAnalysis/evtool_doc.html}} using the option "{\tt gti=FLAREGTI}" to remove periods containing flares. Subsection~\ref{sec:xrayanalysis} describes the rest of the analysis steps. Although this is a newer processing version than the primary dataset presented for \erass, the minor differences are insignificant for this analysis. We apply the same mask used for consistency in the \chandra\ data. These data are fit with \mbproj, assuming the same model as the \chandra\ data. However, to avoid issues due to temperature bias, that is, different temperature assumptions might change the luminosity measurements, we fit only the surface brightness in a single energy band (0.2-2.3~keV band). This approach does not allow temperature determination of ICM. We, therefore, assume a log-normal prior on the \chandra\ temperature posterior probability distribution when measuring the luminosity. The background is assumed to be flat over the fitted region.

\begin{figure}
    \centering
    \includegraphics[width=\columnwidth]{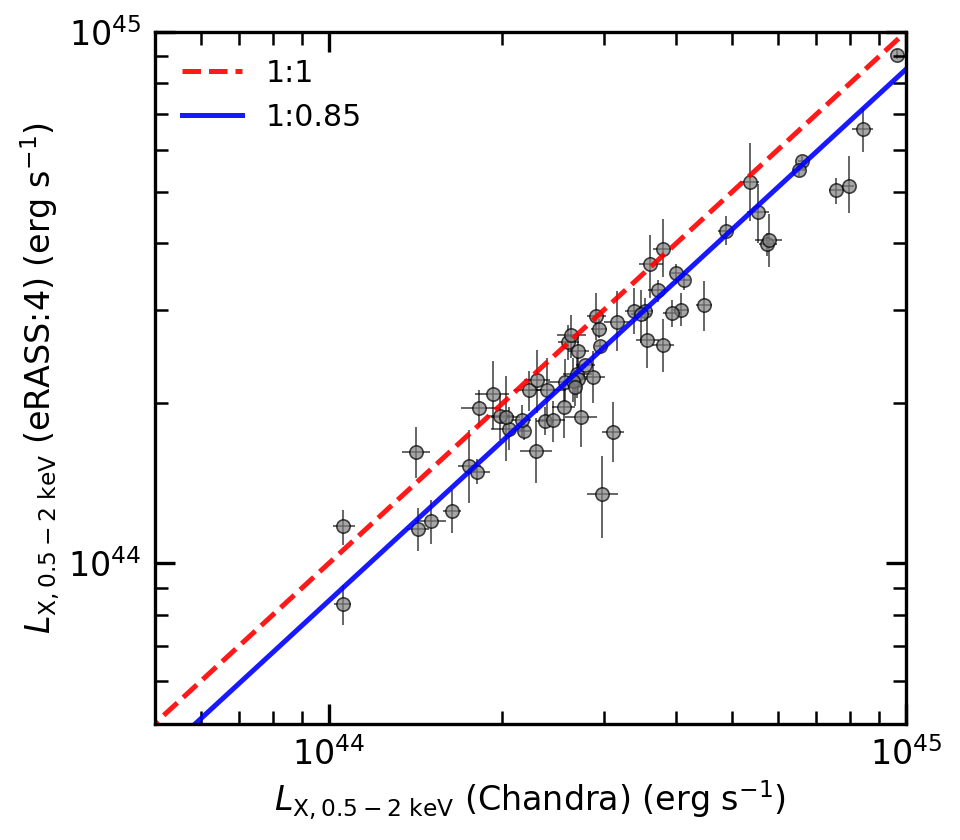}
    \caption{
    \chandra\ and \rosi\ luminosities of a sample of SPT-selected galaxy clusters. Luminosities are calculated in the 0.5--2.0 keV rest-frame band within a projected radius of 600~kpc. In this plot, only those clusters with an uncertainty of less than 10\% from \chandra\ and 20\% from \rosi\ are shown (66 out of 83). The median ratio between the values suggests that the \rosi\ luminosities are lower than the \chandra\ ones by around 15\%.
    }
    \label{fig:chandra_spt_comparison}
\end{figure}

Fig.~\ref{fig:chandra_spt_comparison} shows a comparison between the luminosities in the 0.5--2.0 keV rest frame band obtained by \rosi\ and \chandra, for a 600~kpc aperture in the clusters. Overall, the \rosi\ and \chandra\ luminosities scale together. However, the \rosi\ luminosities are, on average, around 15\% lower than the \chandra\ ones. We also made a more detailed analysis by doing a joint fit of the \chandra\ and \rosi\ data, allowing the overall cluster luminosity to vary between the telescopes. In this analysis, both data sets are fitted by the same density profile and temperature. In addition, we modeled the point sources near the cluster center in the \rosi\ data rather than masking them. This analysis shows fluxes around 15\% lower for \rosi\ compared to \chandra, consistent with the above analysis. In addition, no evidence for any flux-dependent differences is seen. Some fraction of this comes from the threshold issue, which causes some events to be lost in the processing \citep[][and K. Dennerl, private communication]{Merloni2024} although this effect is minor. We will investigate whether calibration differences such as effective area and vignetting could contribute to the remaining difference in the future. In-depth comparisons of cross-calibration between \rosi\ and  \chandra, \xmm\ measured temperatures are performed in \citep{Migkas2024}. We note that the discrepancy in the flux does not affect the cosmology results presented in \citet{Ghirardini2024} as the scaling relations are internally calibrated for all clusters. However, it should be accounted for in the systematic uncertainties in all other applications. 

\section{Population properties of the \erass\ clusters}
\label{sec:sampleprop}

In the previous sections, we provided the X-ray properties of the clusters and groups in the \erass\ sample. This section will compare the population with previous cluster surveys performed in the X-ray band. The completeness and purity of the \erass\ sample should be accounted for to make meaningful and unbiased comparisons with the literature. Our selection function is constructed based solely on the X-ray detection properties. It cannot replicate the $P_{\rm cont}$ measurements we apply to the sample due to the lack of richness information in the \rosi's digital sky \citep{Seppi2022}.
Accordingly, for population studies, we use only a subsample of 6,756 clusters selected with the \extlike$>6$ threshold, where we are confident about the selection process. The results in this section are provided for this subsample after correcting for the completeness and purity of the sample.

\subsection{\lognlogs}

\begin{figure}
\includegraphics[width=0.49\textwidth,]{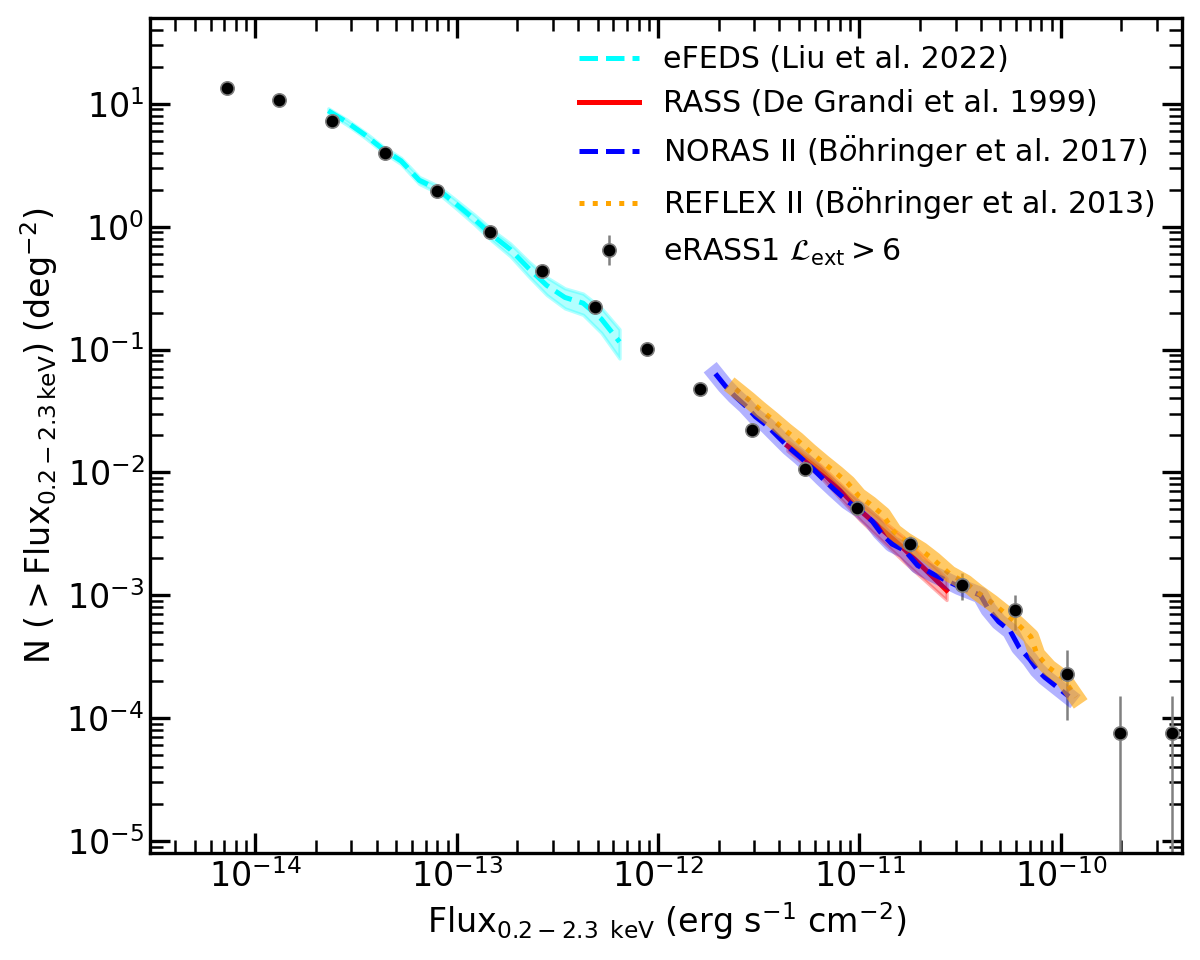} 
\caption{Cumulative cluster number counts as a function of X-ray flux within R$_{500}$ (\lognlogs) in the 0.2--2.3~keV band for the \erass\ cluster sample. A selection of \extlike$>6$ is applied to remove contaminants. Completeness is corrected based on the results of \citet{Seppi2022}. Also plotted are the eFEDS result \citep{Liu2022} and the results from REFLEX II and NORAS II cluster samples based on the \rosat\ All-Sky Survey \citep{Boehringer2013, Boehringer2017_norasII}.\label{fig:lognlogs}}
\end{figure}

 The observed cluster number counts ($N$) above a certain flux limit ($S$), defined as log$N$~-~log$S$ or d$N$/d$S$, is used to characterize and compare flux-limited cluster surveys \citep[e.g.,][]{Rosati1995, Reiprich2002}. In Fig.~\ref{fig:lognlogs}, we present the \lognlogs\ for the \erass\ clusters in the primary catalog in the \rosi\ reference band 0.2--2.3~keV, after accounting for completeness at each flux limit \citep{Seppi2022}. We base our calculations on the flux measurements within $R_{500}$, given in the catalog's $F_{X}\,[R500]$ column.

The \lognlogs\ distribution of the \erass\ sample covers more than four orders of magnitude in flux. Higher sensitivity combined with regions with deep exposure in the survey allows constraining the faint end of the distribution down the flux limits of $\sim 5\times 10^{-15}$~erg~s$^{-1}$~cm$^{-2}$. We have significant constraints in the brightest end of \lognlogs. We measure the slope to be $-1.2$ at high fluxes, while the slope turns shallower in the low flux end ($\sim 10^{-14}$~erg~s$^{-1}$~cm$^{-2}$) where the samples reach high completeness.

 We first compare our \erass\ measurements with the eFEDS \lognlogs\ measurements in the 0.5--2.0~keV band reported in \citet{Liu2022}. To convert eFEDS flux measurements to our default 0.2--2.3~keV band, we use the average conversion factor of 1.32. In general, the results of \erass\ are consistent with eFEDS. The X-ray-selected galaxy clusters in the REFLEX~II survey in the southern sky \citep{Boehringer2017_reflexII} and NORAS~II in the northern \citep{Boehringer2017_norasII} extragalactic sky can constrain the bright end of the \lognlogs\ curve. Both surveys have slightly steeper slopes ($-1.36\pm0.07$ and $-1.4$) compared to the \erass. The \lognlogs\ distribution is slightly under-predicted in both \rosat\ surveys compared to the \erass\ measurements at the bright end of the \lognlogs\ distribution. However, the discrepancy is only at the 1-2$\sigma$ level.

\subsection{The X-ray luminosity function}
Early structure formation models predict a significant increase in the number density of clusters and their evolution per unit volume \citep{Kaiser1986}. This prediction allowed the cluster luminosity and mass functions to be used as tools for constraining cosmology. The cluster X-ray luminosity function (XLF) is also used to study the large-scale structure for cosmological studies in the literature. It is studied in depth for decades in X-ray surveys \citep[see, e.g.,][]{Rosati1998,Vikhlinin1998,DeGrandi1999, Mullis2004, Koens2013,Pacaud2016,Boehringer2014_IV, Liu2022}. For instance, substantial negative evolution in the luminosity function is reported in the {\it Einstein} X-ray observatory-based cluster samples \citep[e.g.][]{Henry1992}. Although we do not use the XLF as a tool for cosmology, we derive the XLF of \erass\ clusters and compare it with the published catalogs in this section. 

The XLF of a galaxy cluster sample can be computed by counting the number of clusters per effective survey volume in different luminosity bins. This is also called the differential XLF and can be written as;

\begin{equation}
\frac{{\rm d} N}{{\rm d} L}(\langle L_i\rangle)=\frac{1}{\Delta L_i} \sum_{j} \frac{1-P{\rm cont}_j}{V_{\rm eff }[L_j, F_{\rm lim}, A]/P{\rm det}(L_j, z_j)},
\end{equation}

\noindent where $\langle L_i\rangle$ and $\Delta L_i$ are the center luminosity (we use $L_{X}$ within $R_{500}$ in 0.2--2.3~keV) and the width of the $i^{th}$ luminosity bin, $L_j$ is the luminosity of the $j^{th}$ cluster in the $i^{th}$ bin. $P_{\rm det}(L_j, z_j)$ is the detection probability of a cluster with luminosity $L_j$ at redshift $z_j$, obtained from the selection function. $V_{{\rm eff},j}[L_{j}, F_{\rm lim}, A]$ is the survey-effective volume corresponding to cluster $j$, as a function of its luminosity $L_j$, and the flux limit and sky coverage of the survey. To be able to compute $V_{\rm eff}$, we set a flux limit of $5\times10^{-14}$ erg~s$^{-2}$~cm$^{-2}$, and the corresponding survey area is 13,116~deg$^{2}$. The survey depth inhomogeneity is accounted for in the computation of the selection function, where $P_{\rm det}$ is given as a sky-average value. The \extlike$>6$ sample has $<5$\% contamination and the stable selection function; therefore, we only compare this sample with the literature.

\begin{figure*}
\begin{center}
\includegraphics[width=0.49\textwidth]{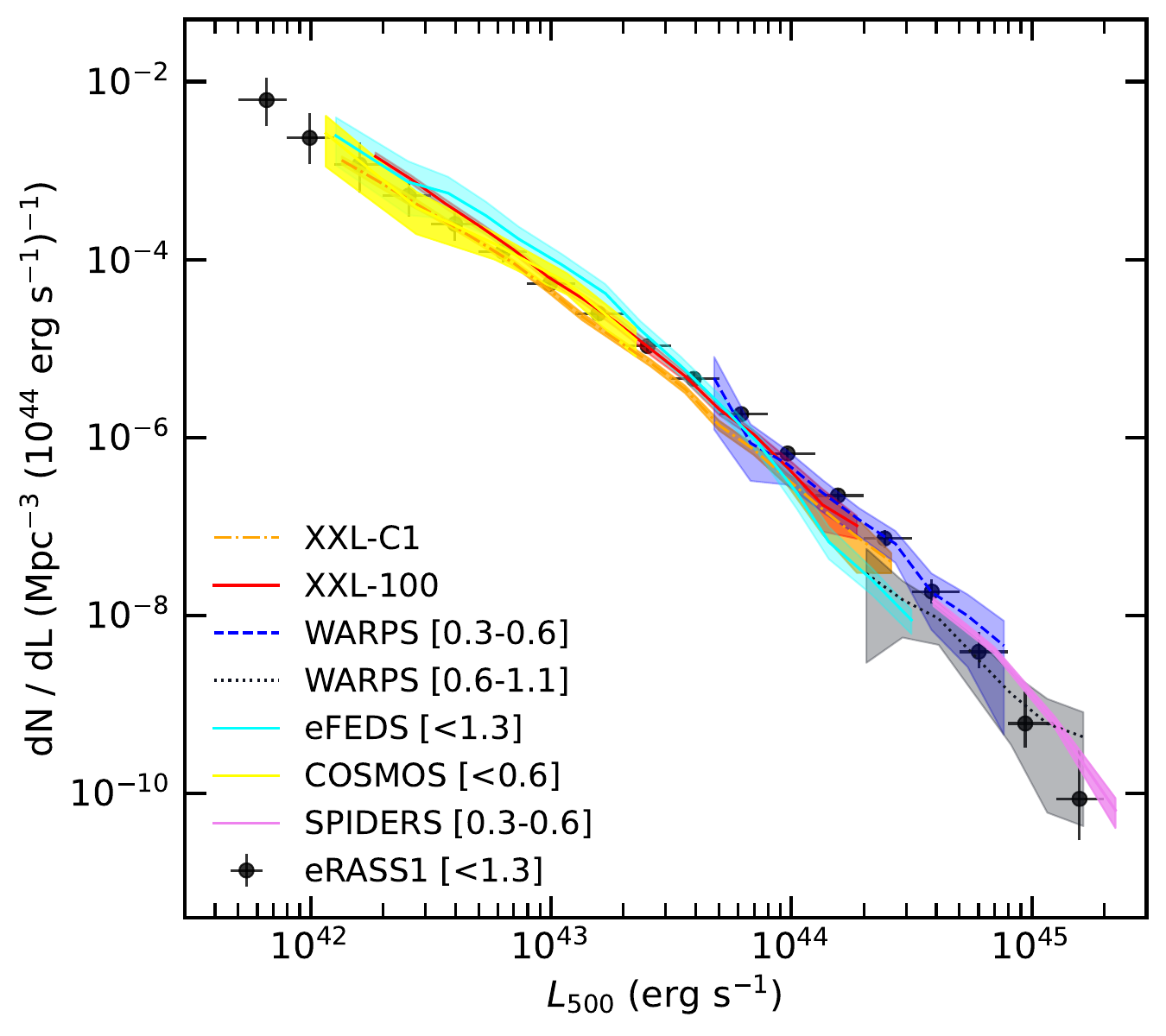}
\includegraphics[width=0.49\textwidth]{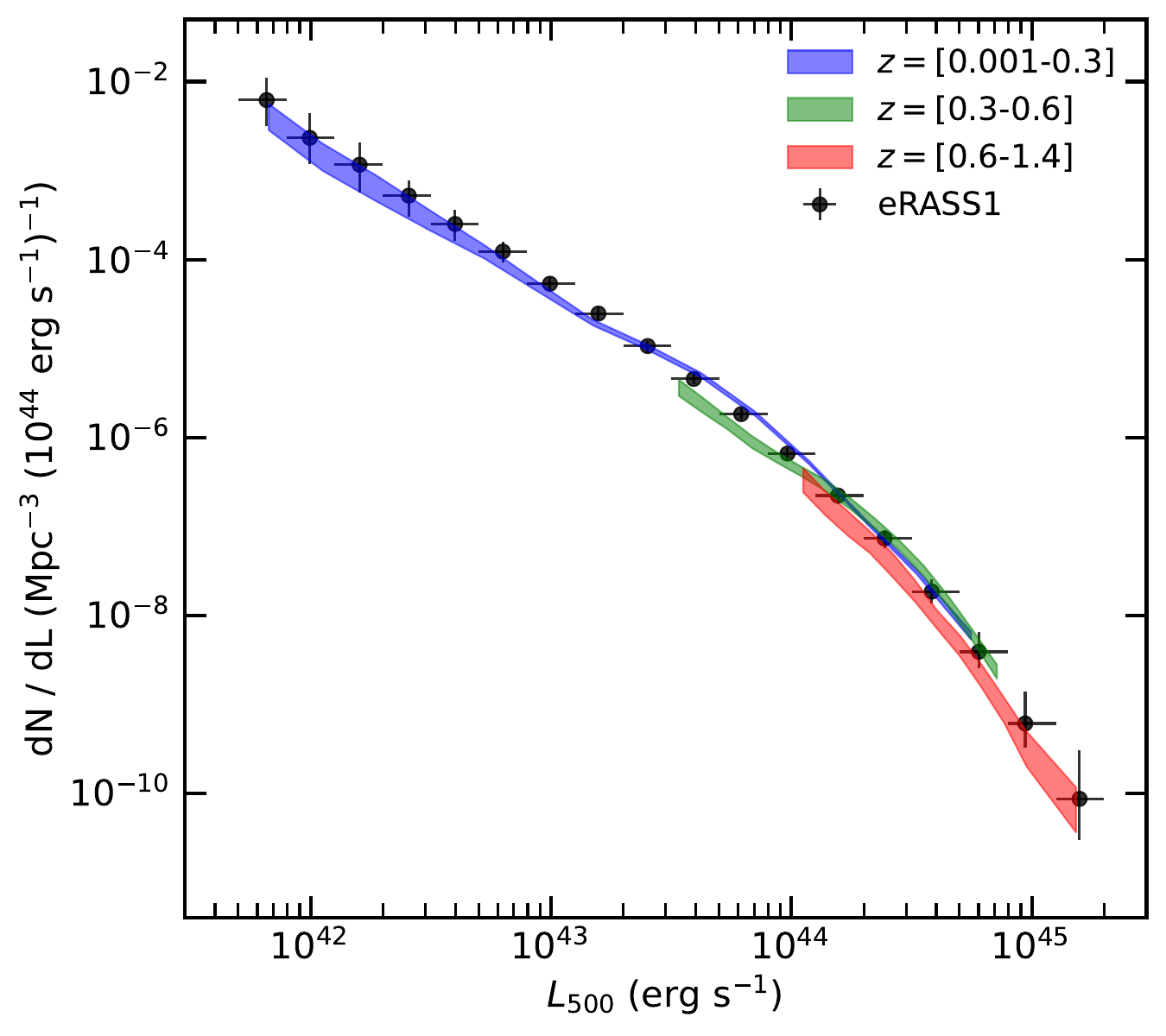}
\caption{X-ray luminosity function of the \erass\ cluster sample. {\sl Left panel:} Results for the full sample. Literature results are plotted as the shaded area for comparison: COSMOS \citep{Finoguenov2007}, WARPS \citep{Koens2013}, XXL-100 \citep{Pacaud2016}, XXL-C1 \citep{Adami2018}, SPIDERS \citep{Clerc2020} surveys. {\sl Right panel:} Results in three redshift bins. No significant XLF evolution is detected in \erass\ clusters.}
\label{fig:xlf}
\end{center}
\end{figure*}

For the full \erass\ cluster sample, we divide the luminosity range [$5\times10^{41}$--$2\times 10^{45}$]~erg~s$^{-1}$ into 18 bins with equal logarithmic width. Each bin contains at least 10 clusters. The uncertainty in ${\rm d}N/{\rm d}L$ is computed by randomly varying the luminosity for each cluster 1000 times over its statistical uncertainty. This error is then added in quadrature with the Poisson error for the number of clusters in each bin following the approach reported in \citet{Liu2022}. The XLF of the primary \erass\ sample is shown in the left panel of Fig.~\ref{fig:xlf}. We compare our results with a survey based on pointed deep \rosat\ PSPC observations of the \rosat\ All-Sky Survey selected clusters, that is, the Wide Angle \rosat\ Pointed Survey (WARPS) \citep{Scharf1997}. WARPS covers a similar luminosity and redshift range to \erass. Broad agreement between \erass\ and WARPS XLF \citep{Koens2013} can be seen in the left panel of Fig.~\ref{fig:xlf}. The XLF measured from the two samples in the XXL Survey, containing 365 and 100 clusters, is also consistent with our measurements at the 2$\sigma$ level. We also show the overall agreement between our results and the COSMOS survey in the redshift range of $0<z<0.6$ \citep{Finoguenov2007} and the SPectroscopic IDentification of \rosi\ Sources (SPIDERS) survey in the range $0.3 < z < 0.6$ \citep{Clerc2020}. The XLF for the SPIDERS and COSMOS surveys are adopted from \citet{Clerc2023}. We also compare our results with the eFEDS clusters sample presented in \citet{Liu2022}. We find that in the low-L$_{500}$ regime, the XLF in the eFEDS sample is overestimated and underestimated compared to the \erass sample. This difference could be due to the differences in the selection function \citep[studied in detail in][]{Clerc2024} or the high contamination level ($\sim20$\%) in the eFEDS sample. 

We also investigate the evolution of the XLF by dividing the sample into three redshift bins, $0.003<z_{\rm best}<0.3$, $0.3<z_{\rm best}<0.6$, and $z_{\rm best}>0.6$. The results are shown in the right panel of Fig.~\ref{fig:xlf}. We find that The XLFs in the three redshift bins are consistent with each other within $2\sigma$, indicating no significant evolution in cluster XLF, in agreement with the results of eFEDS \citep{Liu2022} and other literature \citep[see, e.g.,][]{Koens2013, Clerc2023}.

\section{Conclusions and summary}

We here present the catalog of clusters of galaxies in the Western Galactic hemisphere detected by the first \rosi\ All-Sky Survey in the soft energy band (0.2--2.3~keV). 

The \erass\ sample, with 12,247 confirmed galaxy clusters selected in the effective survey area of 13,116~deg$^2$, represents the largest statistically well-defined sample of galaxy clusters and groups based on the ICM selection to date. 

The primary cluster catalog is constructed based on a low extent likelihood criterion (\extlike$>3$) to maximize completeness and the detection rate of compact high-$z$ clusters and groups \citep{Merloni2024}. The estimated purity of this sample is 86\%, while the approximate flux limit is $4\times10^{-14}$~ergs~cm$^{-2}$~s$^{-1}$. The sample comprises clusters and groups in a wide redshift range from 0.003 to 1.32 with a sample median $z_{\rm best, med}=0.31$. The masses of clusters in the catalog span a large interval from $10^{12}\, M_{\rm sun}$ to $4\times10^{15}\ M_{\rm sun}$, with a median value of $3.2\times 10^{14}\ M_{\rm sun}$. We introduce a new cleaning strategy, the mixture model method, based on the X-ray and optical detection properties, which can increase the purity level to 95\% in the primary cluster catalog. When cross-matched with the other cluster catalogs compiled from X-ray, SZ, and optical surveys, we find that the \erass\ primary cluster catalog contains 8,361 unique detections (68\% of the sample), namely, newly discovered clusters or groups.

Cosmological studies benefit from the use of relatively pure galaxy cluster samples. An efficient way of reaching any desired purity levels is to select a sample with a strict X-ray extent selection. Cleaning the primary cluster sample with \extlike\ cuts ensures that the projection issues in the optical data through richness do not impact the final sample. With a stricter X-ray selection on the primary sample, for instance, taking \extlike\ $>10$ in the same footprint, cluster samples with 98\% purity can be obtained at the cost of reducing the number of clusters by around a factor of 3. 
On the other hand, increasing the \extlike\ cut to $>6$ in the 12,791~deg$^{2}$ LS~DR10-south footprint for the cosmology sample gives purity levels on the order of 95\% in a sample of 5,259 clusters. 

To provide precise X-ray properties of the sample, we reprocess the \erass\ X-ray data for the clusters in the sample. The improvements include a careful treatment of the X-ray background with advanced modeling of the cluster emission, $K$-factor, and galactic absorption corrections. In the primary catalog, we provide the resulting X-ray properties of each cluster with a redshift measurement. These properties, including (but not limited to) counts, count-rate, luminosity, flux, ICM temperature, gas mass, total mass, gas mass fraction, and mass proxy $Y_{X}$, are provided in the primary catalog. For the cosmology sample, we provide the count rate, the primary X-ray mass proxy used in cosmological analyses, with the same advanced X-ray analysis applied to the primary sample to reduce the scatter between the mass proxy and cluster mass.

We find consistent results when we compare the reported observed properties, such as the band averaged luminosity, of 62 clusters commonly detected in eFEDS and \erass\ in the overlapping footprint. However, in a sample of massive clusters observed with \chandra, we find that the luminosity measured by \rosi\ in this work is, on average, $\sim$15\% lower than the \chandra\ measurement. This calibration difference is an active area of investigation and will be addressed in future work.

The \erass\ cluster sample properties, such as \lognlogs\ and the luminosity function, offer an opportunity to compare the cluster population with previous X-ray cluster catalogs in the literature. The catalog's selection function, purity, and completeness are considered when studying both \lognlogs\ and the XLF of the sample. While we extend the measurements of \lognlogs\ and XLF to five orders of magnitude, we find good agreement between the \lognlogs\ and XLF with previous X-ray surveys based on \rosat\ and \xmm\ data, including the XXL, REFLEX~II, NORAS~II, and WARPS surveys in the respective flux and luminosity ranges covered by these surveys. We find no significant evolution of the cluster luminosity function in the redshift range of 0.003 and 1.3, consistent with our previous eFEDS results.

The \erass\ primary cluster and group catalogs provide the largest ICM-based sample of galaxy groups and clusters of galaxies with a well-defined selection function and high purity. This sample will be employed in various astrophysical, large-scale structure formation, and cosmological studies and will result in novel results when the selection effects are adequately accounted for. We confirm the excellent performance of \rosi\ for clusters and groups science and show that pre-launch predictions for detecting 100,000 clusters and groups will be met when the final survey depth (eRASS:8) is reached at the end of the survey program.

\begin{acknowledgement}
The authors thank the referee for helpful and constructive comments on the draft. The authors thank Johannes Buchner and J. Michael Burgess for insightful discussions on the statistical methods of mass calculations. This work is based on data from \rosi, the soft X-ray instrument aboard SRG, a joint Russian-German science mission supported by the Russian Space Agency (Roskosmos), in the interests of the Russian Academy of Sciences represented by its Space Research Institute (IKI), and the Deutsches Zentrum f{\"{u}}r Luft und Raumfahrt (DLR). The SRG spacecraft was built by Lavochkin Association (NPOL) and its subcontractors and is operated by NPOL with support from the Max Planck Institute for Extraterrestrial Physics (MPE).

\\
The development and construction of the \rosi\ X-ray instrument were led by MPE, with contributions from the Dr. Karl Remeis Observatory Bamberg \& ECAP (FAU Erlangen-Nuernberg), the University of Hamburg Observatory, the Leibniz Institute for Astrophysics Potsdam (AIP), and the Institute for Astronomy and Astrophysics of the University of T{\"{u}}bingen, with the support of DLR and the Max Planck Society. The Argelander Institute for Astronomy of the University of Bonn and the Ludwig Maximilians Universit{\"{a}}t Munich also participated in the science preparation for \rosi.

\\

The eROSITA data shown here were processed using the eSASS/NRTA software system developed by the German eROSITA consortium.

\\
E. Bulbul, A. Liu, V. Ghirardini, C. Garrel, S. Zelmer, and X. Zhang acknowledge financial support from the European Research Council (ERC) Consolidator Grant under the European Union’s Horizon 2020 research and innovation program (grant agreement CoG DarkQuest No 101002585). N. Clerc was financially supported by CNES. A. Veronica acknowledges funding by the Deutsche Forschungsgemeinschaft (DFG, German Research Foundation) -- 450861021. T. Schrabback and F. Kleinebreil acknowledge support from the German Federal Ministry for Economic Affairs and Energy (BMWi) provided
through DLR under projects 50OR2002, 50OR2106, and 50OR2302, as well as the support provided by the Deutsche Forschungsgemeinschaft (DFG, German Research Foundation) under grant 415537506.

\\

The Legacy Surveys consist of three individual and complementary projects: the Dark Energy Camera Legacy Survey (DECaLS; Proposal ID \#2014B-0404; PIs: David Schlegel and Arjun Dey), the Beijing-Arizona Sky Survey (BASS; NOAO Prop. ID \#2015A-0801; PIs: Zhou Xu and Xiaohui Fan), and the Mayall z-band Legacy Survey (MzLS; Prop. ID \#2016A-0453; PI: Arjun Dey). DECaLS, BASS and MzLS together include data obtained, respectively, at the Blanco telescope, Cerro Tololo Inter-American Observatory, NSF’s NOIRLab; the Bok telescope, Steward Observatory, University of Arizona; and the Mayall telescope, Kitt Peak National Observatory, NOIRLab. Pipeline processing and analyses of the data were supported by NOIRLab and the Lawrence Berkeley National Laboratory (LBNL). The Legacy Surveys project is honored to be permitted to conduct astronomical research on Iolkam Du’ag (Kitt Peak), a mountain with particular significance to the Tohono O’odham Nation.

\\

This work made use of SciPy \citep{jones_scipy_2001}, matplotlib, a Python library for publication-quality graphics \citep{Hunter2007}, Astropy, a community-developed core Python package for Astronomy \citep{Astropy2013}, NumPy \citep{van2011numpy}. 
\end{acknowledgement}

\bibliography{erass1cat} % your references Yourfile.bib     

\begin{appendix} %First appendix

\section{MBProj2D}
\label{sec:mbproj2d}

MBProj2D\footnote{Source code can be found at \url{https://github.com/jeremysanders/mbproj2d} and documentation at \url{https://mbproj2d.readthedocs.io/en/latest/}}, a Multi-Band Projector in 2D, is a software code for fitting the X-ray images of clusters to determine their physical properties.
By fitting simultaneously in multiple energy bands, the software is sensitive to the temperature and metallicity of the gas, in addition to the density.
As the number of energy bands increases, the accuracy for temperature measurements should tend toward that of the spectral fitting.
The code is forward-fitting, where a physical model of the cluster is used to produce model images.
These images are then compared to the observed X-ray data to obtain a Poisson likelihood.
MCMC, using this likelihood and incorporating priors, can then be used to obtain posterior probability distributions of the model parameters.
The MCMC chains are post-processed to produce profiles of physical quantities and their uncertainties.
 \begin{figure*}[h!]
    \centering
    \includegraphics[width=\textwidth]{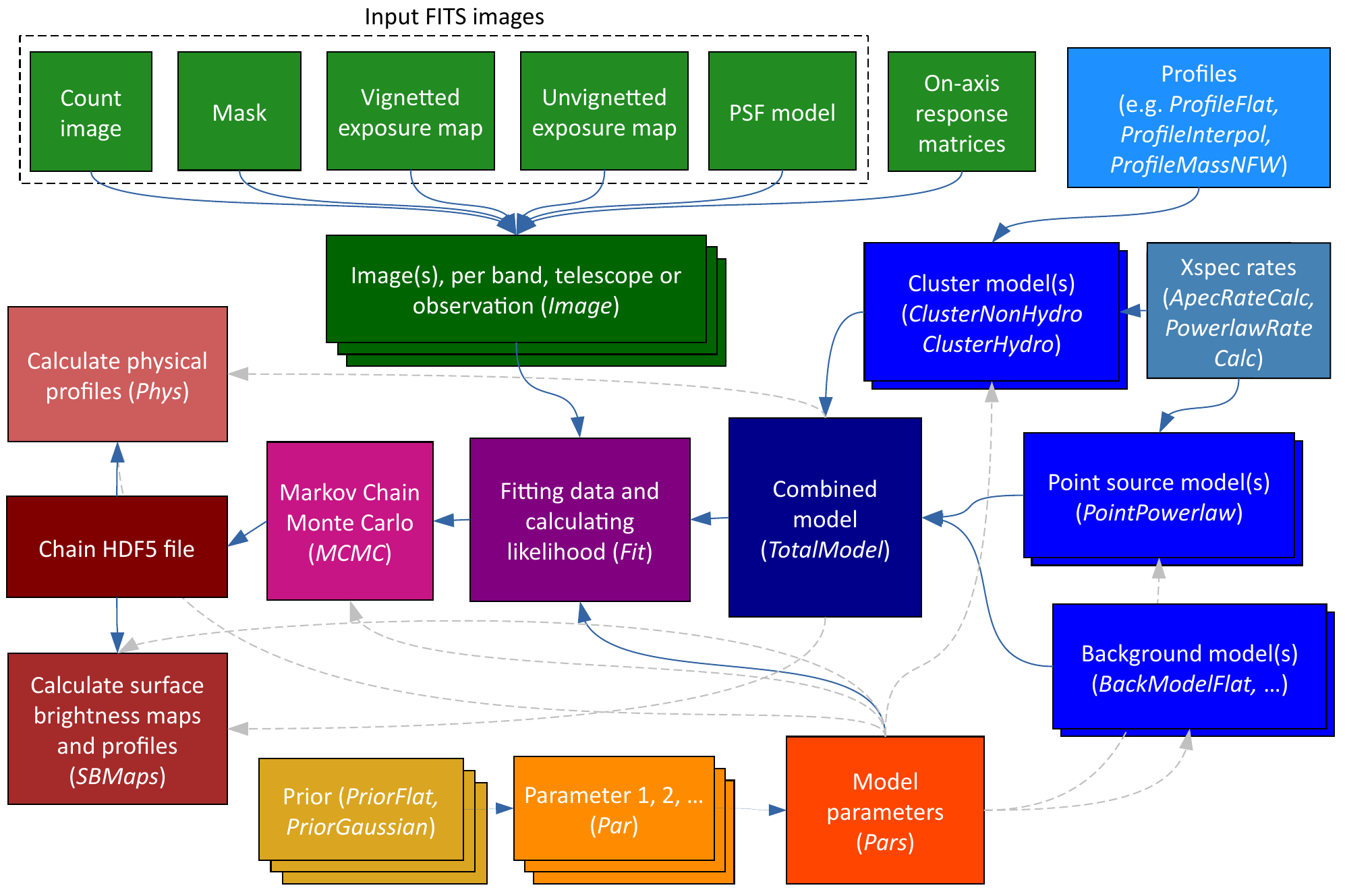}
    \caption{Block diagram showing the components of MBProj2D and their relationships.
    Names that correspond to classes defined in MBProj2D are shown in italics.
    Solid arrows show direct connections, while weaker connections are dashed.}
    \label{fig:mbproj2d_block}
\end{figure*}
MBProj2D builds upon the heritage of its predecessors, MBProj2 \citep{Sanders2018} and MBProj \citep{Sanders2014}, using the same primary method of fitting in multiple energy bands, although much of the code has been rewritten or restructured.
The main difference is that it now models clusters in two dimensions rather than using radial profiles.
This extension into 2D has the advantage that multiple possibly overlapping clusters can be simultaneously fitted by the code, which is necessary for a good description of \rosi\ data.
Point sources can also be included in the model rather than masked out of the images, essential if they are bright or near the cluster center.
Although two dimensions more easily allow the incorporation of the properties of the telescope and detector, such as the PSF or the background, a disadvantage is that it is more computationally expensive to model images rather than profiles.
The new version also removes several restrictions, including the cluster model being able to extend beyond the region examined.

The input data are statistically independent X-ray count images of the cluster in multiple energy bands.
A restriction removed in this version is that the data can come from different observations or X-ray telescopes, even if the pixel sizes differ.
As for the previous versions of the code, these bands should be chosen to cover the energy range of the instrument, taking into account the variation in the effective area as a function of the area and providing sensitivity to the gas temperature.
In addition, exposure maps are required in each band, where the value at a position is the exposure time in that band appropriate for a provided response matrix and ancillary response matrix. Currently, MBProj2D assumes that the same response matrix is valid over the field, which is correct for \rosi—however, the exposure map accounts for ancillary response variations such as vignetting.
The user can also provide a 2D PSF model for the telescope, which is assumed to be constant across the field.

MBProj2D is a flexible framework written in Python\footnote{\url{https://www.python.org/}} for describing a cluster field, fitting this model, obtaining uncertainties on it using MCMC, and then turning the acquired chains into physical profiles or model images.
The user writes a Python script to load the data, describe the model, parameters, and priors, and then conduct the analysis.
A block diagram of the various components is shown in Fig.~\ref{fig:mbproj2d_block}.
The code is highly extensible, allowing users to add multiple source and background components, parametrizations of radial properties, hydrostatic mass models, or Bayesian priors.
Some more computationally expensive routines are written in C++, including code for projecting the one-dimensional cluster profiles onto the sky and calculating model likelihoods.

MBProj2D can either fit clusters in which the temperature, gas density, and metallicity profiles are described using arbitrary parametrizations, or it can fit clusters under the assumption of hydrostatic equilibrium, where the user provides parametrized dark matter or total mass profiles and gas density and metallicity profiles.
In the hydrostatic case, the equation of hydrostatic equilibrium is used to calculate a pressure profile, and its temperature can be computed using gas density.

In detail, the parametrized cluster profiles are converted into 3D emissivity profiles for each energy band, using Xspec \citep{Arnaud1996} with the APEC model \citep{Smith2001} to convert between temperature, density, metallicity to emissivity, given absorption by a fixed Galactic HI column, calculated using TBABS \citep{Wilms2000}.
The profiles are projected onto the sky to make a rate image, given a parametrized cluster central position, out to some fixed physical cluster radius.
Although not used in this paper, the code can also include ellipticity on the sky in the model.
Added to the image are any point sources, where a power-law spectral index, normalization, and position parameters are used to describe the source.
The total model image is convolved with the instrument PSF using a Fast Fourier Transform to produce a smoothed model image.
This rate image is multiplied by the exposure maps to make a count image.
Background components are added to the source image to make a total model counts image.
As an observed image can include contaminating point sources or detector features, the user provides a mask image that selects the valid regions, for example.
Given the combined model count image, the mask image, and the observed count image, a Poisson likelihood is computed.
The likelihoods for each input band and image are combined to give a total Poisson likelihood.
Any Bayesian priors on the model parameters or components are added to this likelihood.

Typically, the user fits the model to the data to find the parameters to start an MCMC analysis.
MCMC is used to obtain a chain of parameters, optionally using multiple CPU cores to speed up the analysis, here using the emcee affine invariant ensemble sampler \citep{ForemanMackey2013}.
This posterior probability distribution of the chain parameters can be examined directly.
Alternatively, the chains can be post-processed to create other outputs.
These include X-ray images in each band and residual maps and profiles, showing the distribution of the difference between the data and models extracted from the chain.
The maps can also be binned using Voronoi tessellation for clarity by adjusting for the local signal-to-noise ratio.
The parameters in the chain can be converted into profiles of physically interesting quantities and their uncertainties for each cluster.
The median profiles and their range can be calculated for the primary parametrization (gas density and temperature). Many further derived quantities can be calculated, including pressure, entropy, luminosity, flux, X-ray count rates, mean radiative cooling time, gas mass, total hydrostatic mass (if fitting a hydrostatic model), cooling flow mass deposition rate, and $Y_X$.

\end{appendix}
\end{document}